\documentclass[fleqn,usenatbib]{mnras}
\pdfoutput=1

\usepackage[T1]{fontenc}
\usepackage{ae,aecompl}
\usepackage{color, soul}
\usepackage{verbatim}
\usepackage{units}
\usepackage{subfig}
\usepackage{pdflscape}
\usepackage{svg}
\usepackage{bm}
\usepackage{stfloats}
\usepackage{float}
\usepackage{multicol}
\usepackage{hyperref}

\usepackage{graphicx}	
\usepackage{amsmath}	
\usepackage{amssymb}	
\usepackage{epstopdf}

\newcommand\clearrow{\global\let\rowmac\relax}
\newcommand{\cmmnt}[1]{\ignorespaces}

\defcitealias{Renzo2019}{R19}
\defcitealias{Fryer2012}{F12}

\title[Hyper-Runaway Stars from Binary Supernovae]{Core-Collapse Supernovae in Binaries as the Origin of Galactic Hyper-Runaway Stars}

\author[Evans et al.]{
F. A. Evans$^{1}$\thanks{E-mail: evans@strw.leidenuniv.nl},
M. Renzo$^{2}$,
E. M. Rossi$^{1}$ \\
$^{1}$Leiden Observatory, Leiden University, PO Box 9513, NL-2300 RA Leiden, The Netherlands\\
$^{2}$Center for Computational Astrophysics, Flatiron Institute, 162 5th Ave., New York, NY 10010 USA \\
}

\date{Accepted XXX. Received YYY; in original form ZZZ}

\pubyear{2020}

\begin{document}
\label{firstpage}
\pagerange{\pageref{firstpage}--\pageref{lastpage}}
\maketitle

\begin{abstract}
  Several stars detected moving at velocities near to or exceeding the
  Galactic escape speed likely originated in the Milky Way disc. We
  quantitatively explore the `binary supernova scenario' hypothesis,
  wherein these `hyper-runaway' stars are ejected at large peculiar velocities when
  their close, massive binary companions undergo a core-collapse
  supernova and the binary is disrupted. We perform an extensive suite
  of binary population synthesis simulations evolving massive systems
  to determine the assumptions and parameters which most impact the
  ejection rate of fast stars. In a simulation tailored to eject fast
  stars, we find the most likely hyper-runaway star progenitor binary is 
  composed of a massive ($\sim$$30\,M_{\odot}$) primary and a
  $\sim$$3-4\,M_{\odot}$ companion on an orbital period that shrinks to
  $\lesssim$1\,day prior to the core collapse following a common
  envelope phase. The black hole remnant formed from the primary must
  receive a natal kick $\gtrsim$1000 $\mathrm{km\ s^{-1}}$ to disrupt the binary
  and eject the companion at a large velocity. We compare the fast
  stars produced in these simulations to a contemporary census of
  early-type Milky Way hyper-runaway star candidates. We find that
  these rare objects may be produced in sufficient number only when
  poorly-constrained binary evolution parameters related to the
  strength of post-core collapse remnant natal kicks and common
  envelope efficiency are adjusted to values currently unsupported --
  but not excluded -- by the literature. We discuss observational
  implications that may constrain the existence of these putative
  progenitor systems.
  
\end{abstract}

\begin{keywords} stars: kinematics and dynamics, massive -- supernovae: general -- binaries: general
\end{keywords}



\section{Introduction}

In recent years, works have reported with an increasing frequency
detections of early-type main sequence (MS) stars in the Galactic halo
moving at very high velocities, near to or exceeding the Galactic
escape velocity at their position \citep[e.g.,][and references
therein]{Brown2005, Hirsch2005, Edelmann2005, Brown2006, Brown2009,
  Brown2012, Brown2014, Zhong2014, Huang2017, Irrgang2019,
  Koposov2019}. For reference, the Galactic escape velocity in the
Solar Neighbourhood is $\sim$530 $\mathrm{km\ s^{-1}}$ \citep{Piffl2014,
  Williams2017} and falls to $\lesssim$400 $\mathrm{km\ s^{-1}}$ beyond 50 kpc
from the centre of the Milky Way \citep{Williams2017}. The population
of high velocity stars is increasing further in the \textit{Gaia}
era \citep{Gaia2016,Gaia2018}, with
the
European Space Agency satellite
providing precise astrometry for billions of Milky Way sources. The
short-lived nature of early-type stars and the dearth of star formation
in the Galactic stellar halo suggest that these fast stars were not likely formed in-situ in the halo; rather, they were likely accelerated and ejected from their primal birthplaces. They therefore flag extreme astrophysical and dynamical processes occurring
in the Milky Way. These processes as well as the uncertain initial conditions
and physics governing them can be elucidated by studying the
properties and kinematics of these fast-moving stars. See \citet{Brown2015rev}
for a recent review on these objects.
    
There exist a number of mechanisms to accelerate stars to such extreme
velocities. \citet{Hills1988} predicted that tight stellar binaries in the
Galactic Centre (GC) could be tidally disrupted by a supermassive
black hole (SMBH) lurking in the centre of the Milky Way. One member
of the binary is captured in orbit around the SMBH while its companion
is ejected with a very high velocity, $\sim$1000 $\mathrm{km\ s^{-1}}$, giving
rise to a population of so-called `hyper-velocity stars'
(HVS). Variations on this mechanism include the interaction of a
single star with a binary system consisting of two SMBHs or a SMBH
and an intermediate mass black hole
\citep[e.g.][]{Yu2003,Gualandris2005,Sesana2006,Guillochon2015}, or the disruption of
a globular cluster by a single SMBH or binary massive black hole
pair in the GC \citep{Capuzzo2015, Fragione2016}. Regardless, a GC
origin for hyper-velocity stars is a shared element of the above
mechanisms. Theoretical estimates place the HVS ejection rate from the
GC on the order of 10$^{-4}\,\mathrm{yr}^{-1}$ \citep{Hills1988, Perets2007,
  Zhang2013}. 
  
With perhaps the notable exception of S5-HVS1
\citep{Koposov2019}, the GC cannot be identified indisputably as the
origin of many HVS candidates with contemporary astrometric
measurements. However, especially given the high-precision astrometry
provided by the second data release (DR2) of the \textit{Gaia} mission
\citep{Gaia2016,Gaia2018}, the GC can be ruled \textit{out} as the
spatial origin of many HVS candidates \citep[see
e.g.][]{Irrgang2018}. Other mechanisms must therefore be invoked to
explain the extreme velocity of these stars. While tidal disruption of
infalling dwarf galaxies \citep{Abadi2009} or ejection from the Large
Magellanic Cloud \citep{Przybilla2008HVS3, Boubert2016, Boubert2017,
  Lennon2017, Erkal2019} or M31 \citep{Sherwin2008} can explain
extreme velocity stars whose past trajectories seem to point towards
extragalactic space, the most plausible origin for many high velocity star candidates
seems to be the Milky Way disc \citep[e.g.,][]{Heber2008,Silva2011, Palladino2014, Irrgang2018, Irrgang2019, Marchetti2018}. This is in
spite of the fact that theoretical studies predict GC-ejected
high-velocity stars to far outnumber disc-ejected high-velocity stars
\citep{Bromley2009,
  Kenyon2014}.

A number of processes can be invoked to explain the existence of these
disc-ejected high velocity stars. In a tight white dwarf/helium star
binary, the deposition of a critical amount of helium onto the
accreting white dwarf can result in the thermonuclear detonation of
the white dwarf -- a proposed progenitor for Type Ia supernovae
\citep[e.g.][]{Wang2009SNIa, Justham2009}. The donor star can be
ejected at a velocity unbound to the Milky Way and therefore be
observed as a hyper-velocity helium star or (eventually) a
hyper-velocity white dwarf \citep{Hansen2003, Wang2009HVS, Geier2013,
  Geier2015, Bauer2019, Neuntefel2020}. Surviving white dwarf donor companions can also be ejected at extreme velocities in the dynamically-driven double-degenerate double detonation Type Ia supernovae scenario \citep[D$^6$;][]{Shen2018}.

For \textit{main sequence} high-velocity stars seemingly ejected from the Galactic disc, on the other hand, two main processes are typically blamed. In the dynamical ejection scenario
\citep[DES, e.g.][]{Poveda1967, Leonard1990, Leonard1991, Perets2012,
  Oh2016}, exchange encounters in dense stellar systems
\citep{Aarseth1974} may eject stars at high velocities. In the binary
supernova scenario \citep[BSS; e.g.][]{Blaauw1961, Boersma1961,
  Tauris1998, Portegies2000, Tauris2015, Renzo2019}, the massive
primary in a binary system explodes in a core-collapse (CC) supernova,
disrupting the binary and ejecting its main-sequence companion with a
velocity comparable to its pre-CC orbital velocity. Both processes are
known to occur in the Milky Way \citep[][]{Hoogerwerf2001,
  Jilinski2010} and are generally thought to be responsible for the
known sample of `runaway stars' with ejection velocities $\geq 30-40\,\mathrm{km\ s^{-1}}$
\citep{Blaauw1961}, though their relative contribution is not yet well-constrained \citep[see][]{Hoogerwerf2001,Renzo2019}. With
characteristic ejection speeds on the order of a few tens of $\mathrm{km\ s^{-1}}$, it is not yet known whether these mechanisms can eject stars
on the order of hundreds of $\mathrm{km\ s^{-1}}$ with sufficient frequency to
explain the current known sample of `hyper-runaway stars' (HRSs) -
runaway stars ejected near to or above the Galactic escape velocity at
their location. While ejection velocities in the neighbourhood of
$\sim$1000 $\mathrm{km\ s^{-1}}$ are possible in both the DES
\citep{Leonard1991} and BSS \citep[e.g.,][]{Tauris1998, Tauris2015}
scenarios, these situations are thought to be rare. Recent N-body
simulations of young star clusters have found that ejections in excess
200 $\mathrm{km\ s^{-1}}$ are very rare \citep{Perets2012, Oh2016}. Binary
population synthesis models simulating a large number of binary
systems show that ejections above 200 $\mathrm{km\ s^{-1}}$ are vastly
outnumbered by ejections on the order of $\sim$10 $\mathrm{km\ s^{-1}}$
\citep{Portegies2000, Eldridge2011, Renzo2019}.
    
Extreme velocity stars are interesting beyond their status as
astrophysical oddities - the violent and uncertain physical processes
that generate them leave an imprint on their kinematics and
properties. The population of stars ejected via the BSS and their mass
and velocity distributions provide
constraints on many uncertain parameters governing binary evolution
and core-collapse supernovae - in particular the debated physics of
the common envelope phase and the nature of the natal `kicks' imparted
on compact objects produced following CC events. Stars ejected via the
DES can reveal information about the initial conditions describing
their parent clusters. Genuine hyper-velocity stars ejected from the
GC offer a convenient `back door' into studying the dust-obscured and
source-crowded GC environment \citep[e.g.,][]{Zhang2013, Madigan2014,
  Rossi2017}, in particular the origin and nature of the Milky Way's
nuclear star cluster \citep[see][for a review]{Boker2010}, the
interplay between Sgr A* and its environment \citep[see][for a
review]{Genzel2010}, and the growth of Sgr A* via tidal disruption of
former HVS companions \citep{Bromley2012}. The long journey from GC to
outer halo makes GC-ejected hyper-velocity stars intriguing dynamical
tracers for studying the size, mass and shape of the Galactic dark
matter halo \citep[e.g.,][]{Gnedin2005, Yu2007, Kenyon2008,
  Kenyon2014, Rossi2017, Contigiani2018}.

In this paper, we focus on HRSs ejected following a core-collapse event occurring
in a massive binary system, building on
the recent work of \citet{Renzo2019}, hereafter \citetalias{Renzo2019}.
They use a rapid binary population synthesis code to examine the
kinematic signatures of ejected stars. To account for uncertainties
in the initial conditions and physical processes important for binary evolution, they ran an extensive grid of simulations varying relevant assumptions. This allowed them to i) make concrete
observational predictions by showing which aspects of the kinematic
signatures are robust to model variations and ii) identify which assumptions and processes most strongly impact the kinematic observables, so that they
may be constrained by future observations. They find in general across all
model variations that the majority of binaries disrupted following the CC of the primary eject slow-moving `walkaway stars'
($v \lesssim 30 \, \mathrm{km \ s^{-1}}$), a term first introduced in
\citet{deMink2012}. They also find, however, that the absolute number of ejected companions and their velocity distribution depend intimately on assumptions about the natal kick delivered to remnant neutron stars (NSs) and black holes (BHs) due to asymmetries in the supernova explosion. While high-velocity ejections were not the main focus of the work, \citetalias{Renzo2019} remark on the presence of a minor peak in the ejection velocity distribution in the vicinity of $100 \, \mathrm{km\ s^{-1}} \lesssim v \lesssim 400 \, \mathrm{km\ s^{-1}}$ (c.f. Fig. C.2 in \citetalias{Renzo2019}), resulting from binaries which experienced a common envelope phase without significant accretion by the secondary.

In this work we employ the same simulation framework and approach as
\citetalias{Renzo2019} to focus further on these fastest ejections: we
study whether the BSS is able to eject fast stars with sufficient
frequency to match the current sample of identified HRS candidates in
the Milky Way. We vary relevant uncertain physical parameters and
determine which most strongly affect the likelihood of a system ejecting a companion at a very high velocity.

This paper is organized as follows. In Sec. \ref{sec:binary} we describe our binary population synthesis code, the physical assumptions we use in all our models and how we compare the simulation results to current HRS candidates. In Sec. \ref{sec:observations} we outline how we build and characterize a sample of known HRS candidates from the literature. In Sec. \ref{sec:results} we show how the number of observable HRSs predicted from our model runs differ from the observed sample. In Sec. \ref{sec:discussion} we discuss the implications and limitations of our findings, and our conclusions can be found in Sec. \ref{sec:conclusions}.

\section{Binary Population Synthesis Models} \label{sec:binary}

We generate and evolve mock populations of massive binary systems
using the population synthesis and evolution code \texttt{binary\_c}
\citep{Izzard2004, Izzard2006, Izzard2009}, 
based on the \texttt{binary star evolution (BSE)} code of \citet{Hurley2002} and the algorithms of \citet{Tout1997}. \texttt{BSE} itself uses the analytical fits of \citet{Hurley2000} to evolutionary models of single
stars from \citet{Pols1998}. The current version of \texttt{binary\_c} also includes
updates from \citet{deMink2013}, \citet{Schneider2014}, and
\citet{Claeys2014} for improved models of stellar rotation, stellar
lifetimes, and Roche lobe overflow mass transfer rates
respectively.

We first generate a mock fiducial population, taking the same initial
conditions and free physical parameters of \citetalias{Renzo2019}'s
fiducial model\footnote{This simulation is available at \url{http://cdsarc.u-strasbg.fr/ftp/J/A+A/624/A66/files/A_fiducial.sn.dat.gz}.}, based on reasonable assumptions consistent with the
current theoretical and observational understanding of the field. We
briefly explain these parameters in the next subsection and refer the
reader to \citetalias{Renzo2019} for more information. Following the
approach of \citetalias{Renzo2019}, we vary relevant physical
parameters one-by-one to examine their impact on the number of ejected
high-velocity stars. As we will see, however, the number of
high-velocity ejections in the fiducial simulation is quite
small. While varying individual parameters can provide valuable hints
towards which parameters are most important for high-velocity
ejections, the impacts of varying individual parameters are difficult
to verify robustly when comparing such small populations. We therefore
instead run an `optimized' simulation with all relevant free
parameters optimistically tuned in directions which favour
high-velocity ejections of companions. By setting parameters back to
their fiducial values one-by-one, we can more confidently assess their
effect on the frequency of high-velocity ejections. See Table \ref{tab:sims} for a summary of the initial
conditions and physical parameters assumed in each simulation.

\subsection{Fiducial Model Initial Conditions and Physical Assumptions}
\label{sec:fiducial_m}

Here we briefly define and state our assumptions for a number of initial conditions and binary evolution parameters in the fiducial model.

Each binary in our fiducial simulation is described by a zero-age main
sequence (ZAMS) mass of the primary ($M_1$), a ZAMS mass ratio
between stars ($q \equiv M_2 / M_1$) and a ZAMS orbital period
$P$. We assume that these parameters are independent from each other,
although see
\citet{Moe2017}.
We generate systems on a grid in this space and later statistically
weigh each system, assigning each a probability according to our assumed distributions of these
parameters, $p_{\rm i} \propto f(M_1) f(q) f(P)$ (see also
\citetalias{Renzo2019}, Appendix B). We select primary masses
logarithmically-spaced in the range
$7.5\,\rm{M_\odot} \leq M_1\leq100 \, \rm{M_\odot}$ and statistically weigh the
systems by a \citet{Kroupa2001} initial mass function. ZAMS mass
ratios are selected at uniform intervals in the range
$0.1\leq q \leq 1$ assuming a probability distribution
$f(q) \propto q^{\kappa}$. In the fiducial simulation we assume a flat
$q$ distribution, i.e. $\kappa=0$ \citep[see e.g.][]{Kuiper1935,
  Sana2012, Kobulnicky2014}. We draw orbital periods
logarithmically-spaced in the range
$1.41\,{\rm days} \leq P \leq 3.16 \times 10^5$ days with a
probability distribution $f(P) \propto P^{\pi}$. For $M_1<15\,\rm{M_\odot}$
we assume a flat distribution ($\pi=0$) in $\log_{10}{P}$
\citep{Opik1924, Kobulnicky2007} while when $M_{1}\geq15\,\rm{M_\odot}$ we
assume a power law distribution in $\log_{10}{P}$ with a slope
$\pi=-0.55$ \citep{Sana2012}. The lower limit on $\log_{10}P$ is also
derived from \citep{Sana2012}, while we choose an upper limit
sufficiently large to include non-interacting binaries as
effectively single stars \citep{deMink2015}.  The orbits are always assumed to be
initially circular, since close orbits are expected to circularize due
to tides or mass transfer before the first CC event
(\citealt{Belcynski1999, Hurley2002}, although see also
\citealt{Eldridge2009}). For all stars in the fiducial run we assume
a
metallicity of $Z=0.02$ \citep{Anders1989}.

We calculate Roche lobe overflow (RLOF) mass transfer rates using the
algorithm of \citet{Claeys2014}. In this fiducial simulation, the mass transfer
efficiency $\beta_{\rm RLOF}$ -- i.e. the relative ability of an accretor with mass $M_{\rm acc}$ to accept mass lost
by a donor at a rate $\dot{M}_{\rm don}$ -- is set to
$$\beta_{\rm thermal}=  \min\left(\sigma
  \frac{\dot{M}_{\rm KH,acc}}{\dot{M}_{\rm don}}, 1\right),$$ where
$\dot{M}_{\rm KH,acc}$=$M_{\rm acc}/\tau_{\rm KH}$, $\tau_{\rm KH}$
is the Kelvin-Helmholtz timescale, and we set the parameter $\sigma$=10
\citep[see][]{Hurley2002,Claeys2014,Schneider2015}.

The matter that is not accreted during the mass transfer leaves the
system. We assume it carries with it the specific angular momentum $h$
of the accretor,
i.e. $h=\gamma_{\rm RLOF}J_{\rm orb}/(M_{\rm don}+M_{\rm acc})$, where
$J_{\rm orb}$ is the total angular momentum, $M_{\rm don}$ is the
donor mass, and the loss parameter
$\gamma_{\rm RLOF}=M_{\rm don}/M_{\rm acc}$
\citep[see][]{Soberman1997, VandenHeuvel2017}.

By the time the primary first fills its Roche Lobe, we
  assume the system enters a common envelope phase evolution
  \citep[CE;][]{Paczynski1976} if the accretor star is sufficiently
  low-mass compared to the donor star
  ($M_{\rm acc}$/$M_{\rm don}<q_{\rm crit}$). This critical mass
ratio threshold $q_{\rm crit}$ depends on the evolutionary stage of
the donor star. We choose $q_{\rm crit, A}=0.65$, $q_{\rm crit, B}=0.4$
and $q_{\rm crit, RSG}=0.25$ when the donor star is a main sequence
\citep[Case A;][]{deMink2007}, Hertzsprung gap (Case B), 
and red supergiant or core
  helium-burning 
star, respectively. We note that this last $q_{\rm crit, RSG}$ value
might be relatively low \citep{Claeys2014}, i.e. the number of CE
events with giant donors might be overestimated. The conclusions of this work are insensitive to variations of this value.

Once common envelope
evolution begins, we model the evolution of the system using the
$\alpha_{\rm CE} \lambda$ formalism \citep[see e.g.][]{Webbink1984,
  deKool1990, 
  deMarco2011}, wherein a fraction
$\alpha_{\rm CE}$ of the change in orbital energy $\Delta E_{\rm orb}$ from the
inspiraling binaries is used to eject the envelope. For the binding
energy parameter $\lambda$ we use the analytic fit to the $\lambda_g$
values of \cite{Dewi2000}, which do not include the thermal energy of
the envelope and its potential energy due to ionization. We assume in our
fiducial simulation that $\alpha_{\rm CE}$=1, i.e. all the liberated
energy from the orbital inspiral is transferred to the envelope, which
escapes to infinity at precisely the escape velocity from the
system. Values $\alpha_{\rm CE}>1$ can be used to mimic the inclusion
of other sources of energy (such as accretion luminosity,
recombination energy, or nuclear burning) to aid the ejection of the
envelope \citep[e.g.,][]{Ivanova2002,deMarco2011, Ivanova2013}.

We model the reaction of a system to a CC event following
\citet{Tauris1998}. We assume that supernova ejecta leave the
exploding star instantaneously since the ejecta speed is much larger
than the orbital velocity of the binary. The mass loss from the
companion due to stripping and ablation of its envelope as a result of
the ejecta impact is treated following fits to the simulations of
\citet{Liu2015} with a companion mass of
$M_{2}=3.5\, \mathrm{M_\odot}$.

The distribution of radio-pulsar proper motions \citep{Shklovskii1970,
  Gunn1970} was the first piece of indirect evidence suggesting that
asymmetries in supernova explosions could impart ``natal kicks" to
neutron stars at formation
\citep{Shklovskii1970}.
Both asymmetric neutrino fluxes \citep{Woosley1987,Socrates2005} and
large-scale density and velocity asymmetries in the star pre-collapse
\citep[e.g.,][]{Janka1994,Burrows1996} have been invoked to explain
this natal kick, with the hydrodynamic-induced kick explanation
currently more favoured over the asymmetric neutrino emission
explanation \citep{HollandAshford2017, Katsuda2018}.

We draw natal kick magnitudes following \citet{Hobbs2005}, who employ pulsar
proper motions to infer that NS natal kicks follow a
Maxwellian distribution with a root mean squared dispersion
$\sigma_{\rm kick}=265 \, \mathrm{km\ s^{-1}}$. We draw natal kick
directions isotropically in the frame of the collapsing star
\citep[see e.g.][]{Wongwathanarat2013}.

Since a large amount of ejecta can fall back onto the remnant
following the CC, final natal kick magnitudes must be modulated
accordingly, i.e. $v_{\rm kick}\rightarrow v_{\rm kick}(1-f_{\rm b})$,
where $f_{\rm b}$ is the fallback fraction. We calculate $f_{\rm b}$ based on the carbon-oxygen core mass of
the collapsing star following the `rapid supernova' algorithm of \citet[][hereafter
\citetalias{Fryer2012}, see their Eq.~16]{Fryer2012}, which also sets the mass of the
remnant. The remnant type is set
  by its mass; above a maximum NS mass
  $M_{\rm NS,max}=2.5 \, \mathrm{M_\odot}$, the remnant is designated
  a black hole. The \citetalias{Fryer2012} algorithm generally prescribes
  small (large) fallback fractions when the remnant is a neutron star
  (black hole) and therefore large (small) kick velocities.

In the event the natal kick launches the remnant directly
towards the main sequence companion, we assume that the two bodies coalesce if the
remnant enters the envelope of the companion \citep{Leonard1994},
i.e. if the post-CC periastron is less than the companion
star's radius. We calculate the post-CC periastron distance following
Eq. 57 of \citet{Tauris1998}, accounting for the change in
momentum of the companion due to the impact of the expanding supernova
shell and the stripping and ablation of the companion's envelope. We assume a shell velocity of $8200 \, \mathrm{km\ s^{-1}}$ and
an escape velocity from the companion of $800 \, \mathrm{km\ s^{-1}}$
\citep{Tauris1998}. We fit to Table 1 of \citet{Wheeler1975} to
determine ablation and stripping mass fractions, assuming the
companion is an $n=3$ polytrope. In all, the impact of the supernova ejecta on the companion is a relatively small contribution to its total post-CC velocity, $\lesssim 10 \, \mathrm{km\ s^{-1}}$ \citep{Tauris1998, Liu2015}.

In total we draw 50 $M_{1}$'s x 50 $q$'s x 100 $P$'s for a total of 250,000 binary systems. As we also draw 20 natal kick magnitudes and directions as well, we end up with a mock population of $\sim$5,000,000 evolved binaries.

\subsection{Optimised Model Initial Conditions and Physical Assumptions}
\label{sec:optimised_m}
In addition to the fiducial model we also run a simulation with
initial conditions and physics tuned to enhance the occurrence of high velocity ejections of companions from disrupted binaries. We discuss in Sec. \ref{sec:param_discussion} whether these tunings are consistent with contemporary observations and theory. If the size of the currently-known sample of HRS candidates cannot be matched by even this ad-hoc scenario, the BSS mechanism can be conclusively ruled out as a significant provenance for Milky Way HRSs. When a binary system is disrupted, the ejection velocity of the companion star is similar to its pre-CC orbital velocity (\citealt{Blaauw1961,Eldridge2011}; \citetalias{Renzo2019}). The philosophy behind many of our changes to the initial conditions and physical parameters is therefore a) to increase the companion's pre-CC velocity as much as possible, and b) to increase the rate of binary disruption. Increasing the companion's orbital velocity implies producing harder binaries, which decreases the ease with which binaries are ionized. Indeed, tighter binaries may preferentially merge or remain bound and we therefore attempt to enhance the occurrence of physical configurations that specifically avoid those fates.

Stable RLOF mass transfer can lead to orbital widening as the system tends towards equal star mass. We suppress this by setting $\beta_{\rm RLOF}=0$. We also set $\gamma_{\rm RLOF}=1$ to increase the amount of angular momentum sapped from the system by escaping material compared to our fiducial setup.

Since common envelope evolution leads to orbital tightening, we want to encourage the occurrence of this phase but simultaneously avoid binary mergers. We achieve
this by choosing $\alpha_{\rm CE}=10$ -- the maximum of the
$\alpha_{\rm CE}$ range suggested by \citet{Hurley2002} -- and also
choosing $q_{\rm crit}=1$ regardless of donor type. A
CE efficiency greater than 1 might be achieved by
tapping extra energy sources, e.g. in the thermal motions and
ionization state of the envelope \citep[][see also
Sec. \ref{sec:disc:alpha}]{deMarco2011, Ivanova2013,
  Ivanova2013b}.

Another physical ingredient that can in principle lead to tighter binaries is metallicity.  At fixed stellar mass, stellar radii are smaller in metal-poor stars \citep[see e.g.][]{Burrows1993,
    Pols1998}. Binary systems composed of smaller stars enter the mass transfer stage later and experience less orbital widening, and are also able to orbit at smaller separations without merging both of which lead to faster orbital velocities (\citetalias{Renzo2019}). In the optimized simulation we set the total stellar metallicity to $Z=0.0063$, one-half dex lower than our fiducial value of $Z=0.02$.
  
  Finally, to increase the frequency of binary disruption, we boost the Maxwellian kick distribution dispersion to $\sigma_{\rm kick}=1000\, \mathrm{km\ s^{-1}}$ and set all fallback fractions to zero. We also set both the ZAMS  mass ratio and initial log-period distribution slopes to $\pi=\kappa=-1$ to enhance the frequency of configurations where the companion has a large orbital velocity at ZAMS.

\subsection{Other Model Variations} \label{sec:othermodels}

Our predictions depend on initial conditions for the binary systems
and parametrized assumptions for binary evolution physics. In addition
to the fiducial and optimized simulations, we run simulations where we
vary these free parameters one-by-one\footnote{Upon publication, all
  simulations will be publicly available at the DOI:
  \href{https://dx.doi.org/10.5281/zenodo.3860055}{10.5281/zenodo.3860055}}. We set one parameter to its fiducial value
while keeping all others set to their optimized value to explore the
impact each parameter has on the rate of high-velocity star ejections.
  
  To determine whether a common envelope efficiency $1<\alpha_{\rm CE}<10$ is more effective than $\alpha_{\rm CE}=10$ at encouraging orbital tightening while avoiding mergers, we additionally perform a run with all parameters set to their optimized value except $\alpha_{\rm CE}$ is set to 5.
  
  To explore the possibility  that the low-mass cores of less massive NSs experience smaller natal kicks as suggested by some authors \citep[e.g.,][]{Katz1975, Arzoumanian2002, Pfahl2002, Podsiadlowski2004, Verbunt2017, VignaGomez2018},
  we additionally perform a
  run with all parameters set to their optimized value except the
  natal kick is drawn from a double-peaked Maxwellian distribution similar to \citet{VignaGomez2018} (see also \citetalias{Renzo2019}). Following \citet{Pfahl2002} and \citet{Podsiadlowski2004}, we draw from a kick distribution with $\sigma_{\rm kick, low}=30 \,\mathrm{km\ s^{-1}}$ for remnants less massive than 1.35 $\rm{M}_\odot$ \citep{Schwab2010, Knigge2011}  while the kick distribution for more massive remnants remains at $\sigma_{\rm kick, high}=265 \, \mathrm{km\ s^{-1}}$.

Table \ref{tab:sims} summarizes the initial conditions and parameters assumed in each simulation.

\subsection{Comparisons to Data}
Each simulation contains a number $N_{\rm f}$ of  systems which eject a companion at a high velocity, here defined as $v_{\rm ej} \geq 400 \, \mathrm{km\ s^{-1}}$. Properly accounting for the probabilities of each system and the finite stellar lifetimes of the ejected companions, we convert this to a number $N_{\rm f, now}$ of stars in the Galaxy today with the same cut in ejection velocity. We estimate the number of systems -- consisting of either isolated stars or binaries -- that are formed per unit time in the Milky Way as $\langle SFR \rangle$/$\langle M\rangle$: we assume a constant star formation rate of $\langle SFR \rangle = 3.5 \, \mathrm{M_{\odot}\ yr^{-1}}$ \citep[see e.g.][]{Dominik2012} and the probability-weighted mean mass of a system is calculated as
\begin{align}
    \langle M\rangle = & \langle M_1(1+q)\rangle = \langle M_1\rangle + \langle qM_1\rangle \\
    \begin{split}
      \langle M\rangle = &\frac{\int_{0.01}^{100}M{_1}f(M_1)dM_1}{\int_{0.01}^{100}f(M_1)dM_1} + \\
            &\frac{\int_{2}^{100}M{_1}f(M_1)dM_1}{\int_{0.01}^{100}f(M_1)dM_1}\frac{\int_{0.1}^{1}qf(q)dq}{\int_{0.1}^{1}f(q)dq} \; ,      
    \end{split}
    \label{eq:meanM}
\end{align}
where $f(M_1)$ is a \citet{Kroupa2001} initial mass function, $f(q) \propto q^{\kappa}$ is the distribution of ZAMS mass ratios, and where we assume a binary fraction of 0 for $M_1<2 \, \rm{M}_\odot$ and 1 for $M_1\geq2 \, \rm{M}_\odot$. Eq. \ref{eq:meanM} gives $\langle M \rangle =0.64 \, \rm{M}_\odot$ for $\kappa=-1$ and 0.68 $\rm{M}_\odot$ for $\kappa=0$. Therefore, the birth rate of systems is $\langle SFR\rangle$/$\langle M\rangle  \approx 5.1-5.5 \, \mathrm{yr^{-1}}$.
A fraction $F_{\rm f}$ of those systems eject fast stars. We calculate this by summing over the statistical weights $p_{\rm i}$ of each sampled system that produces a fast star in our simulation given its initial configuration: $p_{\rm i} \propto  f(M_1) f(q) f(P)$:
\begin{equation}
    F_{\rm f} = \sum_{i}^{N_{\rm f}} p_i \; \text{.}
    \label{eq:Ff}
\end{equation}
These fast-ejected companions survive for a probability-weighted average time $\langle t_{\rm flight}\rangle$,
\begin{equation}
   \langle t_{\rm flight}\rangle = \frac{\sum_{i}^{N_{\rm fast}}\Delta t_{\rm left,i} \: \cdot \: p_i}{\sum_{i}^{N_{\rm fast}} p_i} \: ,
\end{equation}
where $\Delta t_{\rm left, i}$ is the remaining main sequence lifetime of the $i$-th fast-ejected companion.
Putting this all together, the number $N_{\rm f, now}$ of BSS-ejected fast stars in the Milky Way at any given time in the whole sky is therefore
\begin{equation}
    N_{\rm f,now} = F_{\rm f} \frac{\langle SFR\rangle}{\langle M\rangle} \langle t_{\rm flight}\rangle \; .
    \label{eq:nfast}
\end{equation}

To roughly estimate a number $N_{\rm f, obs}$ of fast companions currently \textit{observable} in the whole sky, we cap $\Delta t_{\rm left}$ at $100 \, \mathrm{Myr}$ under the assumption that fast stars with $\Delta t_{\rm left}>100 \, \mathrm{Myr}$ will be too far (and therefore too faint) to be detected at their current distances. This cap at 100 Myr is chosen to match more or less the maximum travel time seen in the current sample of observed hyper-runaway star candidates (see Sec. \ref{sec:observations}).

An alternative method to estimate $N_{\rm f, now}$ would be to assume a CC supernova rate of $\sim$1.9 per century in the Milky Way \citep{Diehl2006}. Assuming all and only $M>8 \, \mathrm{M_\odot}$ stars undergo CC supernovae, we can re-express $F_{\rm f}$ as a fraction only of $M_1>8 \, \mathrm{M_\odot}$ systems. Doing this and swapping $\langle SFR\rangle$/$\langle M\rangle$ for the CC supernova rate in Eq. \ref{eq:nfast} yields very similar estimates for $N_{\rm f, now}$.

\begin{landscape}
\begin{table}
    \centering
    \caption{Summary of the input parameters to our binary synthesis
      models - the zero age main sequence (ZAMS) mass ratio
      distribution slope ($\kappa$), the ZAMS log-period distribution
      slope ($\pi$), the Roche lobe overflow (RLOF) mass transfer
      efficiency ($\beta_{\rm RLOF}$), the RLOF angular momentum loss
      parameter ($\gamma_{\rm RLOF}$), the critical mass ratio for
      common envelope evolution ($q_{\rm crit}$), the common envelope
      efficiency ($\alpha_{\rm CE}$), the RMS dispersion of the natal
      kick Maxwellian distribution ($\sigma_{\rm kick}$), the
      post-kick fallback fraction ($f_{\rm b}$) and the total stellar
      metallicity ($Z$). See Sec. \ref{sec:binary} for
      details.
    }
    \begin{tabular}{lccccccccc}
        \hline \hline 
        Model & $\kappa$ & $\pi$ & $\beta_{\rm RLOF}$ & $\gamma_{\rm RLOF}$ & $q_{\rm crit}$ & $\alpha_{\rm CE}$ & $\sigma_{\rm kick}$ [$\mathrm{km\ s^{-1}}$]& $f_{\rm b}$ & $Z$\\ \hline \hline 

        Fiducial & 0 & \scriptsize{\begin{tabular}{@{}c@{}} 0 for $M_1<15 \, \mathrm{M_\odot}$ \\ -0.55 for $M_1\geq15 \, \mathrm{M_\odot}$\end{tabular}} & $\beta_{\rm thermal}$ & $M_{\rm don}/M_{\rm acc}$ & \scriptsize{\begin{tabular}{@{}c@{}} 0.65 for case A\\ 0.4 for case B\\0.25 for giant donors\end{tabular}} & 1 & 265 & {\citetalias{Fryer2012}} & 0.02 \\ \hline
        
        Optimized & -1 & -1 & 0 & 1 & 1 & 10 & 1000 & 0  & 0.0063  \\ \hline
        
        Opt. BUT $\kappa$ fiducial & 0 & -1 & 0 & 1 & 1 & 10 & 1000 & 0  & 0.0063  \\ \hline
        Opt. BUT $\pi$ fiducial & -1 & \scriptsize{\begin{tabular}{@{}c@{}} 0 for $M_1<15 \, \mathrm{M_\odot}$ \\ -0.55 for $M_1\geq15 \, \mathrm{M_\odot}$ \end{tabular}} & 0 & 1 & 1 & 10 & 1000 & 0  & 0.0063  \\ \hline
        Opt. BUT $\beta_{\rm RLOF}$ fiducial & -1 & -1 & $\beta_{\rm thermal}$ & 1 & 1 & 10 & 1000 & 0  & 0.0063  \\ \hline
        Opt. BUT $\gamma_{\rm RLOF}$ fiducial & -1 & -1 & 0 & $M_{\rm don}/M_{\rm acc}$ & 1 & 10 & 1000 & 0  & 0.0063  \\ \hline
        Opt. BUT $q_{\rm crit}$ fiducial & -1 & -1 & 0 & 1 & \scriptsize{\begin{tabular}{@{}c@{}}0.65 for case A\\ 0.4 for case B\\0.25 for giant donors\end{tabular}} & 10 & 1000 & 0  & 0.0063  \\ \hline
        Opt. BUT $\alpha_{\rm CE}$ fiducial & -1 & -1 & 0 & 1 & 1 & 1 & 1000 & 0  & 0.0063  \\ \hline
        Opt. BUT $\alpha_{\rm CE}=5$ & -1 & -1 & 0 & 1 & 1 & 5 & 1000 & 0 & 0.0063  \\ \hline
        Opt. BUT $\sigma_{\rm kick}$ fiducial & -1 & -1 & 0 & 1 & 1 & 10 & 265 & 0  & 0.0063  \\ \hline
        Opt. BUT $f_{\rm b}$ fiducial  & -1 & -1 & 0 & 1 & 1 & 10 & 1000 & {\citetalias{Fryer2012}}  & 0.0063  \\ \hline
        Opt. BUT $Z$ fiducial & -1 & -1 & 0 & 1 & 1 & 10 & 1000 & 0 & 0.02  \\ \hline
        Opt. BUT bimodal kick & -1 & -1 & 0 & 1 & 1 & 10 & \scriptsize{\begin{tabular}{@{}c@{}}30 for $M_{\rm remnant}\leq1.35 \, \mathrm{M_\odot}$ \\ 265 for $M_{\rm remnant}>1.35 \, \mathrm{M_\odot}$\end{tabular}} & 0  & 0.0063  \\ \hline
        \hline
    \end{tabular}
    \label{tab:sims}
\end{table}
\end{landscape}
\section{Sample of Galactic Hyper-runaway Star Candidates} \label{sec:observations}

Our sample of Milky Way hyper-runaway candidates is derived from the Open Fast Stars Catalog\footnote{\url{https://faststars.space/}}  \citep[OFSC; see][]{Bourbert2018}, constructed using the AstroCats framework \citep{Guillochon2017}. The OFSC is a curated catalog of $\sim$500 hyper-velocity star candidates of all types, combining \textit{Gaia} astrometry and photometry with supplementary properties such as stellar types and radial velocities from the literature. Spectra for each object are provided when available, from e.g. SDSS \citep{Abolfathi2018} or LAMOST \citep{Luo2016}. 
    
    We extract full 6D observations (position, proper motions, spectroscopic distance and radial velocity)
    from the OFSC for the 68 high-velocity star candidates of spectral type A or B. For spectral
    types we refer only to the most recently published type for each
    star. We remove stars identified as subdwarfs. For some high-velocity stars identified via colour-colour or $T_{\rm eff}$ vs. log[g] selection cuts, it can be difficult to distinguish between early-type main sequence stars and lower-mass blue horizontal branch (BHB) stars since they populate similar regions of the Hertzsprung-Russel diagram \citep[see e.g.,][]{Brown2006, Heber2008BHB, Brown2009}. Projected rotational velocities and helium abundances derived via high-resolution spectra can break this degeneracy \citep[see][]{Heber2008BHB, Irrgang2018, Hattori2019}. In cases where this ambiguity exists, we assume the star to be main sequence. \textit{Gaia} proper motions are used for every
    star, though for HVS1, HVS10, HVS12 and HVS13 \citep{Brown2009} we
    also explore using the \textit{Hubble} Space Telescope (HST) proper motion
    measurements of \citet{Brown2015}, as they are more precise than
    the \textit{Gaia} measurements even when $\pm 0.5 \, \mathrm{mas \ yr^{-1}}$ is added
    in quadrature to the uncertainties as suggested by
    \citet{Brown2018}. We remove HVS11
    and HVS14 \citep{Brown2009} as well as HVS23 \citep{Brown2014},
    for whom no proper motion measurements exist in the
    literature. When multiple distance or radial velocity measurements
    are provided, we take the value most recently published as of writing. For the
    9 stars without explicit radial velocity uncertainties in
    \citet{Brown2009, Brown2012}, we assume the reported $\pm11 \, \mathrm{km\ s^{-1}}$
    average uncertainty. We do the same for LMST\textunderscore HVS12
    and LMST\textunderscore HVS19, assuming $\pm13 \mathrm{km\ s^{-1}}$ as the
    reported average uncertainty in \citet{Zhong2014}. We keep sky positions for each star fixed and perform Monte Carlo (MC) resamplings over
    the remaining astrometric parameters, assuming Gaussian
    distributions and accounting for correlation between the proper
    motion components. 
    
    We integrate 1000 MC realizations for each of our A- and
    B-type hyper-runaway candidates backwards in time up to a maximum of 1 Gyr at a
    fixed 0.1 Myr timestep in the \texttt{MWPotential2014} potential
    outlined in \citet{Bovy2015}, a three-component potential model
    consisting of a spherical power law bulge potential,
    Miyamoto-Nagai disc \citep{Miyamoto1975} and spherical NFW halo
    \citep{Navarro1996}. \texttt{MWPotential2014} assumes a circular
    velocity at the Solar position of $220 \, \mathrm{km\ s^{-1}}$ and a distance between
    the Sun and the Galactic Centre of $8 \, \mathrm{kpc}$. We assume local standard
    of rest UVW velocities of (11.1, 12.24, 7.25) $\mathrm{km\ s^{-1}}$
    \citep{Schonrich2010} and a height of the Sun above the stellar
    disc of 25 pc \citep{Bland-Hawthorn2016}. We do not consider
    uncertainties in the Solar position or velocity, though we verify
    that these do not meaningfully affect our results. For the orbital
    traceback of each MC realization we determine the location of the last disc crossing, i.e. the ($x_{\rm GC},y_{\rm GC}$) position when $z_{\rm GC}=0$ in a Galactocentric
    Cartesian frame. For each realization we also track $t_{\rm f}$ -- the
    flight time from the disc crossing location to the observed
    position -- and the ejection velocity $v_{\rm ej}$ in both the
    Galactocentric Cartesian frame and in the frame of the rotating
    disc. Ejection velocities in the corotating frame are computed
    using the rotational velocity curve associated to the
    \texttt{MWPotential2014} potential. For MC realizations dynamically bound to
    the Milky Way, we record the location, velocity and flight time of
    each and every disc crossing up until the estimated main sequence lifetime of the star.
    
    For each star in our sample we compute the probability
    $P_{\rm disc}$ of being ejected from the Milky Way disc by taking
    the fraction of MC realizations which cross the disc at a
    Galactocentric distance 1 kpc $\leq r_{\rm GC} \leq$ 25 kpc,
    imposing the cut at 1 kpc to remove stars ejected from
    the Galactic Centre and taking 25 kpc as the edge of the stellar disc \citep{Xu2015}. We also compute the probability $P_{\rm v>400km/s}$ of a `fast'
    ejection by taking the fraction of MC realizations which
    are ejected in the corotating frame at $\geq$400$ \, \mathrm{km\ s^{-1}}$. This cut at
    $400 \, \mathrm{km\ s^{-1}}$ is commonly cited as a limit for classical ejection
    mechanisms \citep[see][and references therein]{Irrgang2018,Irrgang2019}. 
 This is a somewhat arbitrary velocity cut but our conclusions on the nature of the observed HRSs do not depend on the choice of the velocity threshold as long as it remains above a few hundreds $\mathrm{km\ s^{-1}}$.
    
We select our early-type HRS candidates as those stars
in our sample for which $P_{\rm disc}$>70\,\% and
$P_{\rm v>400km/s}$>70\,\%. We additionally require that the
Galactic Centre does not lie within the 1$\sigma$ contour of the disc
crossing location distribution. Our sample with these cuts applied
results in 14 stars. The names, astrometry and spectral types of these stars can be seen in Table \ref{tab:obssample}. We show as well the probability $P_{\rm ub}$ of each HRS candidate to be unbound from the Galaxy, taken as the fraction of MC realizations whose total Galactocentric velocities exceed the \texttt{MWPotential2014} escape velocity at their position. We also include in our
sample as a special exception the oft-cited HRS candidate HD 271791, whose high space velocity, likely disc origin, and $\alpha$-element
enhancement are taken as evidence for a BSS ejection
\citep{Przybilla2008HD}. Note, however, that the natal kick velocity required in this case must be very large \citep{Gvaramadze2009}. With an ejection velocity in the corotating frame of
$390^{+70}_{-30}$ $\mathrm{km\ s^{-1}}$, it just barely fails our
$P_{\rm v>400km/s}$>70\,\% cut. 

\begin{landscape}
\begin{table}
    \centering
    \caption{Observed properties of HRS candidates with a probability
      >70\,\% of being ejected from the Milky Way disc at a
      velocity $>400 \, \mathrm{km\ s^{-1}}$ in the corotating
      frame. $P_{\rm ub}$ gives the probability of each star being
      unbound from the Milky Way (see text).
    $v_{\rm GC}$ is the star's current total velocity in a Galactic Cartesian reference frame. Stars above the solid separating line only cross the Galactic disk within the nominal stellar lifetime of a star of their mass.}
    \begin{tabular}{lccccccccccc}
        \hline
        Name & Spec. Type & (RA, Dec.) ($^\circ$) & $\mu_{\rm \alpha*}$ (mas/yr) & $\mu_{\rm \delta}$ (mas/yr) & d (kpc) & $v_{\rm radial}$ ($\mathrm{km\ s^{-1}}$) & $v_{\rm GC}$ ($\mathrm{km\ s^{-1}}$) & $P_{\rm ub}$ & Ref.\\[-0.3em]
        \hline
        B485 & A0 & (152.578, 30.341) & -0.866 $\pm$ 0.164 & -0.108 $\pm$ 0.159 & $33.0_{-3.0}^{+4.0}$	\cmmnt{33.0 $\pm$ 3.0} & 422.7 $\pm$ 1.8 & 439$_{-13}^{+16}$ & 1.0 & 1, 2 \\
        B733 & B & (222.482, 31.064) & -1.227 $\pm$ 0.080 & -4.521 $\pm$ 0.125 & 12.0 $\pm$ 1.0 & $351.4_{-2.0}^{+2.2}$\cmmnt{351.4 $\pm$ 2.2} & 460$_{-4}^{+5}$ & 0.3 & 2, 3\\    
        HVS5 & B \cmmnt{B8, A} & (139.498,	67.377) & -0.023 $\pm$ 0.195 & -1.179 $\pm$ 0.297 & 37.0 $\pm$ 4.0 & 541.5 $\pm$ 5.9 & 643$_{-6}^{+6}$ & 1.0 & 2, 4 \\
        HVS8 & B& (145.558, 20.056) & -0.972 $\pm$ 0.403 & +0.117 $\pm$ 0.407 & 36.0 $\pm$ 3.0 & $498.9_{-3.1}^{+3.3}$\cmmnt{498.9 $\pm$ 3.3} & 483$_{-32}^{+54}$ & 1.0 & 2, 3 \\
        HVS17 & B & (250.485, 47.396) & -1.034 $\pm$ 0.223 & -1.065 $\pm$ 0.364 & 35.0 $\pm$ 3.0 & 255.5 $\pm$ 3.0 & 448$_{-12}^{+19}$ & 1.0 & 2, 5 \\
        LAMOST-HVS1 & B1 & (138.027, 9.273) & -3.537 $\pm$ 0.213 & -0.62 $\pm$ 0.180 & $13.4_{-1.5}^{+1.7}$\cmmnt{13.4 $\pm$ 1.7} & 611.7 $\pm$ 4.6 & 553$_{-13}^{+13}$ & 1.0 & 6, 7, 8 \\
        LAMOST-HVS2 & B2 & (245.087, 37.794) & -2.563 $\pm$ 0.086 & -0.924 $\pm$ 0.111 & 22.2 $\pm$ 4.6 & 341.1 $\pm$ 7.8 & 506$_{-12}^{+16}$ & 1.0 & 7 \\
        LAMOST-HVS4 & B6 & (344.657, 40.001) & +0.343 $\pm$ 0.110 & -0.288 $\pm$ 0.120 & 27.9 $\pm$ 1.5 & 359.0 $\pm$ 7.0 & 574$_{-7}^{+7}$ & 1.0 & 9\\
        \footnotesize{SDSSJ013655.91+242546.0} & A & (24.233, 24.429) & -1.985 $\pm$ 0.169 & -6.688 $\pm$ 0.116 & 9.9 $\pm$ 2.0 & 300.7 $\pm$ 1.5 &  521$_{-28}^{+30}$ & 0.97 & 10, 11\\
        SDSSJ115245.91‑021116.2 & B\cmmnt{, A, A1} & (178.191, -2.188) & -0.123 $\pm$ 0.555 & +0.171 $\pm$ 0.249	& 30.4 $\pm$ 5.0 & 424.0 $\pm$ 11.0 & 384$_{-23}^{+29}$ & 0.42 & 3, 12 \\
        HD 271791 & B2/B3 & (90.616, -66.791) & -0.619 $\pm$ 0.083 & +4.731 $\pm$ 0.088 & 20.4 $\pm$ 3.9 & 441.0 $\pm$ 1.0 & 524$_{-78}^{+77}$ & 0.88 & 13 \\
        PG1610+062 & B6 & (243.222, 6.094) & -0.616 $\pm$ 0.109 & +0.176 $\pm$ 0.060 & $17.3_{-2.5}^{+2.9}$\cmmnt{17.3 $\pm$ 2.9} & 157.4 $\pm$ 7.7 & 306$_{-8}^{+8}$ & 0.00 & 14\\[-0.3em]
       \hline
        LMST-HVS9 & A6 \cmmnt{, A0} & (206.590, 30.263) & -15.795 $\pm$ 0.115 & -62.689 $\pm$ 0.067 & 1.8 $\pm$ 0.2	& -186.5 $\pm$ 0.7 & 387$_{-62}^{+63}$ & 0.02 & 15, 16\\
        LMST-HVS19 & A0 & (340.533, 7.469) & +63.831 $\pm$ 0.129 & -39.117 $\pm$ 0.086 & 1.2 $\pm$ 0.1 &	 -436.0 $\pm$ 13.0 & 399$_{-23}^{+24}$ & 0.00 & 15, 16\\
        LMST-HVS24 & A7 \cmmnt{, F0} & (320.118, -1.557) & +49.096 $\pm$ 0.145 & -48.834 $\pm$ 0.107 & 1.2 $\pm$ 0.2 & -301.3 $\pm$ 56.4 & 331$_{-47}^{+50}$ & 0.0003 & 15, 16\\[-0.3em]
        \hline
    \end{tabular}
      \scriptsize{  \textbf{References.} (1) \citet{Brown2015}, (2) \citet{Kreuzer2020}, (3) \citet{Brown2007}, (4) \citet{Brown2006}, (5) \citet{Brown2012}, (6) \citet{Zheng2014}, (7) \citet{Huang2017}, (8) \citet{Hattori2019}, (9) \citet{Li2018}, (10) \citet{Tillich2009}, (11) \citet{Alam2015}, (12) \citet{Brown2009}, (13) \citet{Heber2008}, (14) \citet{Irrgang2019}, (15) \citet{Zhong2014}, (16) \citet{Luo2016}}
    \label{tab:obssample}
\end{table} 


\begin{table}
    \centering
    \caption{Disc-crossing properties of 15 HRS candidates with a probability >70\,\% of being ejected from the Galactic disc at a velocity >400 $\mathrm{km\ s^{-1}}$ in the corotating frame. Uncertainties represent 1$\sigma$ error intervals. Stars below the solid line cross the Galactic disc more than once within the estimated lifetime of a star with their mass. For such stars we only show the disc crossing properties for the most recent crossing. $v_{\rm ej, corotating}$ denotes ejection velocities in the corotating frame of the Milky Way disc. $P_{\rm disc}$ denotes the probability that the star is ejected from a Galactocentric distance 1 kpc $\leq r_{\rm GC} \leq $ 25 kpc. $P_{\rm v>400 km/s}$ denotes the probability of that the star is ejected at $v_{\rm ej, corotating}\geq 400 \, \mathrm{km\ s^{-1}}$. }
    \begin{tabular}{llcccccccccc}
        \hline
        Name & \textit{Gaia} DR2 ID & $r_{\rm GC}$ (kpc) & $v_{\rm ej} \, (\mathrm{km\ s^{-1}}$) & $v_{\rm ej, corotating}$ ($\mathrm{km\ s^{-1}}$) & $t_{\rm flight}$ (Myr) & $P_{\rm disc}$ & $P_{\rm v>400 km/s}$\\[-0.3em]
        \hline
        B485 & 742363828436051456 & $13.4_{-2.2}^{+3.9}$ & $520_{-4}^{+4}$ & $410_{-6}^{+12}$ & $86_{-12}^{+8}$ & 0.99 & 0.99\\
        B733 & 1283080527168129536 & $10.8_{-0.5}^{+0.6}$ & $479_{-5}^{+5}$ & $459_{-8}^{+8}$ & $27_{-2}^{+2}$ & 1.00 & 1.00 \\
        HVS17 & 1407293627068696192 & $14.2_{-4.3}^{+3.7}$ & $523_{-8}^{+15}$ & $415_{-16}^{+25}$ & $64_{-5}^{+7}$ & 1.00 & 0.83\\
        HVS5 & 1069326945513133952 &  $8.9_{-3.2}^{+2.4}$ & $733_{-15}^{+30}$ & $674_{-67}^{+88}$ & $54_{-5}^{+6}$ & 1.00 & 1.00 \\ 
        HVS8 & 633599760258827776 & $16.1_{-5.3}^{+9.4}$ & $557_{-10}^{+16}$ & $443_{-22}^{+39}$ & $83_{-11}^{+12}$ & 0.83 & 1.00 \\
        LAMOST-HVS1 & 590511484409775360 & $3.2_{-1.0}^{+2.6}$ & $684_{-30}^{+20}$ & $570_{-22}^{+19}$ & $35_{-6}^{+6}$ & 1.00 & 1.00 \\ 
        LAMOST-HVS2 & 1330715287893559936 & $6.8_{-1.0}^{+2.3}$ & $600_{-11}^{+10}$ & $539_{-41}^{+25}$ & $34_{-5}^{+4}$ & 1.00 & 1.00 \\ 
        LAMOST-HVS4 & 1928660566125735680 & $8.6_{-0.6}^{+0.9}$ & $658_{-9}^{+9}$ & $464_{-9}^{+14}$ & $50_{-4}^{+4}$ & 1.00 & 1.00 \\
        SDSSJ013655.91+242546.0 & 291821209329550464 & $13.6_{-1.7}^{+1.7}$ & $532_{-27}^{+30}$ & $461_{-48}^{+43}$ & $13_{-1}^{+1}$ & 1.00 & 0.91 \\
        SDSSJ115245.91-021116.2 & 3602104614919092736 & $15.7_{-3.0}^{+6.2}$ & $444_{-16}^{+19}$ & $478_{-25}^{+22}$ & $61_{-9}^{+13}$ & 0.94 & 1.00 \\
        HD 271791 & 5284151216932205312 & $12.7_{-4.1}^{+6.4}$ & $545_{-41}^{+60}$ & 390$_{-26}^{+70}$ & 35$_{-6}^{+6}$ & 0.97 & 0.43 & \\
        PG1610+062 & 4450123955938796160 & 6.4$_{-0.9}^{+1.3}$ & 403$_{-10}^{+10}$ & 519$_{-39}^{+27}$ & 44$_{-9}^{+9}$ & 1.00 & 0.99 \\[-0.3em]
        \hline
        LMST\_HVS9 & 1455666400613902976 & $10.1_{-3.4}^{+3.9}$ & $317_{-3}^{+12}$ & $473_{-19}^{+34}$ & $422_{-168}^{+291}$ & 1.00 & 1.00 \\ 
        LMST\_HVS19 & 2715682777307050240 &  $9.6_{-2.2}^{+3.0}$ & $376_{-55}^{+50}$ & $550_{-73}^{+73}$ & $34_{-11}^{+14}$ & 1.00 & 0.99 \\ 
        LMST\_HVS24 & 2686125426556270592 &  $6.6_{-0.3}^{+0.2}$ & $355_{-49}^{+51}$ & $533_{-61}^{+63}$ & $4_{-1}^{+1}$ & 1.00 & 0.99 \\[-0.3em]
        \hline
    \end{tabular}
    \label{tab:crossings}
\end{table} 
\end{landscape}
\newpage
\begin{figure*}
    \includegraphics[width=1.86\columnwidth]{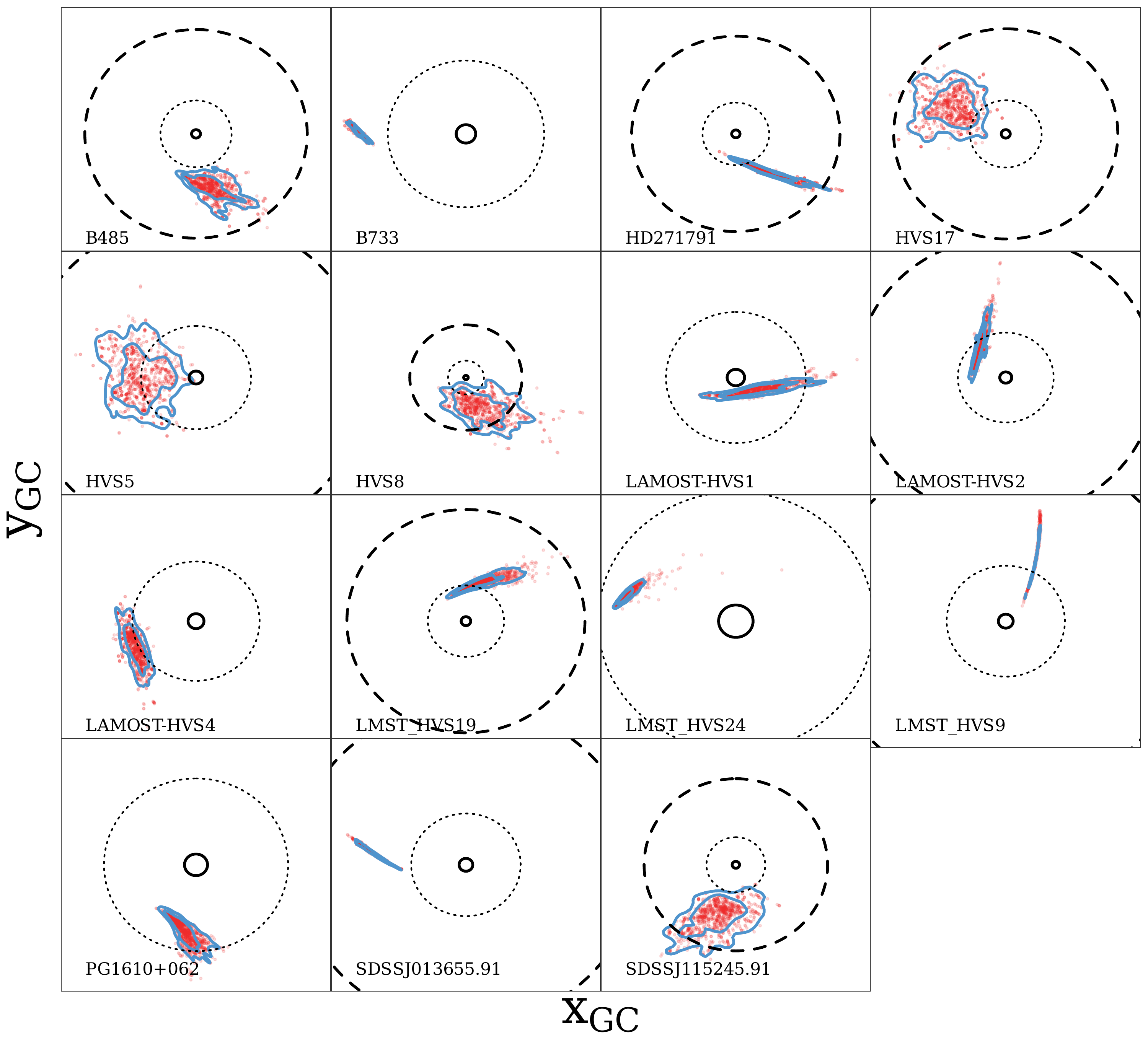}
    \caption{Red points show most recent disc-crossing locations for 1000 Monte Carlo realizations of our
      HRS candidates. Blue contours show 1$\sigma$ and
      2$\sigma$ confidence intervals of the disc-crossing location
      distributions. The solid inner, dotted middle and dashed outer
      rings have radii of 1 kpc, 8 kpc and 25 kpc
      respectively (not shown in all panels).}
    \label{fig:crossings}
\end{figure*}
\newpage
\begin{figure*}
    \includegraphics[width=2\columnwidth]{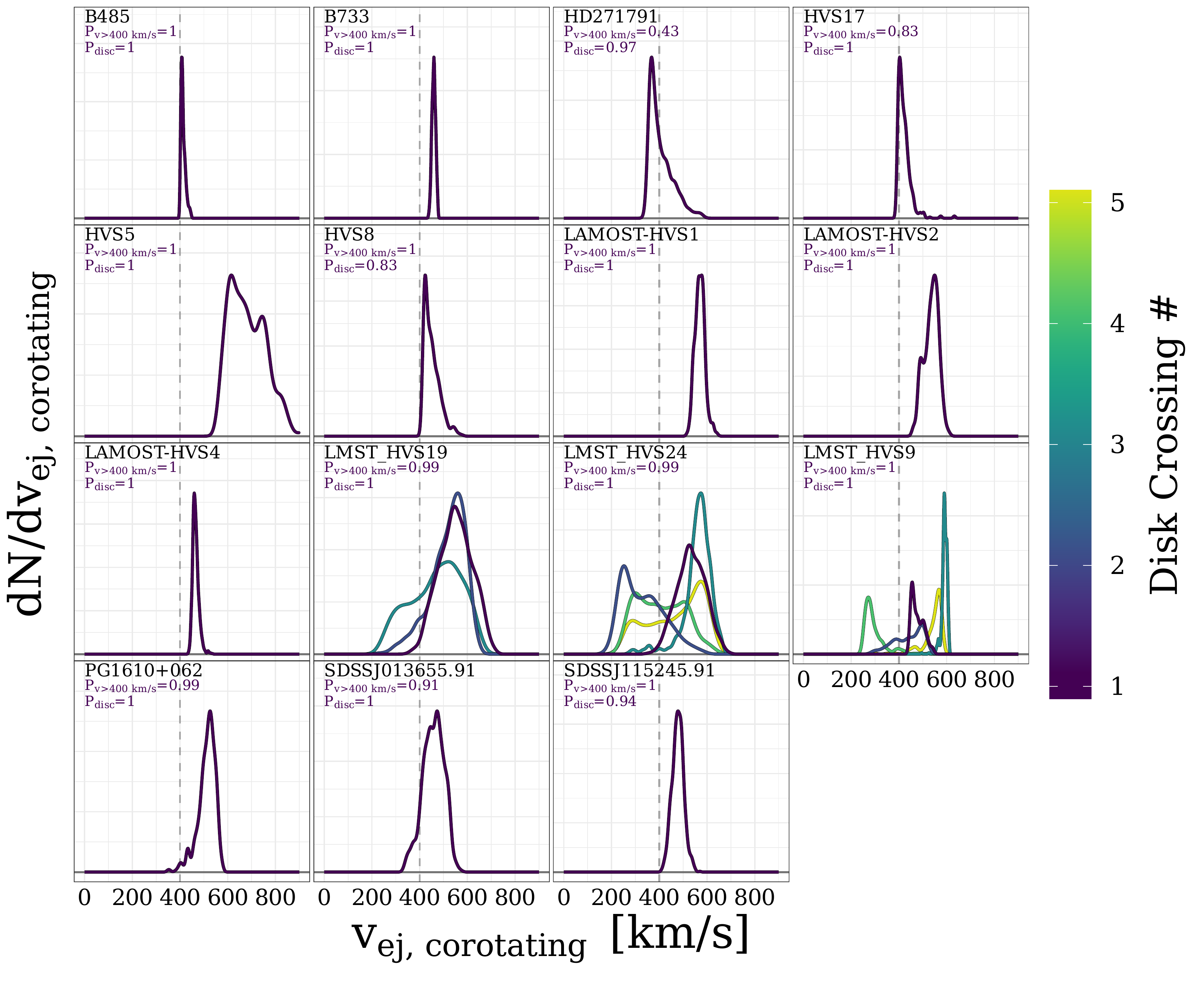}
    \caption{Distribution of ejection velocities in the corotating
      frame from the Milky Way disc for 1000 Monte Carlo (MC)
      realizations of 15 HRS candidates. Different colours denote
      separate disc crossings, starting with the most recent. Shown
      for each star are $P_{\rm v>400 km/s}$, the fraction of MC
      realizations where the star is ejected from the Milky Way disc
      during the most recent disc crossing at >400
      $\mathrm{km\ s^{-1}}$ in the corotation frame, and
      $P_{\rm disc}$, the fraction of MC realizations where the star
      is ejected during the most recent disc crossing from a
      Galactocentric distance 1 kpc $\leq r_{\rm GC} \leq$ 25 kpc. See
      Table \ref{tab:crossings} for more disc-crossing properties for
      each star.}
    \label{fig:vejs}
\end{figure*}
\newpage
\newpage
The Galactocentic Cartesian ($y_{\rm GC}$, $y_{\rm GC}$) locations of the most recent disc crossing for each MC realization of each star can be seen in Fig. \ref{fig:crossings}. They range from stars ejected from a tightly-constrained region of the disc (e.g. B733) to stars with wide spreads of possible birthplaces (e.g. SDSS J115245.91-021116.2). In Fig. \ref{fig:vejs} we show the distributions of corotating ejection velocities for each hyper-runaway candidate. 12 of our 15 stars only cross the Galactic disc once within their estimated main sequence lifetime. The possible ejection locations, velocities and flight times for the first crossing for each star are summarized in Table \ref{tab:crossings}. With the exception of LMST\_HVS9 \citep{Zhong2014, Luo2016} which is currently on an inbound orbit, every star in our sample has crossed the Milky Way disc within the last $\sim$100 Myr.

This sample of HRS candidates is neither volume- nor magnitude-limited. Since our sample is drawn from a variety of surveys and targeted searches with different sky coverage and completeness limits, we do not attempt to correct for these biases to determine a robust, corrected estimate for the number of HRS candidates in the Milky Way. Rather, this sample serves as a useful low-end estimate against which we compare the number of HRSs predicted by our binary population synthesis simulations. 

\section{Results} \label{sec:results}

\begin{figure*}
    \includegraphics[width=0.98\columnwidth]{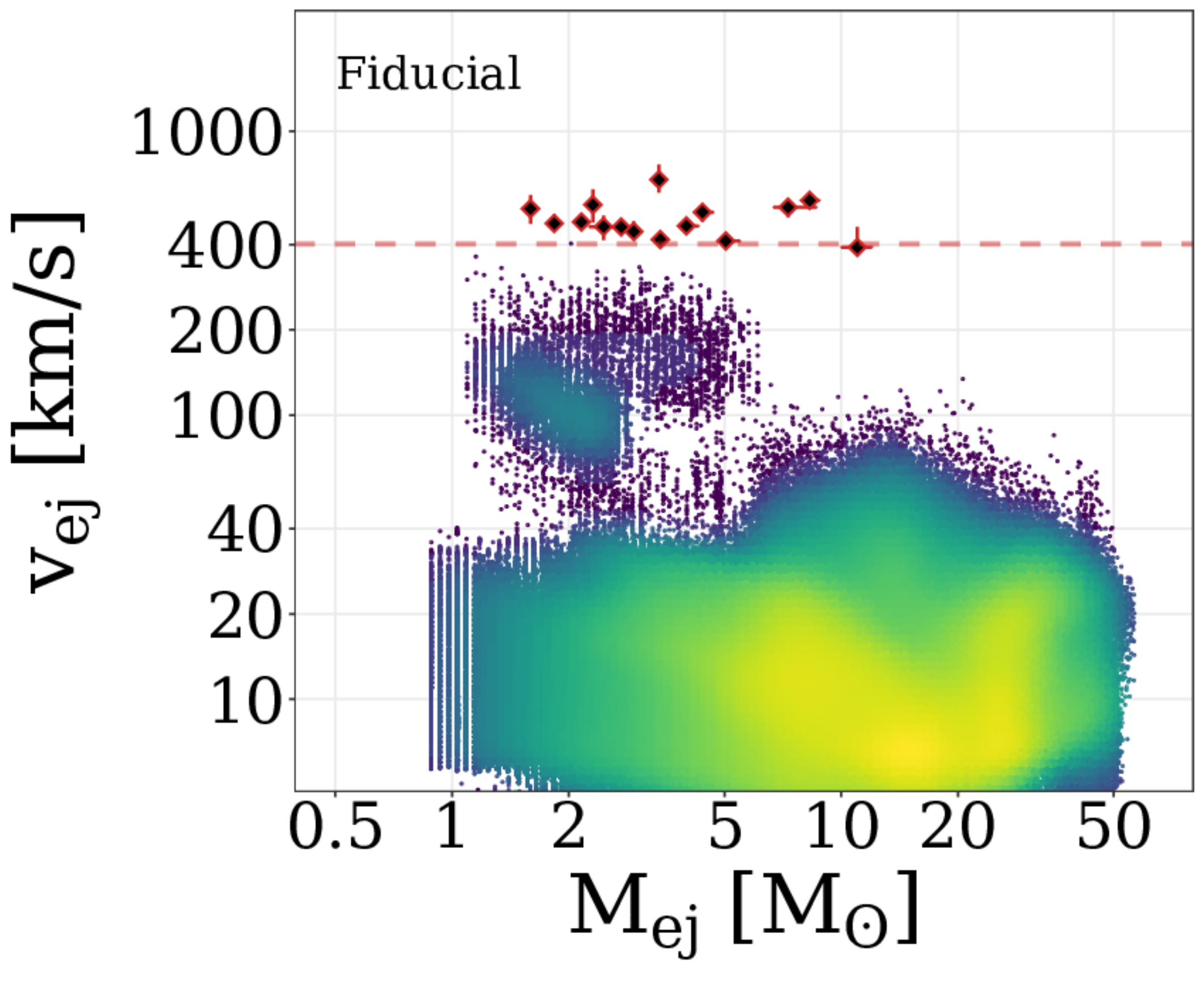}
    \includegraphics[width=0.98\columnwidth]{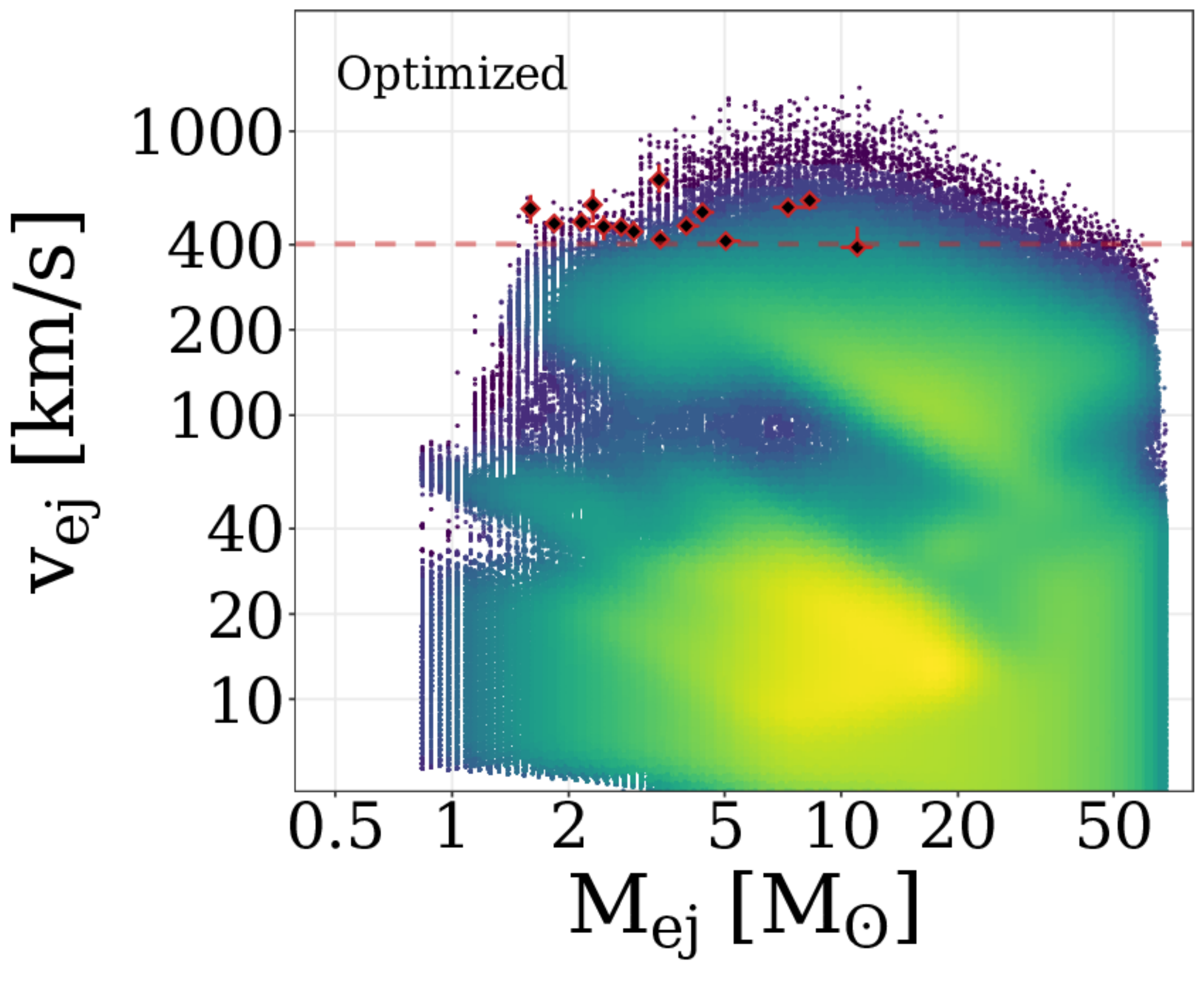}    
    \caption{Ejection velocity $v_{\rm ej}$
      of ejected main sequence companions following a core-collapse
      (CC) event as a function of their mass $M_{\rm ej}$ in our
      fiducial (left) and optimized (right) simulations. Colour is
      proportional to logarithmic number density. Note that features at $M_{\rm ej}$\,$\sim$\,$1 \, \mathrm{M_\odot}$ are artifacts of the \texttt{binary\_c} grid. Horizontal dashed line
      corresponds to $v_{\rm ej}=400 \, \mathrm{km\ s^{-1}}$. Black-and-red points show ejection
      velocities (in the disc corotating frame) and masses of 15 HRS
      candidates. Vertical error bars are 1$\sigma$ range in ejection
      velocities from 1000 Monte Carlo realizations. Horizontal error
      bars are mass uncertainties.}

    \label{fig:vejsm}
\end{figure*}

We select from our simulations the main sequence companions ejected
from disrupted binary systems, i.e. from systems which did not undergo
a merger and did not stay bound after the CC of the primary. We
further remove those systems which experience a collision between the
remnant and the companion star post-CC (see
Sec. \ref{sec:fiducial_m}). We show in Fig. \ref{fig:vejsm} how these
ejected companions populate the $v_{\rm ej}$-$M$ space for both the
fiducial (Sec. \ref{sec:fiducial_m}) and optimized
(Sec. \ref{sec:optimised_m}) simulations, with the other
models described in Sec. \ref{sec:othermodels} not shown for
brevity. We include for reference the ejection velocities (in the
corotating disc frame) and stellar masses of the 15 stars in our sample of observed HRS candidates. Notice in the fiducial model (left
panel) that the bulk of main sequence companions are ejected at
$\sim$10 $\mathrm{km\ s^{-1}}$ regardless of mass, as reported in
\citetalias{Renzo2019}. There is in fact only a single main sequence
companion in the fiducial model that is ejected at $\geq 400 \, \mathrm{km\ s^{-1}}$. By Eq. \ref{eq:nfast}, this model predicts
$N_{\rm f, now}=8$ HRSs currently in the Milky Way and one BSS-ejected HRS currently observable, clearly at odds with the current known population of HRS
candidates without even accounting for magnitude limits and sky
coverage.

For the optimized model shown in the right panel of
Fig. \ref{fig:vejsm}, notice that although the main peak of the
$v_{\rm ej}$ distribution is still at $\sim$10 $\mathrm{km\ s^{-1}}$,
there is a second peak near
v$_{\rm ej}\approx 100\, \mathrm{km\ s^{-1}}$ with a high-velocity
tail extending past our cut at
v$_{\rm ej}=400 \, \mathrm{km\ s^{-1}}$. The probability $F_{\rm f}$
described by Eq. \ref{eq:Ff} of ejecting a fast companion is
1.77$\times10^{-5}$ in the optimized simulation,
$\sim$1.4$\times$10$^{4}$ times more likely than in the fiducial
simulation. This optimized simulation predicts
$N_{\rm f, now}\simeq 1.2\times$10$^{4}$ HRSs in the Milky Way, 5590
ejected less than 100 Myr ago -- many more than the 15 HRS candidates
in our observed sample. These typically have masses greater than
$\sim$2$\, \mathrm{M_{\odot}}$, with a broad peak around
$\sim$10$\, \mathrm{M_{\odot}}$ that extends to the highest masses we
probe. At the lower end, the masses of HRSs might be in part
  limited by our choice of minimum ZAMS mass ratio
  ($q=M_2/M_1\geq0.1$, see also
  Sec.~\ref{sec:disc:caveats}). We point out that the former companions of these stars are even
more massive, with a pre-CC mass distribution peaking at
$\sim$30$\, \mathrm{M_\odot}$. The typical progenitor binaries of fast
stars are therefore quite massive overall. We come back to this later
in this section.

\begin{table*}
    \centering
    \caption{Summary of the frequency of fast ejections seen in the
      simulations described in Sec. \ref{sec:binary} and summarized in
      Table. \ref{tab:sims}. We include the per-system probability
      $F_{\rm f}$ of a $\geq400 \, \mathrm{km\ s^{-1}}$ ejection, the
      number $N_{\rm f, now}$ of
      $v_{\rm ej} \geq 400 \, \mathrm{km\ s^{-1}}$ hyper-runaway stars
      these simulations predict to be in the Milky Way today (see
      Eq. \ref{eq:nfast}), the predicted number $N_{\rm f, obs}$ of
      stars with a flight time less than 100 Myr, and the predicted
      fraction $f_{15}^{RW}$ of all $M>15\,\mathrm{M_\odot}$ stars
      which are $v_{\rm ej}>30 \, \mathrm{km\ s^{-1}}$ runaways (see
      text).}
    \begin{tabular}{lccccc}
        \hline \hline 
        Model  & $F_{\rm f}$ & $N_{\rm f, now}$ & $N_{\rm f, obs}$ & $f_{15}^{\rm RW}$ \\ \hline
        Fiducial & 1.26x10$^{-9}$\cmmnt{1.57x10$^{-7}$} & 8 & 1 & 0.54\%\\
        \hline
        Optimized & 1.77$\times$10$^{-5}$ \cmmnt{2.5$\times$10$^{-3}$} & 11767 & 5590 & 1.63\% \\
        Opt. BUT $\gamma$ fiducial &  1.79$\times$10$^{-5}$ \cmmnt{2.6$\times$10$^{-3}$} & 12492 & 5728 & 1.60\%\\
        Opt. BUT $\beta$ fiducial & 1.78$\times$10$^{-5}$\cmmnt{2.5$\times$10$^{-3}$} & 13384 & 5671 & 1.46\%\\
        Opt. BUT $\pi$ fiducial & 2.01$\times$10$^{-5}$\cmmnt{2.9$\times$10$^{-3}$} & 14575 & 6694 & 1.95\%\\
        Opt. BUT $\alpha_{\rm CE}=5$ & 1.03$\times$10$^{-5}$ & 10042 & 3739 & 0.99\% \\
        Opt. BUT $q_{\rm crit}$ fiducial & 1.72$\times$10$^{-5}$ & 11782 & 5635 & 1.79\%\\
        Opt. BUT $Z$ fiducial & 8.79$\times$10$^{-6}$\cmmnt{1.3$\times$10$^{-3}$} & 4848 & 2786 & 1.47\%\\
        Opt. BUT $\kappa$ fiducial & 7.57$\times$10$^{-6}$\cmmnt{1.1$\times$10$^{-3}$} & 3625 & 1878 & 1.94\%\\
        Opt. BUT $f_{\rm b}$ fiducial &  2.02$\times$10$^{-6}$ \cmmnt{2.9$\times$10$^{-4}$} & 5574 & 1070 & 0.38\%\\
        Opt. BUT $\alpha_{\rm CE}$ fiducial & 3.72$\times$10$^{-7}$\cmmnt{4.9$\times$10$^{-5}$} & 683 & 180 & 0.38\%\\
        Opt. BUT $\sigma_{\rm kick}$ fiducial & 1.47$\times$10$^{-7}$\cmmnt{2.12$\times$10$^{-5}$} & 226 & 63 & 0.56\%\\
        Opt. BUT bimodal kick & 1.24$\times$10$^{-7}$\cmmnt{1.78$\times$10$^{-5}$} & 119 & 48 & 0.54\%\\
        Opt. BUT $\sigma_{\rm kick}$, $\alpha_{\rm CE}$, $f_{\rm b}$ fiducial & 8.93$\times$10$^{-9}$\cmmnt{1.19$\times$10$^{-6}$} & 109 & 5 & 0.04\%\\
        \hline
        Opt. BUT $\kappa$, $\pi$, $Z$ fiducial & 4.12$\times$10$^{-6}$\cmmnt{6.0$\times$10$^{-4}$} & 1868 & 1092 & 2.19\% \\
        Opt. BUT $\kappa$, $\pi$, $Z$, $f_{\rm b}$ fiducial & 3.30$\times$10$^{-7}$\cmmnt{4.8$\times$10$^{-5}$} & 611 & 147 & 0.83\% \\        
        Opt. BUT $\kappa$, $\pi$, $Z$, $\alpha_{\rm CE}$ fiducial & 1.90$\times$10$^{-7}$\cmmnt{2.6$\times$10$^{-5}$} & 76 & 35 & 0.25\% \\
        Opt. BUT $\kappa$, $\pi$, $Z$, $\sigma_{\rm kick}$ fiducial & 8.44$\times$10$^{-8}$\cmmnt{1.2$\times$10$^{-5}$} & 59 & 27 & 0.87\% \\
        
        \hline
    \end{tabular}
    \label{tab:results}
\end{table*}

The results for the fiducial and optimized simulations as well as all
other simulations described in Sec. \ref{sec:othermodels} are
summarized in Table \ref{tab:results}. We show the
probability $F_{\rm f}$ of systems in the simulation
to eject a companion at $v_{\rm ej} \geq 400 \, \mathrm{km\ s^{-1}}$ (see
Eq. \ref{eq:Ff}) along with the predicted number $N_{\rm f, now}$ of
$v_{\rm ej} \geq 400 \, \mathrm{km\ s^{-1}}$ stars in the Milky Way today and the
number $N_{\rm f, obs}$ with flight times less than 100 Myr (see
Eq.~\ref{eq:nfast}). These results are useful in assessing which
physical parameters and assumptions in our binary synthesis models most strongly affect the rate of
high-velocity ejections. Keeping the notation of \citetalias{Renzo2019}, we also include for reference the fraction $f_{15}^{\rm RW}$ of \textit{all} $M>15\,\mathrm{M_\odot}$ stars predicted by each simulation to be runaway stars ($v_{\rm ej} \geq 400 \, \mathrm{km\ s^{-1}}$), properly accounting for the per-system probability and the duration of each evolutionary stage. This cut at $15\, \mathrm{M_\odot}$ corresponds roughly to O-type stars and makes $f_{15}^{\rm RW}$ a useful prediction against which to compare observations of massive Galactic runaways. In agreement with \citetalias{Renzo2019} we find that $f_{15}^{\rm RW}$ is robust against model variations and is limited to a few\,\% or lower. We defer a deeper discussion of O-type runaway fractions to Sec. \ref{sec:beyond}.

From Table \ref{tab:results}, it is clear that the frequency of fast
star ejections is governed by only a few parameters. Setting the RLOF
angular momentum loss parameter $\gamma$ or the RLOF mass transfer
efficiency $\beta_{\rm RLOF}$ back to their fiducial values does not
significantly change the probability for a companion to be ejected at
a high velocity. The initial stellar properties, instead, are somewhat responsible for
producing high velocity ejections. Notice that a reversion to our fiducial prescription for the ZAMS period power law slope ($\pi=-0.55$ for $M_{1}>15 \, \mathrm{M_\odot}$) actually \textit{increases} the probability to eject fast companions relative to the optimized simulation ($\pi=-1$ for all $M_1$). This is perhaps counter-intuitive, as the steeper $P_{\rm ZAMS}$ slope in the optimized simulation should result in a larger probability for the low-$P_{\rm ZAMS}$, high-$v_{\rm orb}$ systems most likely to eject fast companions. However, it also significantly increases the probability of the $P_{\rm ZAMS}\lesssim10$ day systems most likely to undergo a merger (see Sec. \ref{sec:results:progenitor}). Thus the slightly shallower $P_{\rm ZAMS}$ log-slope provided by the fiducial prescription ends up being more effective at encouraging the low-$P_{\rm ZAMS}$ HRS progenitor systems while not over-producing merging systems. Replacing our optimized $P_{\rm ZAMS}$ prescription with the fiducial prescription in the optimized simulation and all `Opt. BUT' simulations does not change the qualitative results of this study.

Two other stellar initial condition
parameters which affect the probability of a fast ejection are the
ZAMS mass ratio distribution slope $\kappa$ and total stellar
metallicity $Z$. Increasing the total stellar metallicity back to its fiducial value of
$Z=0.02$ decreases $N_{\rm f,obs}$ by 50\,\% relative to the optimized
simulation and returning $\kappa$ from its optimized value of -1 to
its fiducial value of 0 reduces the expected number of observable fast
stars by 66\,\%.

The most impactful parameters controlling the prevalence of
high-velocity ejections are the prescriptions for common envelope
evolution and post-CC natal kicks. A phase of common envelope
evolution appears indeed to be a necessary 
phase in the evolution of binaries that produce fast ejections: with
the efficiency $\alpha_{\rm CE}$ set back to its fiducial value of
unity, $N_{\rm f,obs}$ is reduced by
97\,\%. Setting $\alpha_{\rm CE}$ to 5 reduces
  $N_{\rm f,obs}$ by 33\,\%. This implies that the number of
  observable HRS $N_{\rm f,obs}$ either a)
  increases monotonically with $\alpha_{\rm CE}$ up to and possibly
  beyond $\alpha_{\rm CE}$=10, or b) peaks somewhere in the range
  $5<\alpha_{\rm CE}<10$. The distinction between these cases is of
  little consequence, both in terms of the impact on our results and
  the feasibility of having an $\alpha_{\rm CE}$ significantly above 1
  for a large population of CE systems (see
  Sec. \ref{sec:disc:alpha}). Note however, that $q_{\rm crit}$ --
also related to the CE evolution -- does not significantly affect
$N_{\rm f,now}$ when set back to its fiducial
donor-dependent value (see Table \ref{tab:sims}).
This is because systems capable of ejecting a main sequence companion
at $v_{\rm ej}\geq 400 \, \mathrm{km\ s^{-1}}$ are already biased
towards small mass ratios to maximize the orbital velocity of the
companion. 94\,\% of systems ejecting a companion at
$v_{\rm ej}\geq 400 \, \mathrm{km\ s^{-1}}$ in the optimized
simulation have an initial $q<0.4$.

With the natal kick dispersion $\sigma_{\rm kick}$ set back to its fiducial value of 265 $\mathrm{km\ s^{-1}}$ (but fallback fractions still set to zero), $N_{\rm f, obs}$ decreases by 99\,\% relative to
the optimized simulation. Implementing the double-peaked Maxwellian natal
kick distribution described in Sec. \ref{sec:othermodels} decreases
$N_{\rm f, obs}$ by a similar amount. Mediating the natal kicks by
returning to the \citetalias{Fryer2012} prescription for ejecta
fallback fractions decreases $N_{\rm f, obs}$ by 81\,\%. Together, these results demonstrate the relevance of black hole kicks for fast ejections of
companions -- black holes constitute 96\,\% of the remnants left behind by
CC events in systems ejecting a companion at
$v_{\rm ej}\geq 400 \, \mathrm{km\ s^{-1}}$ in the optimized simulation. Only 8\% of such systems would be unbound due to mass loss alone at CC \citep[see][]{Blaauw1961} -- significant natal kicks are required. The \citetalias{Fryer2012} prescription in the fiducial simulation assumes large
fallback fractions for black holes and thus very small natal
kicks. We note, however, that as long as the natal
  kick delivered to the remnant is large, our main results are
  agnostic towards the nature of the remnant (see also Sec. \ref{sec:disc:kicks}.)

Small ZAMS mass ratios allow for larger companion orbital velocities and low stellar metallicities result in small stellar radii and therefore tighter orbits, but
these contributions are minor when compared to the role of natal kicks and
CE evolution. Included in Table \ref{tab:results} is a simulation with
$\sigma_{\rm kick}$, $\alpha_{\rm CE}$ and $f_{\rm b}$ reset to their
fiducial prescriptions. Relative to the optimized simulation, the
number $N_{\rm f, obs}$ of $t_{\rm flight}<100 \, \mathrm{Myr}$ stars observable in the sky is reduced by over 99.9\,\% and clearly demonstrates the importance of
these parameters. The impact of natal kicks and CE evolution is also
independent of $\kappa$, $\pi$ and $Z$. We include in Table \ref{tab:results} a
simulation with all of $\kappa$, $\pi$ and $Z$ set to their fiducial
values. Relative to this simulation, additionally returning $\alpha_{\rm CE}$,
$f_{\rm b}$ or $\sigma_{\rm kick}$ to their fiducial prescriptions
still results in a massive reduction in the number of companions
ejected with $v_{\rm ej}\geq 400 \, \mathrm{km\ s^{-1}}$. Note as well that the
`Opt. BUT $\alpha_{\rm CE}$ fiducial' and `Opt. BUT
$\sigma_{\rm kick}$ fiducial' simulations predict a number of
$t_{\rm flight}$<100 Myr hyper-runaway stars similar to our sample of 15 observed HRS candidates, and are thus not likely to be consistent with
observations when accounting for completion and sky coverage
limits. This indicates that \textit{all} of $\alpha_{\rm CE}$,
$\sigma_{\rm kick}$ and $f_{\rm b}$ must be set significantly far from
their fiducial prescriptions for the BSS to be consistent with current
observations, as any single one set individually to its fiducial value
is inconsistent.

\subsection{The properties of progenitor binaries to fast
  ejections} \label{sec:results:progenitor} The results presented
above motivate the following picture for the production of an HRS
through isolated binary evolution. Massive binaries with quite unequal
mass ratios achieve small pre-CC orbital periods via CE evolution. As
a result, the companion to obtains a very high pre-CC orbital
velocity. RLOF does not appear to play a role for the
formation of HRS.
In the majority of cases, the binary also has an unequal pre-CC
mass ratio, with the dying star massive than its
companion. This results in the latter having a larger orbital
velocity that can more easily exceed our 400 $\mathrm{km\ s^{-1}}$ cut.
The massive primaries undergo core-collapse events,
most often leaving black hole remnants. Subsequently,
strong natal kicks are required to disrupt the binaries. The main
sequence companion is then ejected with a velocity similar to its
pre-CC orbital velocity.

\begin{figure*}
    \centering
    \includegraphics[width=\columnwidth]{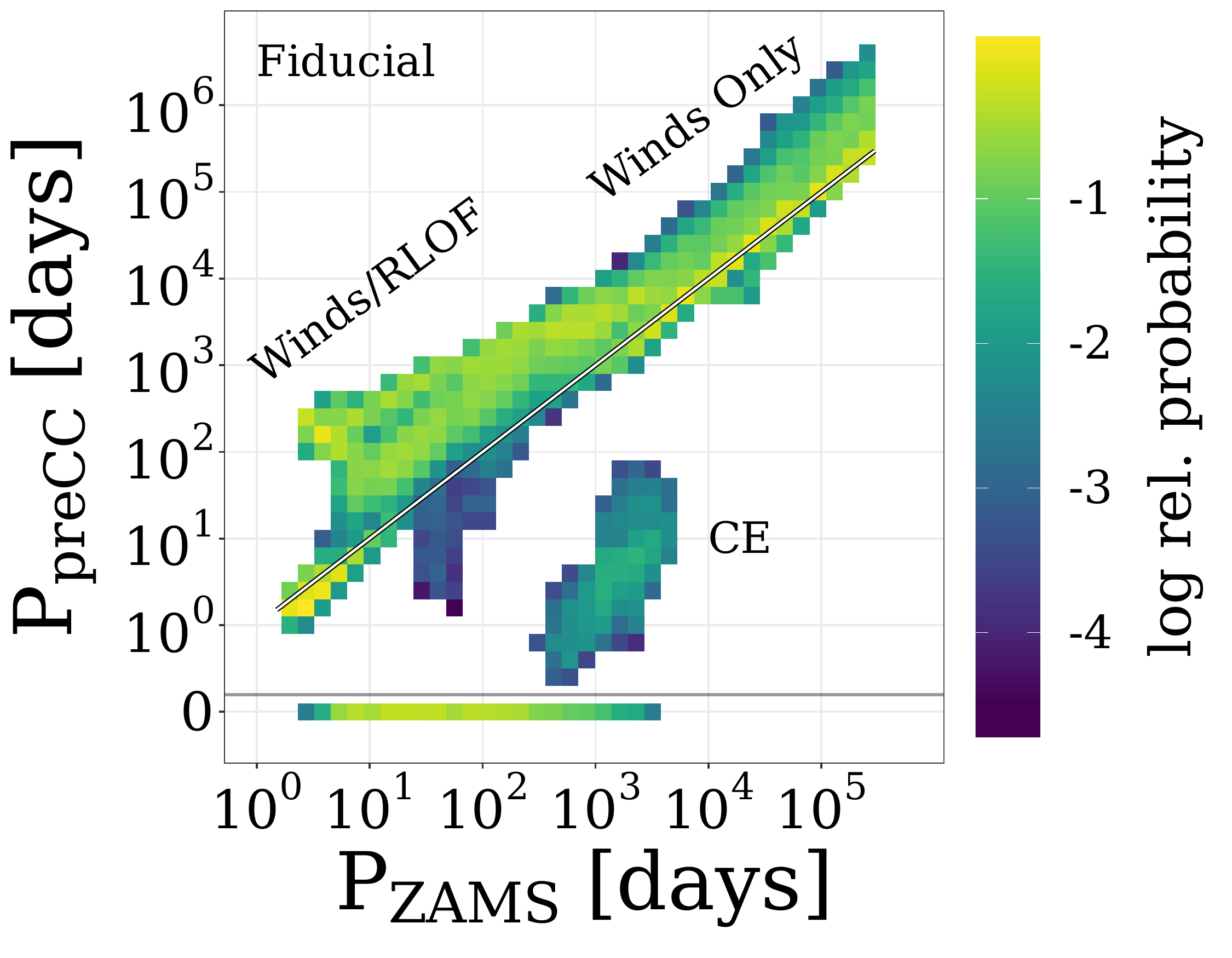}
        \includegraphics[width=\columnwidth]{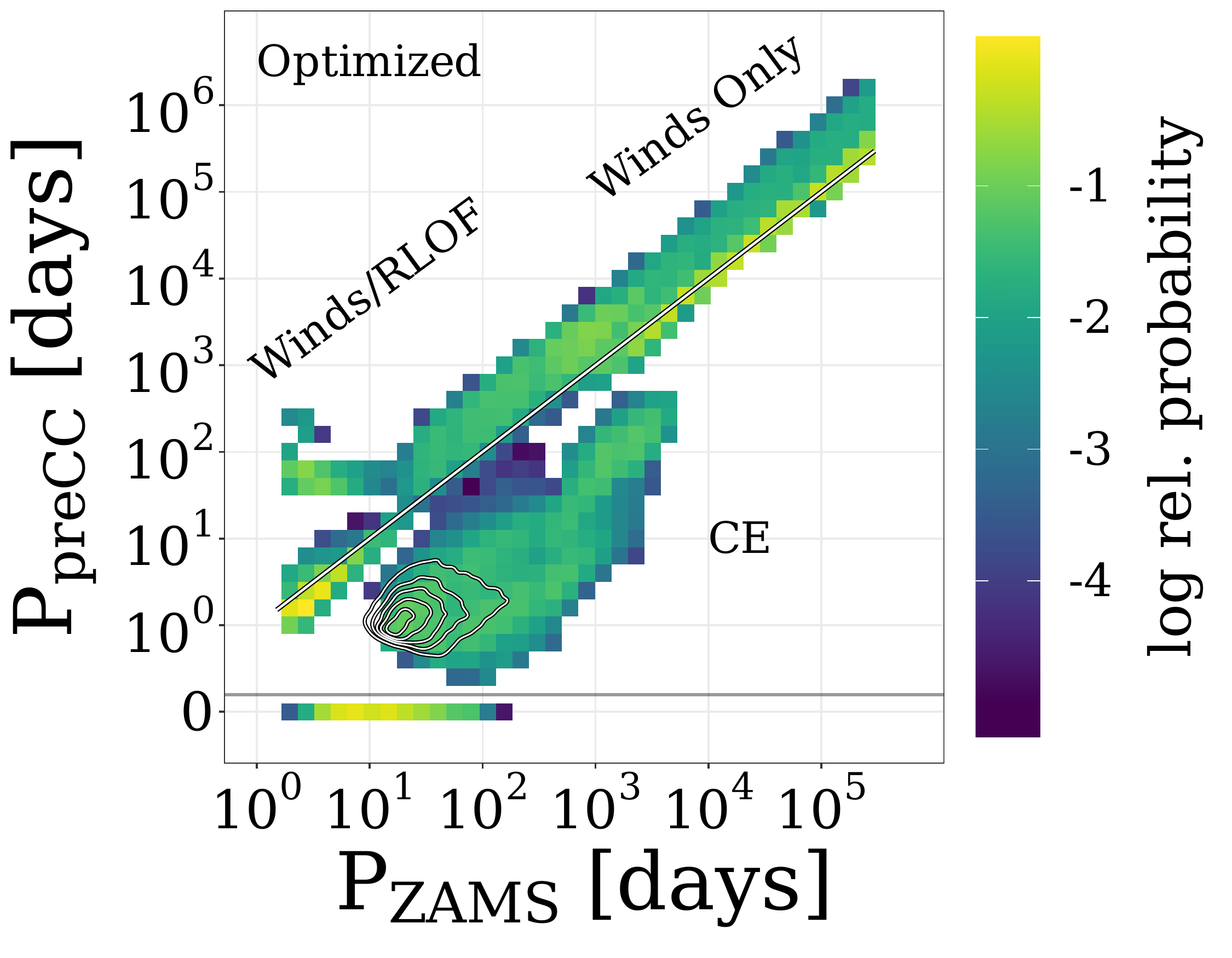}
    \caption{Distribution in the pre-core collapse period $P_{\rm preCC}$
      vs. zero age main sequence period $P_{\rm ZAMS}$ plane for binary systems in the
      fiducial (left) and optimised (right) simulations. Bins are coloured by the total probability of all systems within the bin, scaled relative to the most likely bin. Solid white
      line shows a 1:1 correspondence. Systems lying above the white line experience orbital widening via wind mass loss or Roche lobe overflow mass transfer. Systems below the line experience orbital tightening through common envelope evolution. Systems at $P_{\rm preCC}=0$ are
      systems that undergo a merger before the first core collapse
      event. White contours in the right panel show the
      10\%/30\%/50\%/70\%/90\% contours of the distribution of systems
      that eject companions at
      $v_{\rm ej} \geq 400 \, \mathrm{km\ s^{-1}}$} 
    \label{fig:Periods}
\end{figure*}

In Fig. \ref{fig:Periods} we illustrate the importance of tight binary
orbits. In this plot we show how binaries in both the fiducial (left)
and optimized (right) simulations populate the period pre-CC
vs. the period at ZAMS plane. By the
time of the first CC event, systems above the solid white line
have had their orbits widened, generally via mass loss from stellar
winds for large $P_{\rm ZAMS}$ and wind mass loss plus RLOF at
shorter $P_{\rm ZAMS}$. Systems below
the solid line have experienced orbital tightening, mainly via common envelope evolution. Systems at
$P_{\rm preCC}=0$ in this figure have merged prior to the first
CC event. The contours in the right panel show the
distribution only for systems which eject a companion at
$v_{\rm ej} \geq 400 \, \mathrm{km\ s^{-1}}$. These systems have periods of
$\sim10^1-10^2$ days at ZAMS and experience aggressive orbital shrinking due to common envelope evolution, reducing their orbital periods
to $\sim$1 day. Conversely, the
$10 \lesssim P_{\rm ZAMS} \lesssim 100$ day systems in the fiducial
simulation (left panel) either experience orbital widening or have their orbits shrunk to the point of coalescence and
experience a merger. Regardless, it is clear that the fiducial model does not possess the same reservoir of $P_{\rm preCC}{\sim}$1 day systems from which to eject
companions at large velocities.

As the common envelope efficiency $\alpha_{\rm CE}$ is increased, less orbital inspiral is required to fully eject the common envelope. It may seem counter-intuitive, then, to increase $\alpha_{\rm CE}$ in the optimized simulation, as it results in wider post-CE systems. In this study we are concerned with the very high tail of the velocity distribution of ejected companions, ejected from systems which may have otherwise merged if $\alpha_{\rm CE}$ were set lower. The priority for setting the common envelope efficiency in the optimized simulation is allowing systems to tighten while avoiding a merger. This is clear in Fig. \ref{fig:Periods}, where systems at much higher $P_{\rm ZAMS}$ experience a merger in the fiducial simulation than in the optimized simulation.

\begin{figure}
    \centering
    \includegraphics[width=\columnwidth]{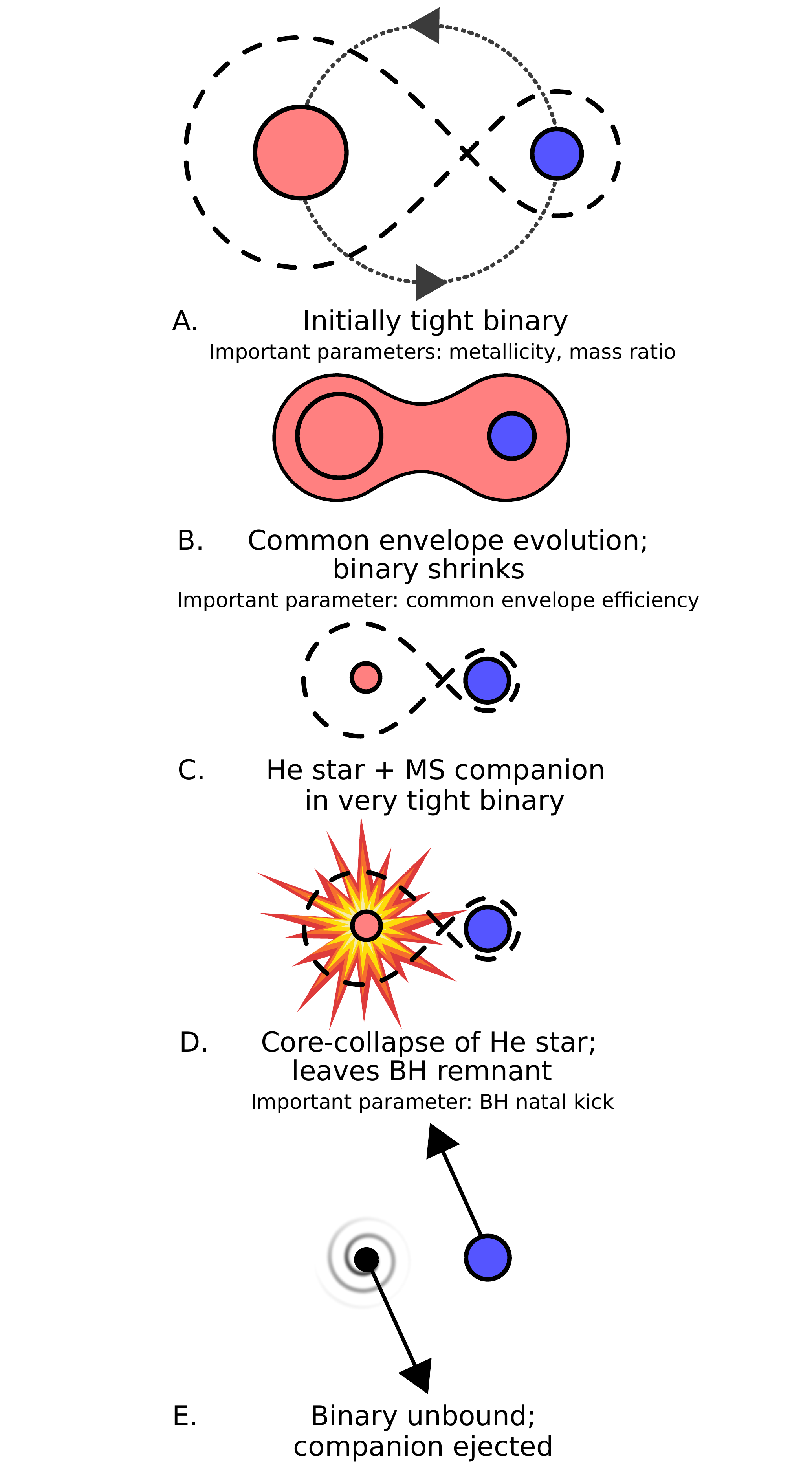}
    \caption{Schematic depicting the evolution of a typical system in the optimized simulation which ejects an HRS.
      An unequal-mass binary has a short period at zero age main sequence.
      Once the massive primary fills its Roche lobe, the system enters a common envelope phase, further tightening the orbit and increasing the velocity of the companion.
      When the primary explodes in a core-collapse supernova, the natal kick applied to the remnant black hole is strong enough to ionize the binary.
      The companion is ejected with a speed comparable to its pre-CC orbital velocity.}
    \label{fig:Cartoon}
\end{figure}

To further demonstrate the conditions and evolutionary stages required for an HRS ejection via the BSS mechanism, we present in Fig. \ref{fig:Cartoon} a schematic outlining the evolution
of the most common HRS progenitor system in the optimized simulation. Weighting by the per-system probabilities and the remaining lifetimes of ejected hyper-runaway companions, at ZAMS (phase A) the typical binary consists of a massive ($M_{1}$\,$\sim$\,$27_{-10}^{+14} \, \mathrm{M_\odot}$)
primary and comparatively less massive ($M_{2}$\,$ \sim$\,$3.2_{-1.3}^{+2.2} \, \mathrm{M_\odot}$)
companion on a short-period orbit ($P_{\rm ZAMS}$\,$\sim$\,$47_{-18}^{+40}$
days). When the primary star first fills its Roche lobe (phase B), the
companion cannot accept the material lost from the primary and the
system enters a phase of common-envelope evolution. Dynamical friction
within the common envelope tightens the binary, decreasing the
orbital period to $P_{\rm preCC}$\,$\sim$\,$0.72_{-0.21}^{+0.51}$ days. The envelope is
eventually ejected and the core of the primary is exposed as either a
naked helium-burning main sequence star or naked helium-burning
Hertzsprung gap star (phase C). At this time, the orbital velocity of the companion is already quite high ($v_{\rm orb}$\,$\sim$\,$370_{-60}^{+50} \, \mathrm{km\ s^{-1}}$). When the primary
undergoes a core-collapse event (phase D), the natal kick
delivered to the remnant black hole is very large
($v_{\rm kick}$\,$\sim$\,$1700_{-500}^{+600} \, \mathrm{km\ s^{-1}}$). This kick is sufficient to
ionize the binary (phase E) - the companion is ejected with a velocity
comparable to its pre-CC orbital velocity ($v_{\rm ej}$\,$\sim$\,$445_{-35}^{+87} \, \mathrm{km\ s^{-1}}$) while the BH is ejected at an even larger velocity, ($v_{\rm ej}$\,$\sim$\,$1500_{-500}^{+700} \, \mathrm{km\ s^{-1}}$).

\section{Discussion} \label{sec:discussion}
We have shown in the previous section that the binary supernova scenario is unlikely to contribute
significantly to the current population of observed HRS candidates, not unless the model prescriptions for a number of parameters are altered significantly from our fiducial prescriptions. In this section we determine whether these alterations are consistent with current understanding. To offer extra constraints on our models, we explore the predictions our fiducial and optimized models give for the number and kinematics of $v_{\rm ej}>30 \, \mathrm{km\ s^{-1}}$ runaway stars and of binaries that remain bound following the core-collapse of the primary. Finally we explore alternative mechanisms that may contribute to the existing population of Milky Way HRS candidates.

\subsection{Binary evolution parameters that have the largest impact} 
\label{sec:param_discussion}

We demonstrate in the previous section that the abundance of HRSs
ejected through the BSS depends intimately on prescriptions
for remnant formation (natal kick and ejecta fallback fractions), common
envelope evolution, and (to a lesser extent) the binary initial mass
ratio distribution and stellar metallicity. In this subsection, we
explore existing constraints on each of these prescriptions from the
literature, assessing whether the altered prescriptions we assume in
the optimized model are consistent -- or at least not in conflict --
with the current understanding of stellar binary evolution. We discuss
the relevant parameters in `chronological' order in the life of an
evolving binary (see Fig. \ref{fig:Cartoon}).

\subsubsection{Initial conditions: the stellar metallicity and binary initial mass ratio}

Of the relevant parameters we discuss here, a less impactful parameter which affects the frequency of
$v_{\rm ej}\geq 400 \, \mathrm{km\ s^{-1}}$ companions is $\kappa$, the slope of the ZAMS mass ratio distribution. Though progress has been made in the past decade,
attempts at constraining $\kappa$ for massive binaries remain
hamstrung by the difficulty of detecting even intermediate-mass
companions around massive stars \citep[see e.g.][]{Kobulnicky2007, Duchene2013}. Our fiducial model adopts a flat distribution,
i.e. $\kappa=0$, consistent with observations of nearby Galactic
O-type binaries \citep{Sana2012} and O-type binaries in the Cygnus OB2
association \citep{Kobulnicky2014}. However, our choice of $\kappa=-1$
in the optimized simulation is also supported by several recent
studies. \citet{Sana2013} investigate O-type binaries in the
VLT-FLAMES Tarantula Survey of the 30 Doradus region in the LMC
\citep{Evans2011} and derive $\kappa=-1.0 \pm 0.4$. Though they alter
the functional form to allow for an excess of near-equal mass
binaries, \citet{Moe2013} derive $\kappa\approx-1$ in the Milky Way,
SMC and LMC when studying eclipsing O- and B-type binaries found in the
\textit{Hipparcos} \citep{Lefevre2009}, OGLE-II
\citep{Wyrzykowski2004} and OGLE-III \citep{Graczyk2011} data sets,
respectively, though uncertainties remain large except in the case of
the LMC. Even steeper mass ratio distributions have been determined by
e.g. \citet{Dunstall2015}, who investigate B-type binaries in the
VLT-FLAMES Tarantula Survey and fit a mass ratio distribution with a
slope $\kappa=-2.8 \pm 0.8$. Therefore while our choice of $\kappa=-1$
in the optimized simulation is reasonable, larger and more complete
samples are required to determine whether this accurately describes
massive binary systems in the Milky Way.

The initial stellar metallicity also affects the frequency of HRS ejections. The effect of metallicity should not be overlooked, as evidence already exists of metal-poor stars ejected from the edge of the Milky
Way disc \citep[HD 271791;][]{Heber2008,
  Przybilla2008HD}. The total stellar metallicity in our fiducial simulation for all stars is the canonical
Solar metallicity $Z=0.02$ \citep{Anders1989} and one-half dex lower
($Z=0.0063$) in our optimized simulation. While our fiducial $Z$ is
slightly higher than modern estimates \citep[see][and references
therein]{Asplund2009}, the Sun does not appear to have an atypical
metallicity when compared to other G dwarfs
\citep{Fuhrmann2008,Holmberg2009} or B stars
\citep{Przybilla2008Bmetals} in the Solar Neighbourhood. Additionally,
moderately-massive binary systems ejecting fast companions in the
recent past would be expected to be more metal-rich than the Sun if
anything, due to chemical enrichment over the past 4.5 Gyr. In light of this, assuming a metallicity of $Z=0.0063$ over
the entire Milky Way is difficult to justify. However, pockets of
low metallicity in the Milky Way could contribute significantly
to our current observed sample of HRS candidates. Additionally, the negative radial metallicity
gradient of the Milky Way should not be neglected -- although the
stellar density drops off towards the edge of the disc, the stellar
metallicity may reach more than 0.5 dex below Solar by the edge of the
disc \citep[see][and references therein]{Lemasle2018}.

\subsubsection{Late-stage evolution: the common envelope efficiency} \label{sec:disc:alpha}

More impactful than the ZAMS mass ratio or stellar
metallicity, an important physical parameter that governs the predicted number of HRSs is the common envelope efficiency
$\alpha_\mathrm{CE}$. \citetalias{Renzo2019} found that this has
little impact on the bulk population and kinematics of runaways from
the BSS considering all $v_\mathrm{ej}\geq30 \, \mathrm{km\ s^{-1}}$ companions,
since common envelope episodes are rather rare events in massive
binary evolution (at least in comparison to stable RLOF mass
transfer). However, here we focus on a rare outcome of massive binary
evolution -- the ejection of a companion at very large
velocity. Therefore, a rare evolutionary channel can (and in this case
does) play a large role in the formation scenario of the outcome of
interest.  To eject enough companions at large velocities to be
consistent with current observations, we have shown that a common
envelope efficiency significantly larger than unity must be
prescribed, i.e. on top of the liberated orbital energy, other energy
sources must be tapped to assist in ejecting the common envelope. These extra sources may
include thermal energy and recombination within the envelope
\citep[e.g][]{Han2002,Webbink2008, Ivanova2013}, the enthalpy of the
envelope \citep{Ivanova2011} or nuclear fusion \citep{Ivanova2002,
  Podsiadlowski2010}. Due to our presently-insufficient understanding
of the physics involved in the common envelope phase,
$\alpha_{\rm CE}$ is poorly-constrained and likely differs from system
to system as the relevant timescales and energy sinks and sources vary
\citep[see][]{Regos1995, Zorotovic2010, Ivanova2013}. Estimates for
$\alpha_{\rm CE}$ must be inferred from either observations of
individual post-CE systems \citep[e.g.][]{Afsar2008, Zorotovic2010},
via binary population synthesis simulations \citep[see][and references
therein]{Zuo2014} or from detailed hydrodynamical simulations of
individual systems \citep[e.g.][]{Sandquist1998, deMarco2011,
  Ohlmann2016, Fragos2019}. While sample sizes remain low, several
studies \citep[e.g.][]{Zorotovic2010, deMarco2011, Davis2012}
identify systems where $\alpha_{\rm CE}>1$. The $\alpha_{\rm CE}\gg 1$
binary progenitors of HRSs however are rare systems, and observations
of much more common systems may be only partially relevant. Whether $\alpha_{\rm CE}= 10$ is possible in a rare and specific
number of instances throughout the history of the Galaxy is currently
observationally and theoretically unclear. 

\subsubsection{The end: CC natal kicks} \label{sec:disc:kicks}

Among all the physical parameters we explore, the post-CC natal kick
and its mediation by ejecta fallback is most influential on the
frequency of hyper-runaway ejections by the BSS. We find in our
optimized simulation that the primaries of HRS progenitor binaries nearly always leave a black hole remnant
according to the \citetalias{Fryer2012}
rapid supernova algorithm. This is an effect of HRS progenitor binaries
maximizing the orbital velocity of the companion, trending towards
large total masses and small mass ratios. Primaries in these
progenitor binaries will almost certainly be massive enough to leave a
BH remnant in the \citetalias{Fryer2012}
  rapid supernova algorithm. Our fiducial treatment for the natal kick
distribution for all remnants, however, is based not on BHs but on
pulsar velocities. It follows \citet{Hobbs2005} who infer the
distribution of 3D velocities for a sample of young (and therefore
relatively unaffected by the Galactic potential) pulsars and fit a
Maxwellian distribution with a root mean squared dispersion
$\sigma_{\rm kick}=265 \, \mathrm{km\ s^{-1}}$ \citep[see
also][]{Lyne1994}. The magnitudes of black hole natal kicks are
poorly-constrained when compared to pulsars
\citep[e.g.][]{Repetto2012, Janka2013, Mandel2015, Janka2017,
  Renzo2019} -- single BHs can only be detected via microlensing of
background sources \citep{Wyrzykowski2016,Wyrzykowski2020} or
accretion from the ISM \citep[e.g.][]{Fender2013,Gaggero2017}. It is
generally accepted that black hole natal kicks are smaller in
magnitude than pulsar natal kicks \citep[see][]{Janka2017, Atri2019,
  Chan2020}. The $\sigma_{\rm kick}=265 \, \mathrm{km\ s^{-1}}$ kick
distribution for pulsars may already overestimate kick velocities, as
several studies both theoretical and observational favour a
double-peaked Maxwellian distribution with an additional low-velocity
peak with $\sigma_{\rm kick}=30 \, \mathrm{km\ s^{-1}}$
\citep[e.g.][]{Arzoumanian2002, Verbunt2017, VerbuntCator2017,
  VignaGomez2018}. In the fiducial model the natal kick is then
mediated by the fallback of supernova ejecta as prescribed in Eq. 16
of \citetalias{Fryer2012},
i.e. $v_{\rm kick}\rightarrow v_{\rm kick}(1-f_{\rm b})$, where
$f_{\rm b}$ is the fallback fraction. Our fiducial feedback
prescription assumes large fallback fractions for BHs, therefore very
small natal kicks effectively indistinguishable from zero
\citepalias{Renzo2019}.

While a paucity of single BH observations continues to hamstring
efforts to constrain BH natal kicks, surviving binaries containing a
BH can offer some insight, at least on lower-velocity kicks
insufficient to ionize the binaries. Evidence for non-zero BH kicks
can be found in the abundance of nearby black hole X-ray binaries \citep[BHXRBs;][]{Vanbeveren2020} as well as their Galactic latitude distribution and velocity
distribution \cmmnt{of black hole X-ray binaries}
\citep[][]{Repetto2012, Repetto2015, Repetto2017}, in
particular the peculiar velocities of certain BHXRBs, e.g. GRO 1655-40
\citep{Willems2005}, XTE J1118+480 \citep{Fragos2009} and GS 1354-64
\citep{Atri2019}. While at least some BHXRBs may have experienced
kicks similar in magnitude to pulsars, recent works find BH kicks
significantly larger than $\sim$\,$100 \, \mathrm{km\ s^{-1}}$ are not
required to explain current BHXRB observations \citep[e.g.][]{Zuo2015,
  Mandel2015, Atri2019}. Detailed 3D simulations of core-collapse
supernovae accounting for the effects of ejecta fallback find that
kicks $\sim$\,$100 \, \mathrm{km\ s^{-1}}$ are possible in specific and somewhat fine-tuned
fallback scenarios
\citep{Chan2018, Chan2020}, but favor more modest kicks in
general. Naturally, systems which have experienced kicks strong enough
to unbind the binary will of course not be visible as BHXRBs,
therefore stronger kicks are not precluded by these studies. We
explore in Sec. \ref{sec:beyond} whether comparing the velocities of
BH-MS binaries in our simulations which remain bound post-CC to known
BHXRBs can offer useful model constraints.

Additional constraints on BH natal kicks will improve with the
determination of the mass distribution of runaway stars
(e.g., \citetalias{Renzo2019}, and especially
  runaway Wolf-Rayet stars,  \citealt{Dray2005}) and further gravitational wave detections of
binary black hole mergers \citep{Oshaughnessy2017,
  Wysocki2018}. In the isolated binary scenario, non-zero BH kicks must be invoked to explain the
misalignment between the orbital plane of the BH binary and the spin
of the more massive companion \citep[e.g.][]{Kalogera2000,
  Oshaughnessy2017}, which is encoded into the gravitational wave
signal. On the other hand, the BH natal kick is bounded from above by
binary BH merger rates and the mass distribution of runaway
stars. Strong BH kicks would often disrupt massive binaries, skewing
the mass distribution of runaways towards higher masses
\citepalias{Renzo2019} and sharply
reducing coalescence rates for binary BHs from isolated binary
evolution \citep[e.g.][]{Belczynski2002, Belczynski2016}. For a
review of the formation and evolution of isolated compact object
binaries, see \citet{Postnov2014}. The contribution of BH
binaries formed not through isolated binary evolution but via
dynamical assembly in dense stellar environments must be considered as
well \citep[see][]{Benacquista2013}. On the basis of the
observed LIGO binary BH merger rate, \cite{Wysocki2018} disfavor a BH natal
kick distribution with $\sigma_{\rm kick}\gtrsim 200 \, \mathrm{km\ s^{-1}}$
\citep[see also][]{Belczynski2016nat}. To conclude, while the  natal kick prescriptions we assume in the optimized model
  cannot be
  directly ruled out due to a lack of observations of isolated single BHs, indirect evidence from pulsar kicks, BH XRB kinematics and
  BH-BH mergers generally do not offer support for this
  scenario.

  We note here that the `explodability' of massive
    stars and the connection between supernova progenitors and the
    remnants (BH or NS) they leave behind is an open question
    \citep[see][]{OConnor2011, Ugliano2012, Sukhbold2016, Ebinger2019,
      Ertl2020}. While the vast majority of primaries in
    HRS progenitor binaries leave black hole remnants in the
    \citetalias{Fryer2012} rapid algorithm, alternative treatments could
    prescribe a significant proportion of NS remnants and/or
    a different remnant mass function \citep[e.g,][]{Mandel2020}. However, overall our
    qualitative main results -- i.e. that strong remnant natal kicks
    and $\alpha_{\rm CE}\gg1$ are required for the BSS to be a
    significant HRS ejection mechanism -- are agnostic to the specific
    details of the post-SNe treatment applied to the primary. See e.g. \citet{Vanbeveren2020}, who argue that the compact remnants of Wolf-Rayet (WR) stars in WR-OB star binaries receive large natal kicks regardless of whether these remnants are NSs or BHs -- otherwise wind-fed high-mass XRBs would be overabundant in the Solar Neighbourhood.
    
    The core-collapse of a massive star can in principle disrupt
      the binary without a significant natal kick in the case of large
    amounts of (fast) mass loss --
    for a perfectly symmetric explosion in the reference frame of the
    exploding star, a disruption can occur if more than half of the
    total mass of the binary is ejected \citep{Blaauw1961}. While this
    case is rare for the typical initial period distribution of
    massive binaries and produces disruptions only in 16\,\% of the
    binaries avoiding merger in the fiducial simulation (\citetalias{Renzo2019}), it
    might remain relevant for systems with extreme initial mass
    ratios. We return to this issue in Sec. \ref{sec:disc:caveats}.
\subsection{Beyond HRS: Broader Implications of Model Variations} \label{sec:beyond}
The binary evolution model variations we assume in this work impact not only the ejection rate of HRSs in general, but leave an imprint on other observables. In particular, the binaries in our simulations which remain bound post-CC can offer model constraints when compared to observations, as can the population of ejected stars with more moderate velocities ($v_{\rm ej} > 30 \, \mathrm{km\ s^{-1}}$). In this subsection we remark on these populations.

\subsubsection{On surviving binaries}

We discussed in Sec. \ref{sec:disc:kicks} how BH natal kicks can be constrained by BHXRBs and BH-BH mergers. Since we do not model X-ray emission via accretion and do not follow our systems beyond the first CC event, our systems cannot directly offer predictions for BHXRB and binary black hole populations. However, we can make inferences based on the abundance and kinematics of BH-MS binaries in our simulation which remain bound after the first CC. Even if they are not disrupted, the natal kick and the change in gravitational potential due to mass loss from the primary can impart peculiar velocities to these systems, possibly up to and exceeding runaway ($v_{\rm ej} > 30 \, \mathrm{km\ s^{-1}}$) velocities \citep{vanOijen1989}. In the case of Hercules X-1, a pulsar XRB at high Galactic latitude, \citet{Sutantyo1975} shows that the spherically-symmetric (i.e. $v_{\rm kick}\sim0$) explosion of a $\sim4 \, \mathrm{M_{\odot}}$ helium star in orbit with a $\sim2 \, \mathrm{M_{\odot}}$ companion can explain its quite high ejection velocity ($v_{\rm ej}\sim150 \, \mathrm{km\ s^{-1}}$). Runaway stars still in orbit with a compact remnant could be identified as single-line (SB1) spectroscopic binaries. If the orbits are sufficiently tight to allow the black hole remnant to eventually accrete material from the companion, these systems may become visible as BHXRBs.

In the fiducial simulation, 6.7\,\% of systems remain bound as BH-MS
binaries after the first CC event. Our simulations predict the
systemic velocity $v_{\rm sys}$ imparted to these systems, here
defined as the velocity of the centre of mass of the post-CC BH-MS binary in
the frame of reference of its pre-CC centre of mass
\citep{Kalogera1996, Tauris1998}. In agreement with
\citetalias{Renzo2019} (c.f. Fig.~11) we find that $\sim$80\,\% of the BH-MS systems in
the fiducial simulation acquire $v_{\rm sys}\simeq 0 \, \mathrm{km\ s^{-1}}$. This
is due to our fiducial prescription for post-CC ejecta fallback
following \citetalias{Fryer2012} -- many BH progenitors experience
complete ejecta fallback; the remnant experiences no kick at all and
no mass is lost from the system. Only 4\,\% of the BH-MS binaries in our fiducial
simulation achieve systemic velocities in excess of 30 $\mathrm{km\ s^{-1}}$.

The optimized simulation, on the other hand, includes very different
prescriptions for BH natal kicks and ejecta fallback, and therefore
predicts a very different velocity distribution for surviving BH-MS
binaries. Owing to the large disruption fraction, only 0.5\,\% of
systems in the optimized simulation remain bound as a BH-MS binary
after the first CC event. Of these, however, 99.9\,\% achieve
$v_{\rm sys}>30 \,\mathrm{km\ s^{-1}}$ (see Fig. \ref{fig:vNS}). This, combined with the fact that 95\,\% have post-CC separations below
$300 \, \mathrm{R_\odot}$, indicate the optimized simulation would predict runaway
BHXRBs to be quite common. In fact, 45\,\% of BH-MS binaries achieve
hyper-runaway speeds, $v_{\rm sys}\geq400 \, \mathrm{km\ s^{-1}}$. The optimized
simulation predicts that lone hyper-runaway stars outnumber
hyper-runaway stars with BH companions by only a factor of 1.4. To
date, no hyper-runaway BHXRBs or SB1 binaries have been observed. With a
sample of 16 known BHXRBs, \citet{Atri2019} find no BHXRBs that
intersect the Galactic disc with a peculiar velocity above $\sim200 \, \mathrm{km\ s^{-1}}$; a systemic velocity achieved by 83\,\% of the BH-MS
binaries in the optimized simulation. This absence of high-velocity
BHXRBs or SB1 systems disfavours our optimized model. It is
worth mentioning, however, that without accounting for deceleration by
the Galactic potential, BH-MS binaries in our optimized simulation travel for
$\sim$17 kpc on average from their birth locations before the companion star
leaves the main sequence. At such distances any X-ray emission may
be difficult to detect and BHXRB samples would therefore be biased towards low-$v_{\rm sys}$ systems remaining at low Galactic latitude.

Systems that remain bound after the first CC are unlikely to be
  disrupted by the eventual CC of the remaining MS companion
  \citep[e.g.,][]{Oshaughnessy2017} provided they do not merge
  beforehand. If the separation is sufficiently small, the resulting
  BH-BH binary may eventually become a gravitational wave source. Since we do not follow the evolution of our binary systems beyond the
  first CC event, our simulations cannot provide predictions for gravitational waves merger rates. However, from the different binary disruption fractions it is clear that the
  fiducial and optimized simulations would give quite different predictions for
  the rate and characteristics of BH-BH mergers. Therefore, upcoming statistical
  samples of BH-BH mergers will provide model constraints, albeit from
  a different part of the parameter space than what is relevant for
  the formation of hyper-runaway stars (see
  previous Section).

\subsubsection{On massive runaway stars}

Naturally, the model variations we explore here to encourage the ejection of HRSs also strongly influence the fraction of systems which eject companions at $30 \, \mathrm{km\ s^{-1}}< v_{\rm ej} < 400 \, \mathrm{km\ s^{-1}}$ and the velocity distribution of these `runaway' stars. Properly accounting for initial configuration probabilities and the duration of each evolutionary phase, the fiducial simulation predicts that in total, 0.54\,\% of all O-type ($M\gtrsim15 M_{\odot}$) stars in the Milky Way are runaways (see Table \ref{tab:results}). The O-type runaway fraction increases to 1.6\,\% in the optimized simulation but never significantly exceeds $\sim$2\,\% in any simulation presented in Table \ref{tab:results}. Although the likelihood of hyper-runaway ejections varies by four orders of magnitude among our chosen models, the O-type runaway fraction is much more robust against model variation. This consistent with \citetalias{Renzo2019} and the binary synthesis study of \citet{Eldridge2011} but in potential tension with observational findings that $\simeq$10-20\,\% of O-type stars are runaways \citep[e.g.,][]{Blaauw1961, Tetzlaff2011, MaizApellaniz2018} and the suggestion of \citet{Hoogerwerf2001} that $\sim$2/3 of runaway stars are due to disrupted binaries (though see \citet{Jilinski2010} for a challenge to this conclusion).  Unless we have overlooked other mechanisms which greatly shrink binary orbits and eject companions at runaway speeds, the inability of even our optimized simulation to match the above observations suggests that the role of the dynamical ejection scenario has been under-estimated and/or observations of O-type stars are biased towards runaways.

\subsubsection{On neutron star kinematics}

In addition to predicting large BH-MS systemic
  velocities that are in potential tension with observations, the
  optimized simulation also predicts quite large neutron star ejection
  velocities due to the large kicks
  assumed. Using 19 young (and therefore relatively unaffected by the
  Galactic potential) pulsars with proper motion measurements,
  \citet{Verbunt2017} fit the observed
  velocity distribution to a double Maxwellian distribution. Their best-fit model suggests than no more
  than 2\% of Galactic pulsars are moving at
  $v>1000 \, \mathrm{km\ s^{-1}}$. In the optimized simulation, 79\%
  of neutron stars ejected from disrupted binaries are ejected with
  velocities above this threshold (see Fig. \ref{fig:vNS}). Although observations may be biased
  towards close-by, low-velocity pulsars, they nonetheless are in
  contrast with our optimized model.

\subsection{Further caveats}
\label{sec:disc:caveats}
There are caveats which should be considered when comparing the HRS populations predicted by the simulations to
the current sample of observed HRS candidates. 

Notice that we assume in Eq. \ref{eq:nfast} a constant Galactic star
formation rate, a reasonable assumption for at least the last $\sim$2
Gyr \citep{Snaith2014, Snaith2015}. Spatial or temporal variations in
the global star formation rate in the Milky Way will affect the number of observable
HRSs, especially the more massive and thus shorter-lived among these. A recent burst of star formation somewhere in the Milky Way disc will lead to an increase in runaway and hyper-runaway stars ejected from that location after a delay corresponding to at
least the lifetime of the most massive stars we consider
\citep[$\sim$3 Myr,][]{Zapartas2017}. Our assumption
  for the global Galactic star formation rate, 3.5 M$_{\odot} \ \mathrm{yr^{-1}}$, is
  a common choice for binary population synthesis studies
  \citep[e.g.][]{Dominik2012}, but rather optimistic when
  compared to recent results converging on 1-2 M$_{\odot} \ \mathrm{yr^{-1}}$
  \citep[][and references therein]{Chomiuk2011, Licquia2015}. However, replacing
  the star formation rate seen in Eq. \ref{eq:nfast} with a lower value will reduce all $N_{\rm f, obs}$ values in Table \ref{tab:results} by the same factor and would not qualitatively change any results presented in this study
  regarding the conditions and physics required to eject fast stars
  via the BSS mechanism. A lower star formation rate in fact strengthens our conclusion that ejecting enough fast companions to be consistent with current observations of Milky Way HRS candidates requires prescriptions for remnant natal kicks, ejecta fallback and common envelope evolution that significantly differ from our fiducial assumptions for the overall binary population.

  Another consideration is our choice of lower bound
    on the ZAMS mass ratio distribution. In all simulations we
    generate only systems with $0.1<q_{\rm ZAMS}<1$, a commonly-explored interval
    \citep[e.g.][]{Sana2012,Sana2013,Moe2013, Dunstall2015} dictated by
    detectability limits and the assumption that systems with
    $q_{\rm ZAMS}<0.1$ are expected to eventually merge
    \citep[e.g.,][]{deMink2013}. With the somewhat idealized physics
    we assume in the optimized simulation(s), however, this assumption
    may not necessarily hold true. Indeed, re-running the optimized
    simulation with a fixed companion lower mass limit of
    $M_{\rm 2, ZAMS}=0.5 \, \mathrm{M_{\odot}}$ does increase the
    number $N_{\rm f, obs}$ of observable early-type HRSs by 45\,\%, however the qualitative results
    presented in this study do not change.  
    Given the observational
    realities of detecting secondaries around massive stars in highly unequal-mass systems
    \citep[see e.g.][]{Kobulnicky2007, Duchene2013}, it remains
    unclear what fraction of massive binaries have extreme $q<0.1$
    mass ratios or whether such a subpopulation can even be realistically constrained by observations. If future observational constraints suggest a significant population of such system which avoid mergers, further investigation on the evolution of extreme mass ratio binaries would be warranted.
 
  As mentioned in Sec. \ref{sec:disc:alpha}, one limitation of our
  approach is that we assume a single value of $\alpha_{\rm CE}$
  applies to all binaries. In reality there is more likely a
  distribution of values, where even optimistically only for rare systems these are equal
  to our optimized value to $\alpha_{\rm CE}=10$. In the optimized
  simulation this same caveat applies to the total stellar metallicity
  $Z$ and fallback fraction $f_{\rm b}$ -- we apply a single value to
  the entire binary population rather than allowing for multiple
  populations to be present as minority fractions of the broader
  population. A model where $Z$ and $\alpha_{\rm CE}$ in
    particular are allowed to vary is a venue worth exploring in the
    future. 

Finally, it is worth pointing out that the parameters chosen
  for our optimized simulation were chosen based on simple
  assumptions about their role in enhancing the number of
  short-period, high-orbital velocity systems. We do not fully
  explore the highly-dimensional space of initial conditions and
  binary evolution parameters to find the truly optimal set of
  prescriptions. Some of our assumptions may be overly simplistic,
  e.g. see Sec. \ref{sec:results} where the fiducial
  $P_{\rm ZAMS}$ prescription is in fact more effective than the
  ``optimized'' prescription at ejecting fast companions. A more
  complete exploration of the entire parameter space and the
  correlations between parameters is a possible future direction
  for this study.

\subsection{Alternative origins of hyper-runaway stars}
Here we only consider main-sequence stars ejected from a disrupted
binary system. If the BSS mechanism is unable to contribute
significantly to the known population of Milky Way main sequence hyper-runaway candidates,
which mechanism is responsible? One possible mechanism includes the
dynamical ejection scenario (DES) wherein stars are ejected via three-
or four-body interactions in dense clusters \citep{Poveda1967,
  Leonard1990, Leonard1991, Fujii2011, Banerjee2012}. \citet{Oh2016}
explore the ejection of massive stars from moderately massive
($M_{\rm cl}\approx 3000\,M_\odot$) star clusters under diverse
initial conditions. Ejections in excess of 200 $\mathrm{km\ s^{-1}}$ are very rare
except in the model where the cluster consists of only initially
massive ($M_1 \geq 5\,\mathrm{M_\odot}$) binaries on short, circular
orbits. In a suite of similar simulations, \citet{Perets2012} infer a
dynamical ejection rate for $v_{\rm ej}>450 \,\mathrm{km\ s^{-1}}$ B-type HRSs
from young clusters of $\sim$10 Gyr$^{-1}$ in their most realistic
model, and conclude that the DES is unlikely to contribute
significantly to the Galactic population of hyper-runaway stars. Since the majority of ejections from dynamical encounters happen
  in the very early evolution of a cluster \citep[e.g.][]{Oh2016}, observations of the stellar dynamics in very young regions (such as
  the Orion Nebula Cluster, e.g., \citealt{Schoettler2019,
    Schoettler2020}, or R136 in the LMC, e.g.,
  \citealt{Lennon2018, Renzo2019vfts}) can constrain the DES scenario, including for the rare very fast
  ejections.

Beyond the BSS and the classical dynamical ejection scenario, possible but less-studied scenarios have been suggested. A first potential pathway is
ejections which involve three- or four-body interactions involving
massive stars. \citet{Gvaramadze2009} point out three possible
ejection channels: (i) the breakup of an unstable hierarchical
triple composed of massive stars (inner binary of two $\sim$50\,M$_\odot$ stars with a $\sim$10\,M$_\odot$ outer companion, (ii) binary-binary dynamical
interactions where massive stars ($\sim$20-40 M$_\odot$) compose at least one member of the binaries, and iii)
exchange encounters between hard, massive binaries and $\gtrsim200 \, \mathrm{M_\odot}$ very massive stars, which form via runaway collisions of
ordinary stars in the cores of young massive star clusters
\citep{Portegies2002}. All three mechanisms are capable of ejecting
massive companions far in excess of $400 \, \mathrm{km\ s^{-1}}$.

Second, encounters between stellar binaries and an intermediate-mass black hole (IMBH) in analogy to the Hills mechanism may also contribute to the population of high-velocity stars ejected from the Galactic disc or halo \citep{Pfahl2005, Gualandris2007, Hopman2009, Sesana2012}. \citet{Fragione2019} assign realistic IMBH masses to the centres of Milky Way globular clusters and study the ejection of high-velocity stars. They infer a fast star ejection rate from this IMBH channel of $\sim$10$^{-4}\, \mathrm{yr^{-1}}$ with a $v_{\rm ej}$ distribution which peaks at $v_{\rm ej}\approx 300 \, \mathrm{km\ s^{-1}}$ with a tail extending to $\sim$2000 $\mathrm{km\ s^{-1}}$.

Finally, we must also consider the possibility that although the backward trajectories of observed HRSs point towards the Galactic disc, they may have been born elsewhere. Main sequence HRSs originally of extragalactic origins may serendipitously intersect the Milky Way disc and appear as disc-ejected HRSs. These stars could come from the tidal disruption of infalling dwarf galaxies as suggested by \citep{Abadi2009} or from a high-velocity ejection from the Large Magellanic Cloud \citep{Gualandris2007, Przybilla2008HVS3, Boubert2016,Boubert2017, Erkal2019} or M31 \citep{Sherwin2008}, though none of our HRS candidates seem to point from the LMC or M31. Any or all of the above origins and mechanisms could contribute to the Galactic population of hyper-runaways.
  
\section{Summary and Conclusion} \label{sec:conclusions}

We have performed an extensive suite of numerical simulations to ascertain the probability with which disruptions of binary systems by core-collapse supernovae are able to eject main sequence companions at
large velocities ($v_{\rm ej}\geq400 \, \mathrm{km\ s^{-1}}$). With each simulation we predict the number of these
so-called `hyper-runaway stars' (HRSs) in the Milky Way which should
still be reasonably nearby with $t_{\rm flight}<100 \, \mathrm{Myr}$. We systematically
vary parameters which dictate the initial conditions and physics
governing the evolution of binary systems. In doing so, our aim is to
rigorously test under a variety of models whether the binary
supernova scenario contributes significantly to the observed
population of early-type HRS candidates seemingly ejected from
the Milky Way disc at high velocities. By varying parameters
one-by-one, we determine which are most important when considering the
ejection of companions at large velocities. Our findings can be
summarized as follows:

\begin{itemize}
    \item In our fiducial model, which generates and evolves binary systems according to free parameters chosen to be consistent with contemporary observations and theory, core-collapse supernovae in binaries do not eject companions at large velocities often enough for the binary supernova scenario to contribute significantly to the known population of HRS candidates (see Fig \ref{fig:vejsm}).
    \item By varying relevant parameters one-by-one, we find that the probability of fast companion ejections depends most intimately on the common-envelope efficiency $\alpha_{\rm CE}$ and the parameters which dictate the magnitude of the post-CC natal kick imparted to stellar remnants, i.e. the ejecta fallback fraction $f_{\rm b}$ and natal kick distribution dispersion $\sigma_{\rm kick}$ (see Table \ref{tab:results}). The $\alpha_{\rm CE}$ parameter indirectly governs the amount of orbital tightening -- and therefore the increase in orbital velocity -- that occurs during the common-envelope phase. A large orbital velocity in turn results in a greater ejection velocity for the companion if the system is disrupted. The natal kick parameters $f_{\rm b}$ and $\sigma_{\rm kick}$, in turn, dictate the frequency with which binary systems are disrupted immediately following the core-collapse of the primary. 
    \item To a lesser degree, the probability of fast companion ejection increases with decreasing total stellar metallicity and with an increasing proportion of highly-unequal initial mass ratios in the binary population. At fixed stellar mass, stars of lower stellar metallicity are smaller in radius and can therefore orbit at a smaller separation without undergoing a merger. At fixed system mass, a smaller mass ratio between the companion and primary results in a greater orbital velocity for the companion.
    \item While the rate of HRS ejections via the binary supernova
      scenario depends strongly on the above parameters, the runaway fraction
      ($v_{\rm ej}>30 \, \mathrm{km \ s^{-1}}$) among O-type
      stars is robust against model variations and is limited to
      $\lesssim2$\,\%.
    \item The prototypical progenitor binary of an HRS ejected by the binary supernova scenario is composed of a massive, unequal-mass binary on a short initial period. Upon the primary filling its Roche lobe, the system enters a phase of common envelope evolution and hardens further. The core-collapse of the primary leaves a black hole remnant, which experiences an asymmetric post-CC natal kick extreme enough to disrupt the binary. The main sequence companion escapes with an ejection velocity similar in magnitude to its pre-core collapse orbital velocity ($v_{\rm ej}\gtrsim 400 \,\mathrm{km\ s^{-1}}$, see Figs. \ref{fig:vejsm} and \ref{fig:Cartoon}). The remnant black hole is ejected at a much larger velocity, $v_{\rm ej}\simeq 1700 \, \mathrm{km\ s^{-1}}$.
    \item For a model to predict a number of $t_{\rm flight}$<100 Myr HRSs comparable to the current observed population of HRS candidates, it is not sufficient to tune one of the aforementioned parameters in the direction which favours the ejection of fast companions. \textit{All of} the common envelope efficiency $\alpha_{\rm CE}$ and the stellar remnant natal kick parameters $f_{\rm b}$ and $\sigma_{\rm kick}$ must be tuned in favourable directions for our population synthesis model to predict a number of hyper-runaway stars consistent with observations (see Table \ref{tab:results}).
    \item To match observations, the relevant parameters above must be tuned to values or prescriptions which are improbable or disfavoured -- but not ruled out -- by contemporary constraints on these parameters from the literature.
\end{itemize}

From the results summarized above, we may conclude that the binary supernova scenario is unlikely to contribute significantly to the current known population of HRS candidates. This conclusion is broadly consistent with other works exploring the binary supernova scenario \citep[e.g.][]{Portegies2000, Eldridge2011, Renzo2019}. With our detailed quantitative investigation we outline for the first time the evolutionary path of hyper-runaway stars ejected via the binary supernova scenario and the characteristics of their binary progenitors. Future dedicated theoretical and observational works may be used to test the existence of such progenitors.

\section*{Acknowledgements}

We are grateful to the anonymous referee for their prompt and constructive feedback. We thank I.~Mandel for helpful discussions, D.~Hendricks, S.~Jha and  S.~Justham for useful comments and R.~Izzard for letting us use the \texttt{binary\_c} code. FAE acknowledges D.~Boubert and S.~Kreuzer for helpful correspondence regarding the Open Fast Stars Catalog. MR acknowledges M.~Heemskerk for allowing continued
use of the helios cluster at the University of Amsterdam, and S.~de~Mink and E.~Zapartas for useful discussions.
FAE acknowledges funding support from the Natural Sciences and
Engineering Research Council of Canada (NSERC) Postgraduate
Scholarship. 

\vspace{-20pt}
\section*{Data Availability}
The data underlying this article are available via Zenodo at \url{https://dx.doi.org/10.5281/zenodo.3860055} and via the Open Fast Stars Catalog at \url{https://faststars.space/}

\bibliographystyle{mnras}
\bibliography{HRS}

\begin{thebibliography}{}
\makeatletter
\relax
\def\mn@urlcharsother{\let\do\@makeother \do\$\do\&\do\#\do\^\do\_\do\%\do\~}
\def\mn@doi{\begingroup\mn@urlcharsother \@ifnextchar [ {\mn@doi@}
  {\mn@doi@[]}}
\def\mn@doi@[#1]#2{\def\@tempa{#1}\ifx\@tempa\@empty \href
  {http://dx.doi.org/#2} {doi:#2}\else \href {http://dx.doi.org/#2} {#1}\fi
  \endgroup}
\def\mn@eprint#1#2{\mn@eprint@#1:#2::\@nil}
\def\mn@eprint@arXiv#1{\href {http://arxiv.org/abs/#1} {{\tt arXiv:#1}}}
\def\mn@eprint@dblp#1{\href {http://dblp.uni-trier.de/rec/bibtex/#1.xml}
  {dblp:#1}}
\def\mn@eprint@#1:#2:#3:#4\@nil{\def\@tempa {#1}\def\@tempb {#2}\def\@tempc
  {#3}\ifx \@tempc \@empty \let \@tempc \@tempb \let \@tempb \@tempa \fi \ifx
  \@tempb \@empty \def\@tempb {arXiv}\fi \@ifundefined
  {mn@eprint@\@tempb}{\@tempb:\@tempc}{\expandafter \expandafter \csname
  mn@eprint@\@tempb\endcsname \expandafter{\@tempc}}}

\bibitem[\protect\citeauthoryear{{Aarseth}}{{Aarseth}}{1974}]{Aarseth1974}
{Aarseth} S.~J.,  1974, \aap, \href
  {https://ui.adsabs.harvard.edu/abs/1974A&A....35..237A} {35, 237}

\bibitem[\protect\citeauthoryear{{Abadi}, {Navarro}  \& {Steinmetz}}{{Abadi}
  et~al.}{2009}]{Abadi2009}
{Abadi} M.~G.,  {Navarro} J.~F.,   {Steinmetz} M.,  2009, \mn@doi [\apjl]
  {10.1088/0004-637X/691/2/L63}, \href
  {https://ui.adsabs.harvard.edu/abs/2009ApJ...691L..63A} {691, L63}

\bibitem[\protect\citeauthoryear{{Abolfathi} et~al.,}{{Abolfathi}
  et~al.}{2018}]{Abolfathi2018}
{Abolfathi} B.,  et~al., 2018, \mn@doi [\apjs] {10.3847/1538-4365/aa9e8a},
  \href {https://ui.adsabs.harvard.edu/abs/2018ApJS..235...42A} {235, 42}

\bibitem[\protect\citeauthoryear{{Af{\textcommabelow s}ar} \&
  {Ibano{\v{g}}lu}}{{Af{\textcommabelow s}ar} \&
  {Ibano{\v{g}}lu}}{2008}]{Afsar2008}
{Af{\textcommabelow s}ar} M.,  {Ibano{\v{g}}lu} C.,  2008, \mn@doi [\mnras]
  {10.1111/j.1365-2966.2008.13927.x}, \href
  {https://ui.adsabs.harvard.edu/abs/2008MNRAS.391..802A} {391, 802}

\bibitem[\protect\citeauthoryear{{Alam} et~al.,}{{Alam}
  et~al.}{2015}]{Alam2015}
{Alam} S.,  et~al., 2015, \mn@doi [\apjs] {10.1088/0067-0049/219/1/12}, \href
  {https://ui.adsabs.harvard.edu/abs/2015ApJS..219...12A} {219, 12}

\bibitem[\protect\citeauthoryear{{Anders} \& {Grevesse}}{{Anders} \&
  {Grevesse}}{1989}]{Anders1989}
{Anders} E.,  {Grevesse} N.,  1989, \mn@doi [\gca]
  {10.1016/0016-7037(89)90286-X}, \href
  {https://ui.adsabs.harvard.edu/abs/1989GeCoA..53..197A} {53, 197}

\bibitem[\protect\citeauthoryear{{Arzoumanian}, {Chernoff}  \&
  {Cordes}}{{Arzoumanian} et~al.}{2002}]{Arzoumanian2002}
{Arzoumanian} Z.,  {Chernoff} D.~F.,   {Cordes} J.~M.,  2002, \mn@doi [\apj]
  {10.1086/338805}, \href
  {https://ui.adsabs.harvard.edu/abs/2002ApJ...568..289A} {568, 289}

\bibitem[\protect\citeauthoryear{{Asplund}, {Grevesse}, {Sauval}  \&
  {Scott}}{{Asplund} et~al.}{2009}]{Asplund2009}
{Asplund} M.,  {Grevesse} N.,  {Sauval} A.~J.,   {Scott} P.,  2009, \mn@doi
  [\araa] {10.1146/annurev.astro.46.060407.145222}, \href
  {https://ui.adsabs.harvard.edu/abs/2009ARA&A..47..481A} {47, 481}

\bibitem[\protect\citeauthoryear{{Atri} et~al.,}{{Atri}
  et~al.}{2019}]{Atri2019}
{Atri} P.,  et~al., 2019, \mn@doi [\mnras] {10.1093/mnras/stz2335}, \href
  {https://ui.adsabs.harvard.edu/abs/2019MNRAS.489.3116A} {489, 3116}

\bibitem[\protect\citeauthoryear{{Banerjee}, {Kroupa}  \& {Oh}}{{Banerjee}
  et~al.}{2012}]{Banerjee2012}
{Banerjee} S.,  {Kroupa} P.,   {Oh} S.,  2012, \mn@doi [\apj]
  {10.1088/0004-637X/746/1/15}, \href
  {https://ui.adsabs.harvard.edu/abs/2012ApJ...746...15B} {746, 15}

\bibitem[\protect\citeauthoryear{{Bauer}, {White}  \& {Bildsten}}{{Bauer}
  et~al.}{2019}]{Bauer2019}
{Bauer} E.~B.,  {White} C.~J.,   {Bildsten} L.,  2019, \mn@doi [\apj]
  {10.3847/1538-4357/ab4ea4}, \href
  {https://ui.adsabs.harvard.edu/abs/2019ApJ...887...68B} {887, 68}

\bibitem[\protect\citeauthoryear{{Belczy{\'n}ski} \& {Bulik}}{{Belczy{\'n}ski}
  \& {Bulik}}{1999}]{Belcynski1999}
{Belczy{\'n}ski} K.,  {Bulik} T.,  1999, \aap, \href
  {https://ui.adsabs.harvard.edu/abs/1999A&A...346...91B} {346, 91}

\bibitem[\protect\citeauthoryear{Belczynski, Kalogera  \& Bulik}{Belczynski
  et~al.}{2002}]{Belczynski2002}
Belczynski K.,  Kalogera V.,   Bulik T.,  2002, \mn@doi [\apj]
  {10.1086/340304}, 572, 407

\bibitem[\protect\citeauthoryear{{Belczynski}, {Holz}, {Bulik}  \&
  {O'Shaughnessy}}{{Belczynski} et~al.}{2016a}]{Belczynski2016nat}
{Belczynski} K.,  {Holz} D.~E.,  {Bulik} T.,   {O'Shaughnessy} R.,  2016a,
  \mn@doi [\nat] {10.1038/nature18322}, \href
  {https://ui.adsabs.harvard.edu/abs/2016Natur.534..512B} {534, 512}

\bibitem[\protect\citeauthoryear{Belczynski, Repetto, Holz, O'Shaughnessy,
  Bulik, Berti, Fryer  \& Dominik}{Belczynski et~al.}{2016b}]{Belczynski2016}
Belczynski K.,  Repetto S.,  Holz D.~E.,  O'Shaughnessy R.,  Bulik T.,  Berti
  E.,  Fryer C.,   Dominik M.,  2016b, \mn@doi [\apj]
  {10.3847/0004-637x/819/2/108}, 819, 108

\bibitem[\protect\citeauthoryear{{Benacquista} \& {Downing}}{{Benacquista} \&
  {Downing}}{2013}]{Benacquista2013}
{Benacquista} M.~J.,  {Downing} J. M.~B.,  2013, \mn@doi [Living Reviews in
  Relativity] {10.12942/lrr-2013-4}, \href
  {https://ui.adsabs.harvard.edu/abs/2013LRR....16....4B} {16, 4}

\bibitem[\protect\citeauthoryear{{Blaauw}}{{Blaauw}}{1961}]{Blaauw1961}
{Blaauw} A.,  1961, \bain, \href
  {https://ui.adsabs.harvard.edu/abs/1961BAN....15..265B} {15, 265}

\bibitem[\protect\citeauthoryear{{Bland-Hawthorn} \&
  {Gerhard}}{{Bland-Hawthorn} \& {Gerhard}}{2016}]{Bland-Hawthorn2016}
{Bland-Hawthorn} J.,  {Gerhard} O.,  2016, \mn@doi [\araa]
  {10.1146/annurev-astro-081915-023441}, \href
  {https://ui.adsabs.harvard.edu/abs/2016ARA&A..54..529B} {54, 529}

\bibitem[\protect\citeauthoryear{{Boersma}}{{Boersma}}{1961}]{Boersma1961}
{Boersma} J.,  1961, \bain, \href
  {https://ui.adsabs.harvard.edu/abs/1961BAN....15..291B} {15, 291}

\bibitem[\protect\citeauthoryear{{B{\"o}ker}}{{B{\"o}ker}}{2010}]{Boker2010}
{B{\"o}ker} T.,  2010, in {de Grijs} R.,  {L{\'e}pine} J. R.~D.,  eds,  IAU
  Symposium Vol. 266, Star Clusters: Basic Galactic Building Blocks Throughout
  Time and Space. pp 58--63 (\mn@eprint {arXiv} {0910.4863}),
  \mn@doi{10.1017/S1743921309990871}

\bibitem[\protect\citeauthoryear{{Boubert} \& {Evans}}{{Boubert} \&
  {Evans}}{2016}]{Boubert2016}
{Boubert} D.,  {Evans} N.~W.,  2016, \mn@doi [\apjl]
  {10.3847/2041-8205/825/1/L6}, \href
  {https://ui.adsabs.harvard.edu/abs/2016ApJ...825L...6B} {825, L6}

\bibitem[\protect\citeauthoryear{Boubert, Erkal, Evans  \& Izzard}{Boubert
  et~al.}{2017}]{Boubert2017}
Boubert D.,  Erkal D.,  Evans N.~W.,   Izzard R.~G.,  2017, Monthly Notices of
  the Royal Astronomical Society, 469, 2151

\bibitem[\protect\citeauthoryear{{Boubert}, {Guillochon}, {Hawkins},
  {Ginsburg}, {Evans}  \& {Strader}}{{Boubert} et~al.}{2018}]{Bourbert2018}
{Boubert} D.,  {Guillochon} J.,  {Hawkins} K.,  {Ginsburg} I.,  {Evans} N.~W.,
   {Strader} J.,  2018, \mn@doi [\mnras] {10.1093/mnras/sty1601}, \href
  {https://ui.adsabs.harvard.edu/abs/2018MNRAS.479.2789B} {479, 2789}

\bibitem[\protect\citeauthoryear{{Bovy}}{{Bovy}}{2015}]{Bovy2015}
{Bovy} J.,  2015, \mn@doi [\apjs] {10.1088/0067-0049/216/2/29}, \href
  {https://ui.adsabs.harvard.edu/abs/2015ApJS..216...29B} {216, 29}

\bibitem[\protect\citeauthoryear{{Bromley}, {Kenyon}, {Brown}  \&
  {Geller}}{{Bromley} et~al.}{2009}]{Bromley2009}
{Bromley} B.~C.,  {Kenyon} S.~J.,  {Brown} W.~R.,   {Geller} M.~J.,  2009,
  \mn@doi [\apj] {10.1088/0004-637X/706/2/925}, \href
  {https://ui.adsabs.harvard.edu/abs/2009ApJ...706..925B} {706, 925}

\bibitem[\protect\citeauthoryear{Bromley, Kenyon, Geller  \& Brown}{Bromley
  et~al.}{2012}]{Bromley2012}
Bromley B.~C.,  Kenyon S.~J.,  Geller M.~J.,   Brown W.~R.,  2012, \mn@doi
  [\apj] {10.1088/2041-8205/749/2/l42}, 749, L42

\bibitem[\protect\citeauthoryear{{Brown}}{{Brown}}{2015}]{Brown2015rev}
{Brown} W.~R.,  2015, \mn@doi [\araa] {10.1146/annurev-astro-082214-122230},
  \href {https://ui.adsabs.harvard.edu/abs/2015ARA&A..53...15B} {53, 15}

\bibitem[\protect\citeauthoryear{{Brown}, {Geller}, {Kenyon}  \&
  {Kurtz}}{{Brown} et~al.}{2005}]{Brown2005}
{Brown} W.~R.,  {Geller} M.~J.,  {Kenyon} S.~J.,   {Kurtz} M.~J.,  2005,
  \mn@doi [\apjl] {10.1086/429378}, \href
  {https://ui.adsabs.harvard.edu/abs/2005ApJ...622L..33B} {622, L33}

\bibitem[\protect\citeauthoryear{{Brown}, {Geller}, {Kenyon}  \&
  {Kurtz}}{{Brown} et~al.}{2006}]{Brown2006}
{Brown} W.~R.,  {Geller} M.~J.,  {Kenyon} S.~J.,   {Kurtz} M.~J.,  2006,
  \mn@doi [\apjl] {10.1086/503279}, \href
  {https://ui.adsabs.harvard.edu/abs/2006ApJ...640L..35B} {640, L35}

\bibitem[\protect\citeauthoryear{{Brown}, {Geller}, {Kenyon}, {Kurtz}  \&
  {Bromley}}{{Brown} et~al.}{2007}]{Brown2007}
{Brown} W.~R.,  {Geller} M.~J.,  {Kenyon} S.~J.,  {Kurtz} M.~J.,   {Bromley}
  B.~C.,  2007, \mn@doi [\apj] {10.1086/513595}, \href
  {https://ui.adsabs.harvard.edu/abs/2007ApJ...660..311B} {660, 311}

\bibitem[\protect\citeauthoryear{{Brown}, {Geller}  \& {Kenyon}}{{Brown}
  et~al.}{2009}]{Brown2009}
{Brown} W.~R.,  {Geller} M.~J.,   {Kenyon} S.~J.,  2009, \mn@doi [\apj]
  {10.1088/0004-637X/690/2/1639}, \href
  {https://ui.adsabs.harvard.edu/abs/2009ApJ...690.1639B} {690, 1639}

\bibitem[\protect\citeauthoryear{{Brown}, {Geller}  \& {Kenyon}}{{Brown}
  et~al.}{2012}]{Brown2012}
{Brown} W.~R.,  {Geller} M.~J.,   {Kenyon} S.~J.,  2012, \mn@doi [\apj]
  {10.1088/0004-637X/751/1/55}, \href
  {https://ui.adsabs.harvard.edu/abs/2012ApJ...751...55B} {751, 55}

\bibitem[\protect\citeauthoryear{{Brown}, {Geller}  \& {Kenyon}}{{Brown}
  et~al.}{2014}]{Brown2014}
{Brown} W.~R.,  {Geller} M.~J.,   {Kenyon} S.~J.,  2014, \mn@doi [\apj]
  {10.1088/0004-637X/787/1/89}, \href
  {https://ui.adsabs.harvard.edu/abs/2014ApJ...787...89B} {787, 89}

\bibitem[\protect\citeauthoryear{{Brown}, {Anderson}, {Gnedin}, {Bond},
  {Geller}  \& {Kenyon}}{{Brown} et~al.}{2015}]{Brown2015}
{Brown} W.~R.,  {Anderson} J.,  {Gnedin} O.~Y.,  {Bond} H.~E.,  {Geller} M.~J.,
    {Kenyon} S.~J.,  2015, \mn@doi [\apj] {10.1088/0004-637X/804/1/49}, \href
  {https://ui.adsabs.harvard.edu/abs/2015ApJ...804...49B} {804, 49}

\bibitem[\protect\citeauthoryear{{Brown}, {Lattanzi}, {Kenyon}  \&
  {Geller}}{{Brown} et~al.}{2018}]{Brown2018}
{Brown} W.~R.,  {Lattanzi} M.~G.,  {Kenyon} S.~J.,   {Geller} M.~J.,  2018,
  \mn@doi [\apj] {10.3847/1538-4357/aadb8e}, \href
  {https://ui.adsabs.harvard.edu/abs/2018ApJ...866...39B} {866, 39}

\bibitem[\protect\citeauthoryear{{Burrows} \& {Hayes}}{{Burrows} \&
  {Hayes}}{1996}]{Burrows1996}
{Burrows} A.,  {Hayes} J.,  1996, \mn@doi [\prl] {10.1103/PhysRevLett.76.352},
  \href {https://ui.adsabs.harvard.edu/abs/1996PhRvL..76..352B} {76, 352}

\bibitem[\protect\citeauthoryear{{Burrows}, {Hubbard}, {Saumon}  \&
  {Lunine}}{{Burrows} et~al.}{1993}]{Burrows1993}
{Burrows} A.,  {Hubbard} W.~B.,  {Saumon} D.,   {Lunine} J.~I.,  1993, \mn@doi
  [\apj] {10.1086/172427}, \href
  {https://ui.adsabs.harvard.edu/abs/1993ApJ...406..158B} {406, 158}

\bibitem[\protect\citeauthoryear{Capuzzo-Dolcetta \& Fragione}{Capuzzo-Dolcetta
  \& Fragione}{2015}]{Capuzzo2015}
Capuzzo-Dolcetta R.,  Fragione G.,  2015, \mn@doi [\mnras]
  {10.1093/mnras/stv2123}, 454, 2677

\bibitem[\protect\citeauthoryear{Chan, Müller, Heger, Pakmor  \&
  Springel}{Chan et~al.}{2018}]{Chan2018}
Chan C.,  Müller B.,  Heger A.,  Pakmor R.,   Springel V.,  2018, \mn@doi
  [\apj] {10.3847/2041-8213/aaa28c}, 852, L19

\bibitem[\protect\citeauthoryear{{Chan}, {M{\"u}ller}  \& {Heger}}{{Chan}
  et~al.}{2020}]{Chan2020}
{Chan} C.,  {M{\"u}ller} B.,   {Heger} A.,  2020, \mn@doi [\mnras]
  {10.1093/mnras/staa1431}, \href
  {https://ui.adsabs.harvard.edu/abs/2020MNRAS.495.3751C} {495, 3751}

\bibitem[\protect\citeauthoryear{{Chomiuk} \& {Povich}}{{Chomiuk} \&
  {Povich}}{2011}]{Chomiuk2011}
{Chomiuk} L.,  {Povich} M.~S.,  2011, \mn@doi [\aj]
  {10.1088/0004-6256/142/6/197}, \href
  {https://ui.adsabs.harvard.edu/abs/2011AJ....142..197C} {142, 197}

\bibitem[\protect\citeauthoryear{{Claeys}, {Pols}, {Izzard}, {Vink}  \&
  {Verbunt}}{{Claeys} et~al.}{2014}]{Claeys2014}
{Claeys} J.~S.~W.,  {Pols} O.~R.,  {Izzard} R.~G.,  {Vink} J.,   {Verbunt}
  F.~W.~M.,  2014, \mn@doi [\aap] {10.1051/0004-6361/201322714}, \href
  {https://ui.adsabs.harvard.edu/abs/2014A&A...563A..83C} {563, A83}

\bibitem[\protect\citeauthoryear{{Contigiani}, {Rossi}  \&
  {Marchetti}}{{Contigiani} et~al.}{2019}]{Contigiani2018}
{Contigiani} O.,  {Rossi} E.~M.,   {Marchetti} T.,  2019, \mn@doi [\mnras]
  {10.1093/mnras/stz1547}, \href
  {https://ui.adsabs.harvard.edu/abs/2019MNRAS.487.4025C} {487, 4025}

\bibitem[\protect\citeauthoryear{{Davis}, {Kolb}  \& {Knigge}}{{Davis}
  et~al.}{2012}]{Davis2012}
{Davis} P.~J.,  {Kolb} U.,   {Knigge} C.,  2012, \mn@doi [\mnras]
  {10.1111/j.1365-2966.2011.19690.x}, \href
  {https://ui.adsabs.harvard.edu/abs/2012MNRAS.419..287D} {419, 287}

\bibitem[\protect\citeauthoryear{{De Marco}, {Passy}, {Moe}, {Herwig}, {Mac
  Low}  \& {Paxton}}{{De Marco} et~al.}{2011}]{deMarco2011}
{De Marco} O.,  {Passy} J.-C.,  {Moe} M.,  {Herwig} F.,  {Mac Low} M.-M.,
  {Paxton} B.,  2011, \mn@doi [\mnras] {10.1111/j.1365-2966.2010.17891.x},
  \href {https://ui.adsabs.harvard.edu/abs/2011MNRAS.411.2277D} {411, 2277}

\bibitem[\protect\citeauthoryear{{Dewi} \& {Tauris}}{{Dewi} \&
  {Tauris}}{2000}]{Dewi2000}
{Dewi} J.~D.~M.,  {Tauris} T.~M.,  2000, \aap, \href
  {http://adsabs.harvard.edu/abs/2000A%26A...360.1043D} {360, 1043}

\bibitem[\protect\citeauthoryear{{Diehl} et~al.,}{{Diehl}
  et~al.}{2006}]{Diehl2006}
{Diehl} R.,  et~al., 2006, \mn@doi [\nat] {10.1038/nature04364}, \href
  {https://ui.adsabs.harvard.edu/abs/2006Natur.439...45D} {439, 45}

\bibitem[\protect\citeauthoryear{{Dominik}, {Belczynski}, {Fryer}, {Holz},
  {Berti}, {Bulik}, {Mand el}  \& {O'Shaughnessy}}{{Dominik}
  et~al.}{2012}]{Dominik2012}
{Dominik} M.,  {Belczynski} K.,  {Fryer} C.,  {Holz} D.~E.,  {Berti} E.,
  {Bulik} T.,  {Mand el} I.,   {O'Shaughnessy} R.,  2012, \mn@doi [\apj]
  {10.1088/0004-637X/759/1/52}, \href
  {https://ui.adsabs.harvard.edu/abs/2012ApJ...759...52D} {759, 52}

\bibitem[\protect\citeauthoryear{{Dray}, {Dale}, {Beer}, {Napiwotzki}  \&
  {King}}{{Dray} et~al.}{2005}]{Dray2005}
{Dray} L.~M.,  {Dale} J.~E.,  {Beer} M.~E.,  {Napiwotzki} R.,   {King} A.~R.,
  2005, \mn@doi [\mnras] {10.1111/j.1365-2966.2005.09536.x}, \href
  {http://adsabs.harvard.edu/abs/2005MNRAS.364...59D} {364, 59}

\bibitem[\protect\citeauthoryear{{Duch{\^e}ne} \& {Kraus}}{{Duch{\^e}ne} \&
  {Kraus}}{2013}]{Duchene2013}
{Duch{\^e}ne} G.,  {Kraus} A.,  2013, \mn@doi [\araa]
  {10.1146/annurev-astro-081710-102602}, \href
  {https://ui.adsabs.harvard.edu/abs/2013ARA&A..51..269D} {51, 269}

\bibitem[\protect\citeauthoryear{{Dunstall} et~al.,}{{Dunstall}
  et~al.}{2015}]{Dunstall2015}
{Dunstall} P.~R.,  et~al., 2015, \mn@doi [\aap] {10.1051/0004-6361/201526192},
  \href {https://ui.adsabs.harvard.edu/abs/2015A&A...580A..93D} {580, A93}

\bibitem[\protect\citeauthoryear{{Ebinger}, {Curtis}, {Fr{\"o}hlich}, {Hempel},
  {Perego}, {Liebend{\"o}rfer}  \& {Thielemann}}{{Ebinger}
  et~al.}{2019}]{Ebinger2019}
{Ebinger} K.,  {Curtis} S.,  {Fr{\"o}hlich} C.,  {Hempel} M.,  {Perego} A.,
  {Liebend{\"o}rfer} M.,   {Thielemann} F.-K.,  2019, \mn@doi [\apj]
  {10.3847/1538-4357/aae7c9}, \href
  {https://ui.adsabs.harvard.edu/abs/2019ApJ...870....1E} {870, 1}

\bibitem[\protect\citeauthoryear{{Edelmann}, {Napiwotzki}, {Heber},
  {Christlieb}  \& {Reimers}}{{Edelmann} et~al.}{2005}]{Edelmann2005}
{Edelmann} H.,  {Napiwotzki} R.,  {Heber} U.,  {Christlieb} N.,   {Reimers} D.,
   2005, \mn@doi [\apjl] {10.1086/498940}, \href
  {https://ui.adsabs.harvard.edu/abs/2005ApJ...634L.181E} {634, L181}

\bibitem[\protect\citeauthoryear{{Eldridge}}{{Eldridge}}{2009}]{Eldridge2009}
{Eldridge} J.~J.,  2009, \mn@doi [\mnras] {10.1111/j.1745-3933.2009.00753.x},
  \href {https://ui.adsabs.harvard.edu/abs/2009MNRAS.400L..20E} {400, L20}

\bibitem[\protect\citeauthoryear{{Eldridge}, {Langer}  \& {Tout}}{{Eldridge}
  et~al.}{2011}]{Eldridge2011}
{Eldridge} J.~J.,  {Langer} N.,   {Tout} C.~A.,  2011, \mn@doi [\mnras]
  {10.1111/j.1365-2966.2011.18650.x}, \href
  {https://ui.adsabs.harvard.edu/abs/2011MNRAS.414.3501E} {414, 3501}

\bibitem[\protect\citeauthoryear{{Erkal}, {Boubert}, {Gualandris}, {Evans}  \&
  {Antonini}}{{Erkal} et~al.}{2019}]{Erkal2019}
{Erkal} D.,  {Boubert} D.,  {Gualandris} A.,  {Evans} N.~W.,   {Antonini} F.,
  2019, \mn@doi [\mnras] {10.1093/mnras/sty2674}, \href
  {https://ui.adsabs.harvard.edu/abs/2019MNRAS.483.2007E} {483, 2007}

\bibitem[\protect\citeauthoryear{{Ertl}, {Woosley}, {Sukhbold}  \&
  {Janka}}{{Ertl} et~al.}{2020}]{Ertl2020}
{Ertl} T.,  {Woosley} S.~E.,  {Sukhbold} T.,   {Janka} H.~T.,  2020, \mn@doi
  [\apj] {10.3847/1538-4357/ab6458}, \href
  {https://ui.adsabs.harvard.edu/abs/2020ApJ...890...51E} {890, 51}

\bibitem[\protect\citeauthoryear{{Evans} et~al.,}{{Evans}
  et~al.}{2011}]{Evans2011}
{Evans} C.~J.,  et~al., 2011, \mn@doi [\aap] {10.1051/0004-6361/201116782},
  \href {https://ui.adsabs.harvard.edu/abs/2011A&A...530A.108E} {530, A108}

\bibitem[\protect\citeauthoryear{{Fender}, {Maccarone}  \& {Heywood}}{{Fender}
  et~al.}{2013}]{Fender2013}
{Fender} R.~P.,  {Maccarone} T.~J.,   {Heywood} I.,  2013, \mn@doi [\mnras]
  {10.1093/mnras/sts688}, \href
  {https://ui.adsabs.harvard.edu/abs/2013MNRAS.430.1538F} {430, 1538}

\bibitem[\protect\citeauthoryear{{Fragione} \& {Capuzzo-Dolcetta}}{{Fragione}
  \& {Capuzzo-Dolcetta}}{2016}]{Fragione2016}
{Fragione} G.,  {Capuzzo-Dolcetta} R.,  2016, \mn@doi [\mnras]
  {10.1093/mnras/stw531}, \href
  {https://ui.adsabs.harvard.edu/abs/2016MNRAS.458.2596F} {458, 2596}

\bibitem[\protect\citeauthoryear{{Fragione} \& {Gualandris}}{{Fragione} \&
  {Gualandris}}{2019}]{Fragione2019}
{Fragione} G.,  {Gualandris} A.,  2019, \mn@doi [\mnras]
  {10.1093/mnras/stz2451}, \href
  {https://ui.adsabs.harvard.edu/abs/2019MNRAS.489.4543F} {489, 4543}

\bibitem[\protect\citeauthoryear{{Fragos}, {Willems}, {Kalogera}, {Ivanova},
  {Rockefeller}, {Fryer}  \& {Young}}{{Fragos} et~al.}{2009}]{Fragos2009}
{Fragos} T.,  {Willems} B.,  {Kalogera} V.,  {Ivanova} N.,  {Rockefeller} G.,
  {Fryer} C.~L.,   {Young} P.~A.,  2009, \mn@doi [\apj]
  {10.1088/0004-637X/697/2/1057}, \href
  {https://ui.adsabs.harvard.edu/abs/2009ApJ...697.1057F} {697, 1057}

\bibitem[\protect\citeauthoryear{{Fragos}, {Andrews}, {Ramirez-Ruiz}, {Meynet},
  {Kalogera}, {Taam}  \& {Zezas}}{{Fragos} et~al.}{2019}]{Fragos2019}
{Fragos} T.,  {Andrews} J.~J.,  {Ramirez-Ruiz} E.,  {Meynet} G.,  {Kalogera}
  V.,  {Taam} R.~E.,   {Zezas} A.,  2019, \mn@doi [\apjl]
  {10.3847/2041-8213/ab40d1}, \href
  {https://ui.adsabs.harvard.edu/abs/2019ApJ...883L..45F} {883, L45}

\bibitem[\protect\citeauthoryear{{Fryer}, {Belczynski}, {Wiktorowicz},
  {Dominik}, {Kalogera}  \& {Holz}}{{Fryer} et~al.}{2012}]{Fryer2012}
{Fryer} C.~L.,  {Belczynski} K.,  {Wiktorowicz} G.,  {Dominik} M.,  {Kalogera}
  V.,   {Holz} D.~E.,  2012, \mn@doi [\apj] {10.1088/0004-637X/749/1/91}, \href
  {https://ui.adsabs.harvard.edu/abs/2012ApJ...749...91F} {749, 91}

\bibitem[\protect\citeauthoryear{{Fuhrmann}}{{Fuhrmann}}{2008}]{Fuhrmann2008}
{Fuhrmann} K.,  2008, \mn@doi [\mnras] {10.1111/j.1365-2966.2007.12671.x},
  \href {https://ui.adsabs.harvard.edu/abs/2008MNRAS.384..173F} {384, 173}

\bibitem[\protect\citeauthoryear{{Fujii} \& {Portegies Zwart}}{{Fujii} \&
  {Portegies Zwart}}{2011}]{Fujii2011}
{Fujii} M.~S.,  {Portegies Zwart} S.,  2011, \mn@doi [Science]
  {10.1126/science.1211927}, \href
  {https://ui.adsabs.harvard.edu/abs/2011Sci...334.1380F} {334, 1380}

\bibitem[\protect\citeauthoryear{{Gaggero}, {Bertone}, {Calore}, {Connors},
  {Lovell}, {Markoff}  \& {Storm}}{{Gaggero} et~al.}{2017}]{Gaggero2017}
{Gaggero} D.,  {Bertone} G.,  {Calore} F.,  {Connors} R. M.~T.,  {Lovell} M.,
  {Markoff} S.,   {Storm} E.,  2017, \mn@doi [\prl]
  {10.1103/PhysRevLett.118.241101}, \href
  {https://ui.adsabs.harvard.edu/abs/2017PhRvL.118x1101G} {118, 241101}

\bibitem[\protect\citeauthoryear{{Gaia Collaboration} et~al.,}{{Gaia
  Collaboration} et~al.}{2016}]{Gaia2016}
{Gaia Collaboration} et~al., 2016, \mn@doi [\aap]
  {10.1051/0004-6361/201629512}, \href
  {https://ui.adsabs.harvard.edu/abs/2016A%26A...595A...2G} {595, A2}

\bibitem[\protect\citeauthoryear{{Gaia Collaboration} et~al.,}{{Gaia
  Collaboration} et~al.}{2018}]{Gaia2018}
{Gaia Collaboration} et~al., 2018, \mn@doi [\aap]
  {10.1051/0004-6361/201833051}, \href
  {https://ui.adsabs.harvard.edu/abs/2018A&A...616A...1G} {616, A1}

\bibitem[\protect\citeauthoryear{{Geier} et~al.,}{{Geier}
  et~al.}{2013}]{Geier2013}
{Geier} S.,  et~al., 2013, \mn@doi [\aap] {10.1051/0004-6361/201321395}, \href
  {https://ui.adsabs.harvard.edu/abs/2013A&A...554A..54G} {554, A54}

\bibitem[\protect\citeauthoryear{{Geier} et~al.,}{{Geier}
  et~al.}{2015}]{Geier2015}
{Geier} S.,  et~al., 2015, \mn@doi [Science] {10.1126/science.1259063}, \href
  {https://ui.adsabs.harvard.edu/abs/2015Sci...347.1126G} {347, 1126}

\bibitem[\protect\citeauthoryear{{Genzel}, {Eisenhauer}  \&
  {Gillessen}}{{Genzel} et~al.}{2010}]{Genzel2010}
{Genzel} R.,  {Eisenhauer} F.,   {Gillessen} S.,  2010, \mn@doi [Reviews of
  Modern Physics] {10.1103/RevModPhys.82.3121}, \href
  {https://ui.adsabs.harvard.edu/abs/2010RvMP...82.3121G} {82, 3121}

\bibitem[\protect\citeauthoryear{{Gnedin}, {Gould}, {Miralda-Escud{\'e}}  \&
  {Zentner}}{{Gnedin} et~al.}{2005}]{Gnedin2005}
{Gnedin} O.~Y.,  {Gould} A.,  {Miralda-Escud{\'e}} J.,   {Zentner} A.~R.,
  2005, \mn@doi [\apj] {10.1086/496958}, \href
  {https://ui.adsabs.harvard.edu/abs/2005ApJ...634..344G} {634, 344}

\bibitem[\protect\citeauthoryear{{Graczyk} et~al.,}{{Graczyk}
  et~al.}{2011}]{Graczyk2011}
{Graczyk} D.,  et~al., 2011, \actaa, \href
  {https://ui.adsabs.harvard.edu/abs/2011AcA....61..103G} {61, 103}

\bibitem[\protect\citeauthoryear{{Gualandris} \& {Portegies
  Zwart}}{{Gualandris} \& {Portegies Zwart}}{2007}]{Gualandris2007}
{Gualandris} A.,  {Portegies Zwart} S.,  2007, \mn@doi [\mnras]
  {10.1111/j.1745-3933.2007.00280.x}, \href
  {https://ui.adsabs.harvard.edu/abs/2007MNRAS.376L..29G} {376, L29}

\bibitem[\protect\citeauthoryear{{Gualandris}, {Portegies Zwart}  \&
  {Sipior}}{{Gualandris} et~al.}{2005}]{Gualandris2005}
{Gualandris} A.,  {Portegies Zwart} S.,   {Sipior} M.~S.,  2005, \mn@doi
  [\mnras] {10.1111/j.1365-2966.2005.09433.x}, \href
  {https://ui.adsabs.harvard.edu/abs/2005MNRAS.363..223G} {363, 223}

\bibitem[\protect\citeauthoryear{{Guillochon} \& {Loeb}}{{Guillochon} \&
  {Loeb}}{2015}]{Guillochon2015}
{Guillochon} J.,  {Loeb} A.,  2015, \mn@doi [\apj]
  {10.1088/0004-637X/806/1/124}, \href
  {https://ui.adsabs.harvard.edu/abs/2015ApJ...806..124G} {806, 124}

\bibitem[\protect\citeauthoryear{{Guillochon}, {Parrent}, {Kelley}  \&
  {Margutti}}{{Guillochon} et~al.}{2017}]{Guillochon2017}
{Guillochon} J.,  {Parrent} J.,  {Kelley} L.~Z.,   {Margutti} R.,  2017,
  \mn@doi [\apj] {10.3847/1538-4357/835/1/64}, \href
  {https://ui.adsabs.harvard.edu/abs/2017ApJ...835...64G} {835, 64}

\bibitem[\protect\citeauthoryear{{Gunn} \& {Ostriker}}{{Gunn} \&
  {Ostriker}}{1970}]{Gunn1970}
{Gunn} J.~E.,  {Ostriker} J.~P.,  1970, \mn@doi [\apj] {10.1086/150487}, \href
  {https://ui.adsabs.harvard.edu/abs/1970ApJ...160..979G} {160, 979}

\bibitem[\protect\citeauthoryear{{Gvaramadze}}{{Gvaramadze}}{2009}]{Gvaramadze2009}
{Gvaramadze} V.~V.,  2009, \mn@doi [\mnras] {10.1111/j.1745-3933.2009.00648.x},
  \href {https://ui.adsabs.harvard.edu/abs/2009MNRAS.395L..85G} {395, L85}

\bibitem[\protect\citeauthoryear{{Han}, {Podsiadlowski}, {Maxted}, {Marsh}  \&
  {Ivanova}}{{Han} et~al.}{2002}]{Han2002}
{Han} Z.,  {Podsiadlowski} P.,  {Maxted} P.~F.~L.,  {Marsh} T.~R.,   {Ivanova}
  N.,  2002, \mn@doi [\mnras] {10.1046/j.1365-8711.2002.05752.x}, \href
  {https://ui.adsabs.harvard.edu/abs/2002MNRAS.336..449H} {336, 449}

\bibitem[\protect\citeauthoryear{{Hansen}}{{Hansen}}{2003}]{Hansen2003}
{Hansen} B. M.~S.,  2003, \mn@doi [\apj] {10.1086/344782}, \href
  {https://ui.adsabs.harvard.edu/abs/2003ApJ...582..915H} {582, 915}

\bibitem[\protect\citeauthoryear{{Hattori}, {Valluri}, {Castro}, {Roederer},
  {Mahler}  \& {Khullar}}{{Hattori} et~al.}{2019}]{Hattori2019}
{Hattori} K.,  {Valluri} M.,  {Castro} N.,  {Roederer} I.~U.,  {Mahler} G.,
  {Khullar} G.,  2019, \mn@doi [\apj] {10.3847/1538-4357/ab05c8}, \href
  {https://ui.adsabs.harvard.edu/abs/2019ApJ...873..116H} {873, 116}

\bibitem[\protect\citeauthoryear{{Heber}, {Hirsch}, {Edelmann}, {Napiwotzki},
  {O'Toole}, {Brown}  \& {Altmann}}{{Heber} et~al.}{2008a}]{Heber2008BHB}
{Heber} U.,  {Hirsch} H.~A.,  {Edelmann} H.,  {Napiwotzki} R.,  {O'Toole}
  S.~J.,  {Brown} W.,   {Altmann} M.,  2008a, {Hypervelocity Stars: Young and
  Heavy or Old and Light?}.
p.~167

\bibitem[\protect\citeauthoryear{{Heber}, {Edelmann}, {Napiwotzki}, {Altmann}
  \& {Scholz}}{{Heber} et~al.}{2008b}]{Heber2008}
{Heber} U.,  {Edelmann} H.,  {Napiwotzki} R.,  {Altmann} M.,   {Scholz} R.~D.,
  2008b, \mn@doi [\aap] {10.1051/0004-6361:200809767}, \href
  {https://ui.adsabs.harvard.edu/abs/2008A&A...483L..21H} {483, L21}

\bibitem[\protect\citeauthoryear{{Hills}}{{Hills}}{1988}]{Hills1988}
{Hills} J.~G.,  1988, \mn@doi [\nat] {10.1038/331687a0}, \href
  {https://ui.adsabs.harvard.edu/abs/1988Natur.331..687H} {331, 687}

\bibitem[\protect\citeauthoryear{{Hirsch}, {Heber}, {O'Toole}  \&
  {Bresolin}}{{Hirsch} et~al.}{2005}]{Hirsch2005}
{Hirsch} H.~A.,  {Heber} U.,  {O'Toole} S.~J.,   {Bresolin} F.,  2005, \mn@doi
  [\aap] {10.1051/0004-6361:200500212}, \href
  {https://ui.adsabs.harvard.edu/abs/2005A%26A...444L..61H} {444, L61}

\bibitem[\protect\citeauthoryear{{Hobbs}, {Lorimer}, {Lyne}  \&
  {Kramer}}{{Hobbs} et~al.}{2005}]{Hobbs2005}
{Hobbs} G.,  {Lorimer} D.~R.,  {Lyne} A.~G.,   {Kramer} M.,  2005, \mn@doi
  [\mnras] {10.1111/j.1365-2966.2005.09087.x}, \href
  {https://ui.adsabs.harvard.edu/abs/2005MNRAS.360..974H} {360, 974}

\bibitem[\protect\citeauthoryear{{Holland-Ashford}, {Lopez}, {Auchettl},
  {Temim}  \& {Ramirez-Ruiz}}{{Holland-Ashford}
  et~al.}{2017}]{HollandAshford2017}
{Holland-Ashford} T.,  {Lopez} L.~A.,  {Auchettl} K.,  {Temim} T.,
  {Ramirez-Ruiz} E.,  2017, \mn@doi [\apj] {10.3847/1538-4357/aa7a5c}, \href
  {https://ui.adsabs.harvard.edu/abs/2017ApJ...844...84H} {844, 84}

\bibitem[\protect\citeauthoryear{{Holmberg}, {Nordstr{\"o}m}  \&
  {Andersen}}{{Holmberg} et~al.}{2009}]{Holmberg2009}
{Holmberg} J.,  {Nordstr{\"o}m} B.,   {Andersen} J.,  2009, \mn@doi [\aap]
  {10.1051/0004-6361/200811191}, \href
  {https://ui.adsabs.harvard.edu/abs/2009A&A...501..941H} {501, 941}

\bibitem[\protect\citeauthoryear{{Hoogerwerf}, {de Bruijne}  \& {de
  Zeeuw}}{{Hoogerwerf} et~al.}{2001}]{Hoogerwerf2001}
{Hoogerwerf} R.,  {de Bruijne} J.~H.~J.,   {de Zeeuw} P.~T.,  2001, \mn@doi
  [\aap] {10.1051/0004-6361:20000014}, \href
  {https://ui.adsabs.harvard.edu/abs/2001A&A...365...49H} {365, 49}

\bibitem[\protect\citeauthoryear{{Hopman}}{{Hopman}}{2009}]{Hopman2009}
{Hopman} C.,  2009, \mn@doi [\apj] {10.1088/0004-637X/700/2/1933}, \href
  {https://ui.adsabs.harvard.edu/abs/2009ApJ...700.1933H} {700, 1933}

\bibitem[\protect\citeauthoryear{{Huang} et~al.,}{{Huang}
  et~al.}{2017}]{Huang2017}
{Huang} Y.,  et~al., 2017, \mn@doi [\apjl] {10.3847/2041-8213/aa894b}, \href
  {https://ui.adsabs.harvard.edu/abs/2017ApJ...847L...9H} {847, L9}

\bibitem[\protect\citeauthoryear{{Hurley}, {Pols}  \& {Tout}}{{Hurley}
  et~al.}{2000}]{Hurley2000}
{Hurley} J.~R.,  {Pols} O.~R.,   {Tout} C.~A.,  2000, \mn@doi [\mnras]
  {10.1046/j.1365-8711.2000.03426.x}, \href
  {https://ui.adsabs.harvard.edu/abs/2000MNRAS.315..543H} {315, 543}

\bibitem[\protect\citeauthoryear{{Hurley}, {Tout}  \& {Pols}}{{Hurley}
  et~al.}{2002}]{Hurley2002}
{Hurley} J.~R.,  {Tout} C.~A.,   {Pols} O.~R.,  2002, \mn@doi [\mnras]
  {10.1046/j.1365-8711.2002.05038.x}, \href
  {https://ui.adsabs.harvard.edu/abs/2002MNRAS.329..897H} {329, 897}

\bibitem[\protect\citeauthoryear{{Irrgang}, {Kreuzer}  \& {Heber}}{{Irrgang}
  et~al.}{2018}]{Irrgang2018}
{Irrgang} A.,  {Kreuzer} S.,   {Heber} U.,  2018, \mn@doi [\aap]
  {10.1051/0004-6361/201833874}, \href
  {https://ui.adsabs.harvard.edu/abs/2018A&A...620A..48I} {620, A48}

\bibitem[\protect\citeauthoryear{{Irrgang}, {Geier}, {Heber}, {Kupfer}  \&
  {F{\"u}rst}}{{Irrgang} et~al.}{2019}]{Irrgang2019}
{Irrgang} A.,  {Geier} S.,  {Heber} U.,  {Kupfer} T.,   {F{\"u}rst} F.,  2019,
  \mn@doi [\aap] {10.1051/0004-6361/201935429}, \href
  {https://ui.adsabs.harvard.edu/abs/2019A&A...628L...5I} {628, L5}

\bibitem[\protect\citeauthoryear{{Ivanova}}{{Ivanova}}{2002}]{Ivanova2002}
{Ivanova} N.,  2002, PhD thesis, University of Oxford

\bibitem[\protect\citeauthoryear{{Ivanova} \& {Chaichenets}}{{Ivanova} \&
  {Chaichenets}}{2011}]{Ivanova2011}
{Ivanova} N.,  {Chaichenets} S.,  2011, \mn@doi [\apjl]
  {10.1088/2041-8205/731/2/L36}, \href
  {https://ui.adsabs.harvard.edu/abs/2011ApJ...731L..36I} {731, L36}

\bibitem[\protect\citeauthoryear{{Ivanova} et~al.,}{{Ivanova}
  et~al.}{2013a}]{Ivanova2013}
{Ivanova} N.,  et~al., 2013a, \mn@doi [\aapr] {10.1007/s00159-013-0059-2},
  \href {https://ui.adsabs.harvard.edu/abs/2013A&ARv..21...59I} {21, 59}

\bibitem[\protect\citeauthoryear{{Ivanova}, {Justham}, {Avendano Nandez}  \&
  {Lombardi}}{{Ivanova} et~al.}{2013b}]{Ivanova2013b}
{Ivanova} N.,  {Justham} S.,  {Avendano Nandez} J.~L.,   {Lombardi} J.~C.,
  2013b, \mn@doi [Science] {10.1126/science.1225540}, \href
  {https://ui.adsabs.harvard.edu/abs/2013Sci...339..433I} {339, 433}

\bibitem[\protect\citeauthoryear{{Izzard}, {Tout}, {Karakas}  \&
  {Pols}}{{Izzard} et~al.}{2004}]{Izzard2004}
{Izzard} R.~G.,  {Tout} C.~A.,  {Karakas} A.~I.,   {Pols} O.~R.,  2004, \mn@doi
  [\mnras] {10.1111/j.1365-2966.2004.07446.x}, \href
  {https://ui.adsabs.harvard.edu/abs/2004MNRAS.350..407I} {350, 407}

\bibitem[\protect\citeauthoryear{{Izzard}, {Dray}, {Karakas}, {Lugaro}  \&
  {Tout}}{{Izzard} et~al.}{2006}]{Izzard2006}
{Izzard} R.~G.,  {Dray} L.~M.,  {Karakas} A.~I.,  {Lugaro} M.,   {Tout} C.~A.,
  2006, \mn@doi [\aap] {10.1051/0004-6361:20066129}, \href
  {https://ui.adsabs.harvard.edu/abs/2006A&A...460..565I} {460, 565}

\bibitem[\protect\citeauthoryear{{Izzard}, {Glebbeek}, {Stancliffe}  \&
  {Pols}}{{Izzard} et~al.}{2009}]{Izzard2009}
{Izzard} R.~G.,  {Glebbeek} E.,  {Stancliffe} R.~J.,   {Pols} O.~R.,  2009,
  \mn@doi [\aap] {10.1051/0004-6361/200912827}, \href
  {https://ui.adsabs.harvard.edu/abs/2009A&A...508.1359I} {508, 1359}

\bibitem[\protect\citeauthoryear{{Janka}}{{Janka}}{2013}]{Janka2013}
{Janka} H.-T.,  2013, \mn@doi [\mnras] {10.1093/mnras/stt1106}, \href
  {https://ui.adsabs.harvard.edu/abs/2013MNRAS.434.1355J} {434, 1355}

\bibitem[\protect\citeauthoryear{{Janka}}{{Janka}}{2017}]{Janka2017}
{Janka} H.-T.,  2017, \mn@doi [\apj] {10.3847/1538-4357/aa618e}, \href
  {https://ui.adsabs.harvard.edu/abs/2017ApJ...837...84J} {837, 84}

\bibitem[\protect\citeauthoryear{{Janka} \& {Mueller}}{{Janka} \&
  {Mueller}}{1994}]{Janka1994}
{Janka} H.~T.,  {Mueller} E.,  1994, \aap, \href
  {https://ui.adsabs.harvard.edu/abs/1994A&A...290..496J} {290, 496}

\bibitem[\protect\citeauthoryear{{Jilinski}, {Ortega}, {Drake}  \& {de la
  Reza}}{{Jilinski} et~al.}{2010}]{Jilinski2010}
{Jilinski} E.,  {Ortega} V.~G.,  {Drake} N.~A.,   {de la Reza} R.,  2010,
  \mn@doi [\apj] {10.1088/0004-637X/721/1/469}, \href
  {https://ui.adsabs.harvard.edu/abs/2010ApJ...721..469J} {721, 469}

\bibitem[\protect\citeauthoryear{{Justham}, {Wolf}, {Podsiadlowski}  \&
  {Han}}{{Justham} et~al.}{2009}]{Justham2009}
{Justham} S.,  {Wolf} C.,  {Podsiadlowski} P.,   {Han} Z.,  2009, \mn@doi
  [\aap] {10.1051/0004-6361:200810106}, \href
  {https://ui.adsabs.harvard.edu/abs/2009A&A...493.1081J} {493, 1081}

\bibitem[\protect\citeauthoryear{{Kalogera}}{{Kalogera}}{1996}]{Kalogera1996}
{Kalogera} V.,  1996, \mn@doi [\apj] {10.1086/177974}, \href
  {https://ui.adsabs.harvard.edu/abs/1996ApJ...471..352K} {471, 352}

\bibitem[\protect\citeauthoryear{Kalogera}{Kalogera}{2000}]{Kalogera2000}
Kalogera V.,  2000, \mn@doi [\apj] {10.1086/309400}, 541, 319

\bibitem[\protect\citeauthoryear{{Katsuda} et~al.,}{{Katsuda}
  et~al.}{2018}]{Katsuda2018}
{Katsuda} S.,  et~al., 2018, \mn@doi [\apj] {10.3847/1538-4357/aab092}, \href
  {https://ui.adsabs.harvard.edu/abs/2018ApJ...856...18K} {856, 18}

\bibitem[\protect\citeauthoryear{{Katz}}{{Katz}}{1975}]{Katz1975}
{Katz} J.~I.,  1975, \mn@doi [\nat] {10.1038/253698a0}, \href
  {https://ui.adsabs.harvard.edu/abs/1975Natur.253..698K} {253, 698}

\bibitem[\protect\citeauthoryear{{Kenyon}, {Bromley}, {Geller}  \&
  {Brown}}{{Kenyon} et~al.}{2008}]{Kenyon2008}
{Kenyon} S.~J.,  {Bromley} B.~C.,  {Geller} M.~J.,   {Brown} W.~R.,  2008,
  \mn@doi [\apj] {10.1086/587738}, \href
  {https://ui.adsabs.harvard.edu/abs/2008ApJ...680..312K} {680, 312}

\bibitem[\protect\citeauthoryear{{Kenyon}, {Bromley}, {Brown}  \&
  {Geller}}{{Kenyon} et~al.}{2014}]{Kenyon2014}
{Kenyon} S.~J.,  {Bromley} B.~C.,  {Brown} W.~R.,   {Geller} M.~J.,  2014,
  \mn@doi [\apj] {10.1088/0004-637X/793/2/122}, \href
  {https://ui.adsabs.harvard.edu/abs/2014ApJ...793..122K} {793, 122}

\bibitem[\protect\citeauthoryear{{Knigge}, {Coe}  \& {Podsiadlowski}}{{Knigge}
  et~al.}{2011}]{Knigge2011}
{Knigge} C.,  {Coe} M.~J.,   {Podsiadlowski} P.,  2011, \mn@doi [\nat]
  {10.1038/nature10529}, \href
  {https://ui.adsabs.harvard.edu/abs/2011Natur.479..372K} {479, 372}

\bibitem[\protect\citeauthoryear{{Kobulnicky} \& {Fryer}}{{Kobulnicky} \&
  {Fryer}}{2007}]{Kobulnicky2007}
{Kobulnicky} H.~A.,  {Fryer} C.~L.,  2007, \mn@doi [\apj] {10.1086/522073},
  \href {https://ui.adsabs.harvard.edu/abs/2007ApJ...670..747K} {670, 747}

\bibitem[\protect\citeauthoryear{{Kobulnicky} et~al.,}{{Kobulnicky}
  et~al.}{2014}]{Kobulnicky2014}
{Kobulnicky} H.~A.,  et~al., 2014, \mn@doi [\apjs]
  {10.1088/0067-0049/213/2/34}, \href
  {https://ui.adsabs.harvard.edu/abs/2014ApJS..213...34K} {213, 34}

\bibitem[\protect\citeauthoryear{{Koposov} et~al.,}{{Koposov}
  et~al.}{2019}]{Koposov2019}
{Koposov} S.~E.,  et~al., 2019, \mn@doi [\mnras] {10.1093/mnras/stz3081}, \href
  {https://ui.adsabs.harvard.edu/abs/2019MNRAS.tmp.2680K} {p.~2680}

\bibitem[\protect\citeauthoryear{{Kreuzer}, {Irrgang}  \& {Heber}}{{Kreuzer}
  et~al.}{2020}]{Kreuzer2020}
{Kreuzer} S.,  {Irrgang} A.,   {Heber} U.,  2020, \mn@doi [\aap]
  {10.1051/0004-6361/202037747}, \href
  {https://ui.adsabs.harvard.edu/abs/2020A&A...637A..53K} {637, A53}

\bibitem[\protect\citeauthoryear{{Kroupa}}{{Kroupa}}{2001}]{Kroupa2001}
{Kroupa} P.,  2001, \mn@doi [\mnras] {10.1046/j.1365-8711.2001.04022.x}, \href
  {https://ui.adsabs.harvard.edu/abs/2001MNRAS.322..231K} {322, 231}

\bibitem[\protect\citeauthoryear{{Kuiper}}{{Kuiper}}{1935}]{Kuiper1935}
{Kuiper} G.~P.,  1935, \mn@doi [\pasp] {10.1086/124531}, \href
  {https://ui.adsabs.harvard.edu/abs/1935PASP...47...15K} {47, 15}

\bibitem[\protect\citeauthoryear{{Lef{\`e}vre}, {Marchenko}, {Moffat}  \&
  {Acker}}{{Lef{\`e}vre} et~al.}{2009}]{Lefevre2009}
{Lef{\`e}vre} L.,  {Marchenko} S.~V.,  {Moffat} A.~F.~J.,   {Acker} A.,  2009,
  \mn@doi [\aap] {10.1051/0004-6361/200912304}, \href
  {https://ui.adsabs.harvard.edu/abs/2009A&A...507.1141L} {507, 1141}

\bibitem[\protect\citeauthoryear{{Lemasle} et~al.,}{{Lemasle}
  et~al.}{2018}]{Lemasle2018}
{Lemasle} B.,  et~al., 2018, \mn@doi [\aap] {10.1051/0004-6361/201834050},
  \href {https://ui.adsabs.harvard.edu/abs/2018A&A...618A.160L} {618, A160}

\bibitem[\protect\citeauthoryear{{Lennon}, {van der Marel}, {Ramos Lerate},
  {O'Mullane}  \& {Sahlmann}}{{Lennon} et~al.}{2017}]{Lennon2017}
{Lennon} D.~J.,  {van der Marel} R.~P.,  {Ramos Lerate} M.,  {O'Mullane} W.,
  {Sahlmann} J.,  2017, \mn@doi [\aap] {10.1051/0004-6361/201630076}, \href
  {https://ui.adsabs.harvard.edu/abs/2017A&A...603A..75L} {603, A75}

\bibitem[\protect\citeauthoryear{{Lennon} et~al.,}{{Lennon}
  et~al.}{2018}]{Lennon2018}
{Lennon} D.~J.,  et~al., 2018, \href
  {http://adsabs.harvard.edu/abs/2018arXiv180508277L} {}

\bibitem[\protect\citeauthoryear{{Leonard}}{{Leonard}}{1991}]{Leonard1991}
{Leonard} P. J.~T.,  1991, \mn@doi [\aj] {10.1086/115704}, \href
  {https://ui.adsabs.harvard.edu/abs/1991AJ....101..562L} {101, 562}

\bibitem[\protect\citeauthoryear{{Leonard} \& {Duncan}}{{Leonard} \&
  {Duncan}}{1990}]{Leonard1990}
{Leonard} P. J.~T.,  {Duncan} M.~J.,  1990, \mn@doi [\aj] {10.1086/115354},
  \href {https://ui.adsabs.harvard.edu/abs/1990AJ.....99..608L} {99, 608}

\bibitem[\protect\citeauthoryear{{Leonard}, {Hills}  \& {Dewey}}{{Leonard}
  et~al.}{1994}]{Leonard1994}
{Leonard} P. J.~T.,  {Hills} J.~G.,   {Dewey} R.~J.,  1994, \mn@doi [\apjl]
  {10.1086/187225}, \href
  {https://ui.adsabs.harvard.edu/abs/1994ApJ...423L..19L} {423, L19}

\bibitem[\protect\citeauthoryear{{Li} et~al.,}{{Li} et~al.}{2018}]{Li2018}
{Li} Y.-B.,  et~al., 2018, \mn@doi [\aj] {10.3847/1538-3881/aad09a}, \href
  {https://ui.adsabs.harvard.edu/abs/2018AJ....156...87L} {156, 87}

\bibitem[\protect\citeauthoryear{{Licquia} \& {Newman}}{{Licquia} \&
  {Newman}}{2015}]{Licquia2015}
{Licquia} T.~C.,  {Newman} J.~A.,  2015, \mn@doi [\apj]
  {10.1088/0004-637X/806/1/96}, \href
  {https://ui.adsabs.harvard.edu/abs/2015ApJ...806...96L} {806, 96}

\bibitem[\protect\citeauthoryear{{Liu}, {Tauris}, {R{\"o}pke}, {Moriya},
  {Kruckow}, {Stancliffe}  \& {Izzard}}{{Liu} et~al.}{2015}]{Liu2015}
{Liu} Z.-W.,  {Tauris} T.~M.,  {R{\"o}pke} F.~K.,  {Moriya} T.~J.,  {Kruckow}
  M.,  {Stancliffe} R.~J.,   {Izzard} R.~G.,  2015, \mn@doi [\aap]
  {10.1051/0004-6361/201526757}, \href
  {https://ui.adsabs.harvard.edu/abs/2015A&A...584A..11L} {584, A11}

\bibitem[\protect\citeauthoryear{{Luo} et~al.,}{{Luo} et~al.}{2016}]{Luo2016}
{Luo} A.-L.,  et~al., 2016, VizieR Online Data Catalog, \href
  {https://ui.adsabs.harvard.edu/abs/2016yCat.5149....0L} {5149}

\bibitem[\protect\citeauthoryear{{Lyne} \& {Lorimer}}{{Lyne} \&
  {Lorimer}}{1994}]{Lyne1994}
{Lyne} A.~G.,  {Lorimer} D.~R.,  1994, \mn@doi [\nat] {10.1038/369127a0}, \href
  {https://ui.adsabs.harvard.edu/abs/1994Natur.369..127L} {369, 127}

\bibitem[\protect\citeauthoryear{{Madigan}, {Pfuhl}, {Levin}, {Gillessen},
  {Genzel}  \& {Perets}}{{Madigan} et~al.}{2014}]{Madigan2014}
{Madigan} A.-M.,  {Pfuhl} O.,  {Levin} Y.,  {Gillessen} S.,  {Genzel} R.,
  {Perets} H.~B.,  2014, \mn@doi [\apj] {10.1088/0004-637X/784/1/23}, \href
  {https://ui.adsabs.harvard.edu/abs/2014ApJ...784...23M} {784, 23}

\bibitem[\protect\citeauthoryear{{Ma{\'\i}z Apell{\'a}niz}, {Pantaleoni
  Gonz{\'a}lez}, {Barb{\'a}}, {Sim{\'o}n-D{\'\i}az}, {Negueruela}, {Lennon},
  {Sota}  \& {Trigueros P{\'a}ez}}{{Ma{\'\i}z Apell{\'a}niz}
  et~al.}{2018}]{MaizApellaniz2018}
{Ma{\'\i}z Apell{\'a}niz} J.,  {Pantaleoni Gonz{\'a}lez} M.,  {Barb{\'a}}
  R.~H.,  {Sim{\'o}n-D{\'\i}az} S.,  {Negueruela} I.,  {Lennon} D.~J.,  {Sota}
  A.,   {Trigueros P{\'a}ez} E.,  2018, \mn@doi [\aap]
  {10.1051/0004-6361/201832787}, \href
  {https://ui.adsabs.harvard.edu/abs/2018A&A...616A.149M} {616, A149}

\bibitem[\protect\citeauthoryear{{Mandel}}{{Mandel}}{2016}]{Mandel2015}
{Mandel} I.,  2016, \mn@doi [\mnras] {10.1093/mnras/stv2733}, \href
  {https://ui.adsabs.harvard.edu/abs/2016MNRAS.456..578M} {456, 578}

\bibitem[\protect\citeauthoryear{{Mandel} \& {M{\"u}ller}}{{Mandel} \&
  {M{\"u}ller}}{2020}]{Mandel2020}
{Mandel} I.,  {M{\"u}ller} B.,  2020, arXiv e-prints, \href
  {https://ui.adsabs.harvard.edu/abs/2020arXiv200608360M} {p. arXiv:2006.08360}

\bibitem[\protect\citeauthoryear{{Marchetti}, {Rossi}  \& {Brown}}{{Marchetti}
  et~al.}{2019}]{Marchetti2018}
{Marchetti} T.,  {Rossi} E.~M.,   {Brown} A.~G.~A.,  2019, \mn@doi [\mnras]
  {10.1093/mnras/sty2592}, \href
  {https://ui.adsabs.harvard.edu/abs/2019MNRAS.490..157M} {490, 157}

\bibitem[\protect\citeauthoryear{{Miyamoto} \& {Nagai}}{{Miyamoto} \&
  {Nagai}}{1975}]{Miyamoto1975}
{Miyamoto} M.,  {Nagai} R.,  1975, \pasj, \href
  {https://ui.adsabs.harvard.edu/abs/1975PASJ...27..533M} {27, 533}

\bibitem[\protect\citeauthoryear{{Moe} \& {Di Stefano}}{{Moe} \& {Di
  Stefano}}{2017}]{Moe2017}
{Moe} M.,  {Di Stefano} R.,  2017, \mn@doi [\apjs] {10.3847/1538-4365/aa6fb6},
  \href {https://ui.adsabs.harvard.edu/abs/2017ApJS..230...15M} {230, 15}

\bibitem[\protect\citeauthoryear{Moe \& Stefano}{Moe \&
  Stefano}{2013}]{Moe2013}
Moe M.,  Stefano R.~D.,  2013, \mn@doi [\apj] {10.1088/0004-637x/778/2/95},
  778, 95

\bibitem[\protect\citeauthoryear{{Navarro}, {Frenk}  \& {White}}{{Navarro}
  et~al.}{1996}]{Navarro1996}
{Navarro} J.~F.,  {Frenk} C.~S.,   {White} S.~D.~M.,  1996, \mn@doi [\apj]
  {10.1086/177173}, \href
  {https://ui.adsabs.harvard.edu/abs/1996ApJ...462..563N} {462, 563}

\bibitem[\protect\citeauthoryear{{Neunteufel}}{{Neunteufel}}{2020}]{Neuntefel2020}
{Neunteufel} P.,  2020, arXiv e-prints, \href
  {https://ui.adsabs.harvard.edu/abs/2020arXiv200611427N} {p. arXiv:2006.11427}

\bibitem[\protect\citeauthoryear{{O'Connor} \& {Ott}}{{O'Connor} \&
  {Ott}}{2011}]{OConnor2011}
{O'Connor} E.,  {Ott} C.~D.,  2011, \mn@doi [\apj]
  {10.1088/0004-637X/730/2/70}, \href
  {https://ui.adsabs.harvard.edu/abs/2011ApJ...730...70O} {730, 70}

\bibitem[\protect\citeauthoryear{{O'Shaughnessy}, {Gerosa}  \&
  {Wysocki}}{{O'Shaughnessy} et~al.}{2017}]{Oshaughnessy2017}
{O'Shaughnessy} R.,  {Gerosa} D.,   {Wysocki} D.,  2017, \mn@doi [\prl]
  {10.1103/PhysRevLett.119.011101}, \href
  {https://ui.adsabs.harvard.edu/abs/2017PhRvL.119a1101O} {119, 011101}

\bibitem[\protect\citeauthoryear{{Oh} \& {Kroupa}}{{Oh} \&
  {Kroupa}}{2016}]{Oh2016}
{Oh} S.,  {Kroupa} P.,  2016, \mn@doi [\aap] {10.1051/0004-6361/201628233},
  \href {https://ui.adsabs.harvard.edu/abs/2016A&A...590A.107O} {590, A107}

\bibitem[\protect\citeauthoryear{{Ohlmann}, {R{\"o}pke}, {Pakmor}, {Springel}
  \& {M{\"u}ller}}{{Ohlmann} et~al.}{2016}]{Ohlmann2016}
{Ohlmann} S.~T.,  {R{\"o}pke} F.~K.,  {Pakmor} R.,  {Springel} V.,
  {M{\"u}ller} E.,  2016, \mn@doi [\mnras] {10.1093/mnrasl/slw144}, \href
  {https://ui.adsabs.harvard.edu/abs/2016MNRAS.462L.121O} {462, L121}

\bibitem[\protect\citeauthoryear{{{\"O}pik}}{{{\"O}pik}}{1924}]{Opik1924}
{{\"O}pik} E.,  1924, Publications of the Tartu Astrofizica Observatory, \href
  {https://ui.adsabs.harvard.edu/abs/1924PTarO..25f...1O} {25, 1}

\bibitem[\protect\citeauthoryear{{Paczynski}}{{Paczynski}}{1976}]{Paczynski1976}
{Paczynski} B.,  1976, in {Eggleton} P.,  {Mitton} S.,   {Whelan} J.,  eds,
  IAU Symposium Vol. 73, Structure and Evolution of Close Binary Systems. p.~75

\bibitem[\protect\citeauthoryear{{Palladino}, {Schlesinger},
  {Holley-Bockelmann}, {Allende Prieto}, {Beers}, {Lee}  \&
  {Schneider}}{{Palladino} et~al.}{2014}]{Palladino2014}
{Palladino} L.~E.,  {Schlesinger} K.~J.,  {Holley-Bockelmann} K.,  {Allende
  Prieto} C.,  {Beers} T.~C.,  {Lee} Y.~S.,   {Schneider} D.~P.,  2014, \mn@doi
  [\apj] {10.1088/0004-637X/780/1/7}, \href
  {https://ui.adsabs.harvard.edu/abs/2014ApJ...780....7P} {780, 7}

\bibitem[\protect\citeauthoryear{{Perets} \& {{\v{S}}ubr}}{{Perets} \&
  {{\v{S}}ubr}}{2012}]{Perets2012}
{Perets} H.~B.,  {{\v{S}}ubr} L.,  2012, \mn@doi [\apj]
  {10.1088/0004-637X/751/2/133}, \href
  {https://ui.adsabs.harvard.edu/abs/2012ApJ...751..133P} {751, 133}

\bibitem[\protect\citeauthoryear{Perets, Hopman  \& Alexander}{Perets
  et~al.}{2007}]{Perets2007}
Perets H.~B.,  Hopman C.,   Alexander T.,  2007, \mn@doi [\apj]
  {10.1086/510377}, 656, 709–720

\bibitem[\protect\citeauthoryear{{Pfahl}}{{Pfahl}}{2005}]{Pfahl2005}
{Pfahl} E.,  2005, \mn@doi [\apj] {10.1086/430167}, \href
  {https://ui.adsabs.harvard.edu/abs/2005ApJ...626..849P} {626, 849}

\bibitem[\protect\citeauthoryear{{Pfahl}, {Rappaport}  \&
  {Podsiadlowski}}{{Pfahl} et~al.}{2002}]{Pfahl2002}
{Pfahl} E.,  {Rappaport} S.,   {Podsiadlowski} P.,  2002, \mn@doi [\apjl]
  {10.1086/341197}, \href
  {https://ui.adsabs.harvard.edu/abs/2002ApJ...571L..37P} {571, L37}

\bibitem[\protect\citeauthoryear{{Piffl} et~al.,}{{Piffl}
  et~al.}{2014}]{Piffl2014}
{Piffl} T.,  et~al., 2014, \mn@doi [\aap] {10.1051/0004-6361/201322531}, \href
  {https://ui.adsabs.harvard.edu/abs/2014A&A...562A..91P} {562, A91}

\bibitem[\protect\citeauthoryear{{Podsiadlowski}, {Langer}, {Poelarends},
  {Rappaport}, {Heger}  \& {Pfahl}}{{Podsiadlowski}
  et~al.}{2004}]{Podsiadlowski2004}
{Podsiadlowski} P.,  {Langer} N.,  {Poelarends} A.~J.~T.,  {Rappaport} S.,
  {Heger} A.,   {Pfahl} E.,  2004, \mn@doi [\apj] {10.1086/421713}, \href
  {https://ui.adsabs.harvard.edu/abs/2004ApJ...612.1044P} {612, 1044}

\bibitem[\protect\citeauthoryear{{Podsiadlowski}, {Ivanova}, {Justham}  \&
  {Rappaport}}{{Podsiadlowski} et~al.}{2010}]{Podsiadlowski2010}
{Podsiadlowski} P.,  {Ivanova} N.,  {Justham} S.,   {Rappaport} S.,  2010,
  \mn@doi [\mnras] {10.1111/j.1365-2966.2010.16751.x}, \href
  {https://ui.adsabs.harvard.edu/abs/2010MNRAS.406..840P} {406, 840}

\bibitem[\protect\citeauthoryear{{Pols}, {Schr{\"o}der}, {Hurley}, {Tout}  \&
  {Eggleton}}{{Pols} et~al.}{1998}]{Pols1998}
{Pols} O.~R.,  {Schr{\"o}der} K.-P.,  {Hurley} J.~R.,  {Tout} C.~A.,
  {Eggleton} P.~P.,  1998, \mn@doi [\mnras] {10.1046/j.1365-8711.1998.01658.x},
  \href {https://ui.adsabs.harvard.edu/abs/1998MNRAS.298..525P} {298, 525}

\bibitem[\protect\citeauthoryear{{Portegies Zwart}}{{Portegies
  Zwart}}{2000}]{Portegies2000}
{Portegies Zwart} S.~F.,  2000, \mn@doi [\apj] {10.1086/317190}, \href
  {https://ui.adsabs.harvard.edu/abs/2000ApJ...544..437P} {544, 437}

\bibitem[\protect\citeauthoryear{{Portegies Zwart} \& {McMillan}}{{Portegies
  Zwart} \& {McMillan}}{2002}]{Portegies2002}
{Portegies Zwart} S.~F.,  {McMillan} S. L.~W.,  2002, \mn@doi [\apj]
  {10.1086/341798}, \href
  {https://ui.adsabs.harvard.edu/abs/2002ApJ...576..899P} {576, 899}

\bibitem[\protect\citeauthoryear{{Postnov} \& {Yungelson}}{{Postnov} \&
  {Yungelson}}{2014}]{Postnov2014}
{Postnov} K.~A.,  {Yungelson} L.~R.,  2014, \mn@doi [Living Reviews in
  Relativity] {10.12942/lrr-2014-3}, \href
  {https://ui.adsabs.harvard.edu/abs/2014LRR....17....3P} {17, 3}

\bibitem[\protect\citeauthoryear{{Poveda}, {Ruiz}  \& {Allen}}{{Poveda}
  et~al.}{1967}]{Poveda1967}
{Poveda} A.,  {Ruiz} J.,   {Allen} C.,  1967, Boletin de los Observatorios
  Tonantzintla y Tacubaya, \href
  {https://ui.adsabs.harvard.edu/abs/1967BOTT....4...86P} {4, 86}

\bibitem[\protect\citeauthoryear{{Przybilla}, {Nieva}, {Heber}, {Firnstein},
  {Butler}, {Napiwotzki}  \& {Edelmann}}{{Przybilla}
  et~al.}{2008a}]{Przybilla2008HVS3}
{Przybilla} N.,  {Nieva} M.~F.,  {Heber} U.,  {Firnstein} M.,  {Butler} K.,
  {Napiwotzki} R.,   {Edelmann} H.,  2008a, \mn@doi [\aap]
  {10.1051/0004-6361:200809391}, \href
  {https://ui.adsabs.harvard.edu/abs/2008A&A...480L..37P} {480, L37}

\bibitem[\protect\citeauthoryear{{Przybilla}, {Fernanda Nieva}, {Heber}  \&
  {Butler}}{{Przybilla} et~al.}{2008b}]{Przybilla2008HD}
{Przybilla} N.,  {Fernanda Nieva} M.,  {Heber} U.,   {Butler} K.,  2008b,
  \mn@doi [\apjl] {10.1086/592245}, \href
  {https://ui.adsabs.harvard.edu/abs/2008ApJ...684L.103P} {684, L103}

\bibitem[\protect\citeauthoryear{{Przybilla}, {Nieva}  \& {Butler}}{{Przybilla}
  et~al.}{2008c}]{Przybilla2008Bmetals}
{Przybilla} N.,  {Nieva} M.-F.,   {Butler} K.,  2008c, \mn@doi [\apjl]
  {10.1086/595618}, \href
  {https://ui.adsabs.harvard.edu/abs/2008ApJ...688L.103P} {688, L103}

\bibitem[\protect\citeauthoryear{{Reg{\H{o}}s} \& {Tout}}{{Reg{\H{o}}s} \&
  {Tout}}{1995}]{Regos1995}
{Reg{\H{o}}s} E.,  {Tout} C.~A.,  1995, \mn@doi [\mnras]
  {10.1093/mnras/273.1.146}, \href
  {https://ui.adsabs.harvard.edu/abs/1995MNRAS.273..146R} {273, 146}

\bibitem[\protect\citeauthoryear{{Renzo} et~al.,}{{Renzo}
  et~al.}{2019a}]{Renzo2019vfts}
{Renzo} M.,  et~al., 2019a, \mn@doi [\mnras] {10.1093/mnrasl/sly194}, \href
  {http://adsabs.harvard.edu/abs/2019MNRAS.482L.102R} {482, L102}

\bibitem[\protect\citeauthoryear{{Renzo} et~al.,}{{Renzo}
  et~al.}{2019b}]{Renzo2019}
{Renzo} M.,  et~al., 2019b, \mn@doi [\aap] {10.1051/0004-6361/201833297}, \href
  {https://ui.adsabs.harvard.edu/abs/2019A&A...624A..66R} {624, A66}

\bibitem[\protect\citeauthoryear{{Repetto} \& {Nelemans}}{{Repetto} \&
  {Nelemans}}{2015}]{Repetto2015}
{Repetto} S.,  {Nelemans} G.,  2015, \mn@doi [\mnras] {10.1093/mnras/stv1753},
  \href {https://ui.adsabs.harvard.edu/abs/2015MNRAS.453.3341R} {453, 3341}

\bibitem[\protect\citeauthoryear{{Repetto}, {Davies}  \&
  {Sigurdsson}}{{Repetto} et~al.}{2012}]{Repetto2012}
{Repetto} S.,  {Davies} M.~B.,   {Sigurdsson} S.,  2012, \mn@doi [\mnras]
  {10.1111/j.1365-2966.2012.21549.x}, \href
  {https://ui.adsabs.harvard.edu/abs/2012MNRAS.425.2799R} {425, 2799}

\bibitem[\protect\citeauthoryear{{Repetto}, {Igoshev}  \& {Nelemans}}{{Repetto}
  et~al.}{2017}]{Repetto2017}
{Repetto} S.,  {Igoshev} A.~P.,   {Nelemans} G.,  2017, \mn@doi [\mnras]
  {10.1093/mnras/stx027}, \href
  {https://ui.adsabs.harvard.edu/abs/2017MNRAS.467..298R} {467, 298}

\bibitem[\protect\citeauthoryear{{Rossi}, {Marchetti}, {Cacciato}, {Kuiack}  \&
  {Sari}}{{Rossi} et~al.}{2017}]{Rossi2017}
{Rossi} E.~M.,  {Marchetti} T.,  {Cacciato} M.,  {Kuiack} M.,   {Sari} R.,
  2017, \mn@doi [\mnras] {10.1093/mnras/stx098}, \href
  {https://ui.adsabs.harvard.edu/abs/2017MNRAS.467.1844R} {467, 1844}

\bibitem[\protect\citeauthoryear{{Sana} et~al.,}{{Sana}
  et~al.}{2012}]{Sana2012}
{Sana} H.,  et~al., 2012, \mn@doi [Science] {10.1126/science.1223344}, \href
  {https://ui.adsabs.harvard.edu/abs/2012Sci...337..444S} {337, 444}

\bibitem[\protect\citeauthoryear{{Sana} et~al.,}{{Sana}
  et~al.}{2013}]{Sana2013}
{Sana} H.,  et~al., 2013, \mn@doi [\aap] {10.1051/0004-6361/201219621}, \href
  {https://ui.adsabs.harvard.edu/abs/2013A&A...550A.107S} {550, A107}

\bibitem[\protect\citeauthoryear{{Sandquist}, {Taam}, {Chen}, {Bodenheimer}  \&
  {Burkert}}{{Sandquist} et~al.}{1998}]{Sandquist1998}
{Sandquist} E.~L.,  {Taam} R.~E.,  {Chen} X.,  {Bodenheimer} P.,   {Burkert}
  A.,  1998, \mn@doi [\apj] {10.1086/305778}, \href
  {https://ui.adsabs.harvard.edu/abs/1998ApJ...500..909S} {500, 909}

\bibitem[\protect\citeauthoryear{{Schneider} et~al.,}{{Schneider}
  et~al.}{2014}]{Schneider2014}
{Schneider} F.~R.~N.,  et~al., 2014, \mn@doi [\apj]
  {10.1088/0004-637X/780/2/117}, \href
  {https://ui.adsabs.harvard.edu/abs/2014ApJ...780..117S} {780, 117}

\bibitem[\protect\citeauthoryear{{Schneider}, {Izzard}, {Langer}  \& {de
  Mink}}{{Schneider} et~al.}{2015}]{Schneider2015}
{Schneider} F.~R.~N.,  {Izzard} R.~G.,  {Langer} N.,   {de Mink} S.~E.,  2015,
  \mn@doi [\apj] {10.1088/0004-637X/805/1/20}, \href
  {https://ui.adsabs.harvard.edu/abs/2015ApJ...805...20S} {805, 20}

\bibitem[\protect\citeauthoryear{{Schoettler}, {Parker}, {Arnold}, {Grimmett},
  {de Bruijne}  \& {Wright}}{{Schoettler} et~al.}{2019}]{Schoettler2019}
{Schoettler} C.,  {Parker} R.~J.,  {Arnold} B.,  {Grimmett} L.~P.,  {de
  Bruijne} J.,   {Wright} N.~J.,  2019, \mn@doi [\mnras]
  {10.1093/mnras/stz1487}, \href
  {https://ui.adsabs.harvard.edu/abs/2019MNRAS.487.4615S} {487, 4615}

\bibitem[\protect\citeauthoryear{{Schoettler}, {de Bruijne}, {Vaher}  \&
  {Parker}}{{Schoettler} et~al.}{2020}]{Schoettler2020}
{Schoettler} C.,  {de Bruijne} J.,  {Vaher} E.,   {Parker} R.~J.,  2020, arXiv
  e-prints, \href {https://ui.adsabs.harvard.edu/abs/2020arXiv200413730S} {p.
  arXiv:2004.13730}

\bibitem[\protect\citeauthoryear{{Sch{\"o}nrich}, {Binney}  \&
  {Dehnen}}{{Sch{\"o}nrich} et~al.}{2010}]{Schonrich2010}
{Sch{\"o}nrich} R.,  {Binney} J.,   {Dehnen} W.,  2010, \mn@doi [\mnras]
  {10.1111/j.1365-2966.2010.16253.x}, \href
  {https://ui.adsabs.harvard.edu/abs/2010MNRAS.403.1829S} {403, 1829}

\bibitem[\protect\citeauthoryear{{Schwab}, {Podsiadlowski}  \&
  {Rappaport}}{{Schwab} et~al.}{2010}]{Schwab2010}
{Schwab} J.,  {Podsiadlowski} P.,   {Rappaport} S.,  2010, \mn@doi [\apj]
  {10.1088/0004-637X/719/1/722}, \href
  {https://ui.adsabs.harvard.edu/abs/2010ApJ...719..722S} {719, 722}

\bibitem[\protect\citeauthoryear{{Sesana}, {Haardt}  \& {Madau}}{{Sesana}
  et~al.}{2006}]{Sesana2006}
{Sesana} A.,  {Haardt} F.,   {Madau} P.,  2006, \mn@doi [\apj]
  {10.1086/507596}, \href
  {https://ui.adsabs.harvard.edu/abs/2006ApJ...651..392S} {651, 392}

\bibitem[\protect\citeauthoryear{{Sesana}, {Sartore}, {Devecchi}  \&
  {Possenti}}{{Sesana} et~al.}{2012}]{Sesana2012}
{Sesana} A.,  {Sartore} N.,  {Devecchi} B.,   {Possenti} A.,  2012, \mn@doi
  [\mnras] {10.1111/j.1365-2966.2012.21958.x}, \href
  {https://ui.adsabs.harvard.edu/abs/2012MNRAS.427..502S} {427, 502}

\bibitem[\protect\citeauthoryear{{Shen} et~al.,}{{Shen}
  et~al.}{2018}]{Shen2018}
{Shen} K.~J.,  et~al., 2018, \mn@doi [\apj] {10.3847/1538-4357/aad55b}, \href
  {https://ui.adsabs.harvard.edu/abs/2018ApJ...865...15S} {865, 15}

\bibitem[\protect\citeauthoryear{{Sherwin}, {Loeb}  \& {O'Leary}}{{Sherwin}
  et~al.}{2008}]{Sherwin2008}
{Sherwin} B.~D.,  {Loeb} A.,   {O'Leary} R.~M.,  2008, \mn@doi [\mnras]
  {10.1111/j.1365-2966.2008.13097.x}, \href
  {https://ui.adsabs.harvard.edu/abs/2008MNRAS.386.1179S} {386, 1179}

\bibitem[\protect\citeauthoryear{{Shklovskii}}{{Shklovskii}}{1970}]{Shklovskii1970}
{Shklovskii} I.~S.,  1970, \sovast, \href
  {https://ui.adsabs.harvard.edu/abs/1970SvA....13..562S} {13, 562}

\bibitem[\protect\citeauthoryear{{Silva} \& {Napiwotzki}}{{Silva} \&
  {Napiwotzki}}{2011}]{Silva2011}
{Silva} M.~D.~V.,  {Napiwotzki} R.,  2011, \mn@doi [\mnras]
  {10.1111/j.1365-2966.2010.17864.x}, \href
  {https://ui.adsabs.harvard.edu/abs/2011MNRAS.411.2596S} {411, 2596}

\bibitem[\protect\citeauthoryear{{Snaith}, {Haywood}, {Di Matteo}, {Lehnert},
  {Combes}, {Katz}  \& {G{\'o}mez}}{{Snaith} et~al.}{2014}]{Snaith2014}
{Snaith} O.~N.,  {Haywood} M.,  {Di Matteo} P.,  {Lehnert} M.~D.,  {Combes} F.,
   {Katz} D.,   {G{\'o}mez} A.,  2014, \mn@doi [\apjl]
  {10.1088/2041-8205/781/2/L31}, \href
  {https://ui.adsabs.harvard.edu/abs/2014ApJ...781L..31S} {781, L31}

\bibitem[\protect\citeauthoryear{{Snaith}, {Haywood}, {Di Matteo}, {Lehnert},
  {Combes}, {Katz}  \& {G{\'o}mez}}{{Snaith} et~al.}{2015}]{Snaith2015}
{Snaith} O.,  {Haywood} M.,  {Di Matteo} P.,  {Lehnert} M.~D.,  {Combes} F.,
  {Katz} D.,   {G{\'o}mez} A.,  2015, \mn@doi [\aap]
  {10.1051/0004-6361/201424281}, \href
  {https://ui.adsabs.harvard.edu/abs/2015A&A...578A..87S} {578, A87}

\bibitem[\protect\citeauthoryear{{Soberman}, {Phinney}  \& {van den
  Heuvel}}{{Soberman} et~al.}{1997}]{Soberman1997}
{Soberman} G.~E.,  {Phinney} E.~S.,   {van den Heuvel} E.~P.~J.,  1997, \aap,
  \href {https://ui.adsabs.harvard.edu/abs/1997A&A...327..620S} {327, 620}

\bibitem[\protect\citeauthoryear{{Socrates}, {Blaes}, {Hungerford}  \&
  {Fryer}}{{Socrates} et~al.}{2005}]{Socrates2005}
{Socrates} A.,  {Blaes} O.,  {Hungerford} A.,   {Fryer} C.~L.,  2005, \mn@doi
  [\apj] {10.1086/431786}, \href
  {https://ui.adsabs.harvard.edu/abs/2005ApJ...632..531S} {632, 531}

\bibitem[\protect\citeauthoryear{{Sukhbold}, {Ertl}, {Woosley}, {Brown}  \&
  {Janka}}{{Sukhbold} et~al.}{2016}]{Sukhbold2016}
{Sukhbold} T.,  {Ertl} T.,  {Woosley} S.~E.,  {Brown} J.~M.,   {Janka} H.~T.,
  2016, \mn@doi [\apj] {10.3847/0004-637X/821/1/38}, \href
  {https://ui.adsabs.harvard.edu/abs/2016ApJ...821...38S} {821, 38}

\bibitem[\protect\citeauthoryear{{Sutantyo}}{{Sutantyo}}{1975}]{Sutantyo1975}
{Sutantyo} W.,  1975, \aap, \href
  {https://ui.adsabs.harvard.edu/abs/1975A&A....41...47S} {41, 47}

\bibitem[\protect\citeauthoryear{{Tauris}}{{Tauris}}{2015}]{Tauris2015}
{Tauris} T.~M.,  2015, \mn@doi [\mnras] {10.1093/mnrasl/slu189}, \href
  {https://ui.adsabs.harvard.edu/abs/2015MNRAS.448L...6T} {448, L6}

\bibitem[\protect\citeauthoryear{{Tauris} \& {Takens}}{{Tauris} \&
  {Takens}}{1998}]{Tauris1998}
{Tauris} T.~M.,  {Takens} R.~J.,  1998, \aap, \href
  {https://ui.adsabs.harvard.edu/abs/1998A&A...330.1047T} {330, 1047}

\bibitem[\protect\citeauthoryear{{Tetzlaff}, {Neuh{\"a}user}  \&
  {Hohle}}{{Tetzlaff} et~al.}{2011}]{Tetzlaff2011}
{Tetzlaff} N.,  {Neuh{\"a}user} R.,   {Hohle} M.~M.,  2011, \mn@doi [\mnras]
  {10.1111/j.1365-2966.2010.17434.x}, \href
  {https://ui.adsabs.harvard.edu/abs/2011MNRAS.410..190T} {410, 190}

\bibitem[\protect\citeauthoryear{{Tillich}, {Przybilla}, {Scholz}  \&
  {Heber}}{{Tillich} et~al.}{2009}]{Tillich2009}
{Tillich} A.,  {Przybilla} N.,  {Scholz} R.-D.,   {Heber} U.,  2009, \mn@doi
  [\aap] {10.1051/0004-6361/200913173}, \href
  {https://ui.adsabs.harvard.edu/abs/2009A%26A...507L..37T} {507, L37}

\bibitem[\protect\citeauthoryear{{Tout}, {Aarseth}, {Pols}  \&
  {Eggleton}}{{Tout} et~al.}{1997}]{Tout1997}
{Tout} C.~A.,  {Aarseth} S.~J.,  {Pols} O.~R.,   {Eggleton} P.~P.,  1997,
  \mn@doi [\mnras] {10.1093/mnras/291.4.732}, \href
  {https://ui.adsabs.harvard.edu/abs/1997MNRAS.291..732T} {291, 732}

\bibitem[\protect\citeauthoryear{{Ugliano}, {Janka}, {Marek}  \&
  {Arcones}}{{Ugliano} et~al.}{2012}]{Ugliano2012}
{Ugliano} M.,  {Janka} H.-T.,  {Marek} A.,   {Arcones} A.,  2012, \mn@doi
  [\apj] {10.1088/0004-637X/757/1/69}, \href
  {https://ui.adsabs.harvard.edu/abs/2012ApJ...757...69U} {757, 69}

\bibitem[\protect\citeauthoryear{{Vanbeveren}, {Mennekens}, {van den Heuvel}
  \& {Van Bever}}{{Vanbeveren} et~al.}{2020}]{Vanbeveren2020}
{Vanbeveren} D.,  {Mennekens} N.,  {van den Heuvel} E. P.~J.,   {Van Bever} J.,
   2020, \mn@doi [A\&A] {10.1051/0004-6361/201937253}, 636, A99

\bibitem[\protect\citeauthoryear{{Verbunt} \& {Cator}}{{Verbunt} \&
  {Cator}}{2017}]{VerbuntCator2017}
{Verbunt} F.,  {Cator} E.,  2017, \mn@doi [Journal of Astrophysics and
  Astronomy] {10.1007/s12036-017-9474-5}, \href
  {https://ui.adsabs.harvard.edu/abs/2017JApA...38...40V} {38, 40}

\bibitem[\protect\citeauthoryear{{Verbunt}, {Igoshev}  \& {Cator}}{{Verbunt}
  et~al.}{2017}]{Verbunt2017}
{Verbunt} F.,  {Igoshev} A.,   {Cator} E.,  2017, \mn@doi [\aap]
  {10.1051/0004-6361/201731518}, \href
  {https://ui.adsabs.harvard.edu/abs/2017A&A...608A..57V} {608, A57}

\bibitem[\protect\citeauthoryear{{Vigna-G{\'o}mez} et~al.,}{{Vigna-G{\'o}mez}
  et~al.}{2018}]{VignaGomez2018}
{Vigna-G{\'o}mez} A.,  et~al., 2018, \mn@doi [\mnras] {10.1093/mnras/sty2463},
  \href {https://ui.adsabs.harvard.edu/abs/2018MNRAS.481.4009V} {481, 4009}

\bibitem[\protect\citeauthoryear{{Wang} \& {Han}}{{Wang} \&
  {Han}}{2009}]{Wang2009HVS}
{Wang} B.,  {Han} Z.,  2009, \mn@doi [\aap] {10.1051/0004-6361/200913326},
  \href {https://ui.adsabs.harvard.edu/abs/2009A&A...508L..27W} {508, L27}

\bibitem[\protect\citeauthoryear{{Wang}, {Meng}, {Chen}  \& {Han}}{{Wang}
  et~al.}{2009}]{Wang2009SNIa}
{Wang} B.,  {Meng} X.,  {Chen} X.,   {Han} Z.,  2009, \mn@doi [\mnras]
  {10.1111/j.1365-2966.2009.14545.x}, \href
  {https://ui.adsabs.harvard.edu/abs/2009MNRAS.395..847W} {395, 847}

\bibitem[\protect\citeauthoryear{{Webbink}}{{Webbink}}{1984}]{Webbink1984}
{Webbink} R.~F.,  1984, \mn@doi [\apj] {10.1086/161701}, \href
  {https://ui.adsabs.harvard.edu/abs/1984ApJ...277..355W} {277, 355}

\bibitem[\protect\citeauthoryear{{Webbink}}{{Webbink}}{2008}]{Webbink2008}
{Webbink} R.~F.,  2008, {Common Envelope Evolution Redux}.
p.~233, \mn@doi{10.1007/978-1-4020-6544-6_13}

\bibitem[\protect\citeauthoryear{{Wheeler}, {Lecar}  \& {McKee}}{{Wheeler}
  et~al.}{1975}]{Wheeler1975}
{Wheeler} J.~C.,  {Lecar} M.,   {McKee} C.~F.,  1975, \mn@doi [\apj]
  {10.1086/153771}, \href
  {https://ui.adsabs.harvard.edu/abs/1975ApJ...200..145W} {200, 145}

\bibitem[\protect\citeauthoryear{{Willems}, {Henninger}, {Levin}, {Ivanova},
  {Kalogera}, {McGhee}, {Timmes}  \& {Fryer}}{{Willems}
  et~al.}{2005}]{Willems2005}
{Willems} B.,  {Henninger} M.,  {Levin} T.,  {Ivanova} N.,  {Kalogera} V.,
  {McGhee} K.,  {Timmes} F.~X.,   {Fryer} C.~L.,  2005, \mn@doi [\apj]
  {10.1086/429557}, \href
  {https://ui.adsabs.harvard.edu/abs/2005ApJ...625..324W} {625, 324}

\bibitem[\protect\citeauthoryear{{Williams}, {Belokurov}, {Casey}  \&
  {Evans}}{{Williams} et~al.}{2017}]{Williams2017}
{Williams} A.~A.,  {Belokurov} V.,  {Casey} A.~R.,   {Evans} N.~W.,  2017,
  \mn@doi [\mnras] {10.1093/mnras/stx508}, \href
  {https://ui.adsabs.harvard.edu/abs/2017MNRAS.468.2359W} {468, 2359}

\bibitem[\protect\citeauthoryear{{Wongwathanarat}, {Janka}  \&
  {M{\"u}ller}}{{Wongwathanarat} et~al.}{2013}]{Wongwathanarat2013}
{Wongwathanarat} A.,  {Janka} H.~T.,   {M{\"u}ller} E.,  2013, \mn@doi [\aap]
  {10.1051/0004-6361/201220636}, \href
  {https://ui.adsabs.harvard.edu/abs/2013A&A...552A.126W} {552, A126}

\bibitem[\protect\citeauthoryear{{Woosley}}{{Woosley}}{1987}]{Woosley1987}
{Woosley} S.~E.,  1987, in {Helfand} D.~J.,  {Huang} J.~H.,  eds,  IAU
  Symposium Vol. 125, The Origin and Evolution of Neutron Stars. p.~255

\bibitem[\protect\citeauthoryear{{Wyrzykowski} \& {Mandel}}{{Wyrzykowski} \&
  {Mandel}}{2020}]{Wyrzykowski2020}
{Wyrzykowski} {\L}.,  {Mandel} I.,  2020, \mn@doi [\aap]
  {10.1051/0004-6361/201935842}, \href
  {https://ui.adsabs.harvard.edu/abs/2020A&A...636A..20W} {636, A20}

\bibitem[\protect\citeauthoryear{{Wyrzykowski} et~al.,}{{Wyrzykowski}
  et~al.}{2004}]{Wyrzykowski2004}
{Wyrzykowski} L.,  et~al., 2004, \actaa, \href
  {https://ui.adsabs.harvard.edu/abs/2004AcA....54....1W} {54, 1}

\bibitem[\protect\citeauthoryear{{Wyrzykowski} et~al.,}{{Wyrzykowski}
  et~al.}{2016}]{Wyrzykowski2016}
{Wyrzykowski} {\L}.,  et~al., 2016, \mn@doi [\mnras] {10.1093/mnras/stw426},
  \href {https://ui.adsabs.harvard.edu/abs/2016MNRAS.458.3012W} {458, 3012}

\bibitem[\protect\citeauthoryear{{Wysocki}, {Gerosa}, {O'Shaughnessy},
  {Belczynski}, {Gladysz}, {Berti}, {Kesden}  \& {Holz}}{{Wysocki}
  et~al.}{2018}]{Wysocki2018}
{Wysocki} D.,  {Gerosa} D.,  {O'Shaughnessy} R.,  {Belczynski} K.,  {Gladysz}
  W.,  {Berti} E.,  {Kesden} M.,   {Holz} D.~E.,  2018, \mn@doi [\prd]
  {10.1103/PhysRevD.97.043014}, \href
  {https://ui.adsabs.harvard.edu/abs/2018PhRvD..97d3014W} {97, 043014}

\bibitem[\protect\citeauthoryear{{Xu}, {Newberg}, {Carlin}, {Liu}, {Deng},
  {Li}, {Sch{\"o}nrich}  \& {Yanny}}{{Xu} et~al.}{2015}]{Xu2015}
{Xu} Y.,  {Newberg} H.~J.,  {Carlin} J.~L.,  {Liu} C.,  {Deng} L.,  {Li} J.,
  {Sch{\"o}nrich} R.,   {Yanny} B.,  2015, \mn@doi [\apj]
  {10.1088/0004-637X/801/2/105}, \href
  {https://ui.adsabs.harvard.edu/abs/2015ApJ...801..105X} {801, 105}

\bibitem[\protect\citeauthoryear{{Yu} \& {Madau}}{{Yu} \&
  {Madau}}{2007}]{Yu2007}
{Yu} Q.,  {Madau} P.,  2007, \mn@doi [\mnras]
  {10.1111/j.1365-2966.2007.12034.x}, \href
  {https://ui.adsabs.harvard.edu/abs/2007MNRAS.379.1293Y} {379, 1293}

\bibitem[\protect\citeauthoryear{{Yu} \& {Tremaine}}{{Yu} \&
  {Tremaine}}{2003}]{Yu2003}
{Yu} Q.,  {Tremaine} S.,  2003, \mn@doi [\apj] {10.1086/379546}, \href
  {https://ui.adsabs.harvard.edu/abs/2003ApJ...599.1129Y} {599, 1129}

\bibitem[\protect\citeauthoryear{{Zapartas} et~al.,}{{Zapartas}
  et~al.}{2017}]{Zapartas2017}
{Zapartas} E.,  et~al., 2017, \mn@doi [\apj] {10.3847/1538-4357/aa7467}, \href
  {https://ui.adsabs.harvard.edu/abs/2017ApJ...842..125Z} {842, 125}

\bibitem[\protect\citeauthoryear{{Zhang}, {Lu}  \& {Yu}}{{Zhang}
  et~al.}{2013}]{Zhang2013}
{Zhang} F.,  {Lu} Y.,   {Yu} Q.,  2013, \mn@doi [\apj]
  {10.1088/0004-637X/768/2/153}, \href
  {https://ui.adsabs.harvard.edu/abs/2013ApJ...768..153Z} {768, 153}

\bibitem[\protect\citeauthoryear{{Zheng} et~al.,}{{Zheng}
  et~al.}{2014}]{Zheng2014}
{Zheng} Z.,  et~al., 2014, \mn@doi [\apjl] {10.1088/2041-8205/785/2/L23}, \href
  {https://ui.adsabs.harvard.edu/abs/2014ApJ...785L..23Z} {785, L23}

\bibitem[\protect\citeauthoryear{{Zhong} et~al.,}{{Zhong}
  et~al.}{2014}]{Zhong2014}
{Zhong} J.,  et~al., 2014, \mn@doi [\apjl] {10.1088/2041-8205/789/1/L2}, \href
  {https://ui.adsabs.harvard.edu/abs/2014ApJ...789L...2Z} {789, L2}

\bibitem[\protect\citeauthoryear{{Zorotovic}, {Schreiber}, {G{\"a}nsicke}  \&
  {Nebot G{\'o}mez-Mor{\'a}n}}{{Zorotovic} et~al.}{2010}]{Zorotovic2010}
{Zorotovic} M.,  {Schreiber} M.~R.,  {G{\"a}nsicke} B.~T.,   {Nebot
  G{\'o}mez-Mor{\'a}n} A.,  2010, \mn@doi [\aap] {10.1051/0004-6361/200913658},
  \href {https://ui.adsabs.harvard.edu/abs/2010A&A...520A..86Z} {520, A86}

\bibitem[\protect\citeauthoryear{{Zuo}}{{Zuo}}{2015}]{Zuo2015}
{Zuo} Z.-Y.,  2015, \mn@doi [\aap] {10.1051/0004-6361/201424604}, \href
  {https://ui.adsabs.harvard.edu/abs/2015A&A...573A..58Z} {573, A58}

\bibitem[\protect\citeauthoryear{{Zuo} \& {Li}}{{Zuo} \& {Li}}{2014}]{Zuo2014}
{Zuo} Z.-Y.,  {Li} X.-D.,  2014, \mn@doi [\mnras] {10.1093/mnras/stu993}, \href
  {https://ui.adsabs.harvard.edu/abs/2014MNRAS.442.1980Z} {442, 1980}

\bibitem[\protect\citeauthoryear{{de Kool}}{{de Kool}}{1990}]{deKool1990}
{de Kool} M.,  1990, \mn@doi [\apj] {10.1086/168974}, \href
  {https://ui.adsabs.harvard.edu/abs/1990ApJ...358..189D} {358, 189}

\bibitem[\protect\citeauthoryear{{de Mink} \& {Belczynski}}{{de Mink} \&
  {Belczynski}}{2015}]{deMink2015}
{de Mink} S.~E.,  {Belczynski} K.,  2015, \mn@doi [\apj]
  {10.1088/0004-637X/814/1/58}, \href
  {http://adsabs.harvard.edu/abs/2015ApJ...814...58D} {814, 58}

\bibitem[\protect\citeauthoryear{{de Mink}, {Pols}  \& {Hilditch}}{{de Mink}
  et~al.}{2007}]{deMink2007}
{de Mink} S.~E.,  {Pols} O.~R.,   {Hilditch} R.~W.,  2007, \mn@doi [\aap]
  {10.1051/0004-6361:20067007}, \href
  {http://adsabs.harvard.edu/abs/2007A%26A...467.1181D} {467, 1181}

\bibitem[\protect\citeauthoryear{{de Mink}, {Brott}, {Cantiello}, {Izzard},
  {Langer}  \& {Sana}}{{de Mink} et~al.}{2012}]{deMink2012}
{de Mink} S.~E.,  {Brott} I.,  {Cantiello} M.,  {Izzard} R.~G.,  {Langer} N.,
  {Sana} H.,  2012, {Challenges for Understanding the Evolution of Massive
  Stars: Rotation, Binarity, and Mergers}.
p.~65

\bibitem[\protect\citeauthoryear{{de Mink}, {Langer}, {Izzard}, {Sana}  \& {de
  Koter}}{{de Mink} et~al.}{2013}]{deMink2013}
{de Mink} S.~E.,  {Langer} N.,  {Izzard} R.~G.,  {Sana} H.,   {de Koter} A.,
  2013, \mn@doi [\apj] {10.1088/0004-637X/764/2/166}, \href
  {https://ui.adsabs.harvard.edu/abs/2013ApJ...764..166D} {764, 166}

\bibitem[\protect\citeauthoryear{{van Oijen}}{{van Oijen}}{1989}]{vanOijen1989}
{van Oijen} J.~G.~J.,  1989, \aap, \href
  {https://ui.adsabs.harvard.edu/abs/1989A&A...217..115V} {217, 115}

\bibitem[\protect\citeauthoryear{{van den Heuvel}, {Portegies Zwart}  \& {de
  Mink}}{{van den Heuvel} et~al.}{2017}]{VandenHeuvel2017}
{van den Heuvel} E.~P.~J.,  {Portegies Zwart} S.~F.,   {de Mink} S.~E.,  2017,
  \mn@doi [\mnras] {10.1093/mnras/stx1430}, \href
  {https://ui.adsabs.harvard.edu/abs/2017MNRAS.471.4256V} {471, 4256}

\makeatother
\end{thebibliography}

\vspace{-30pt}
\appendix
\section{Kinematics of Unbound Neutron Stars and Bound Black Hole X-ray Binaries}
\begin{figure*}
    \includegraphics[width=1.\columnwidth]{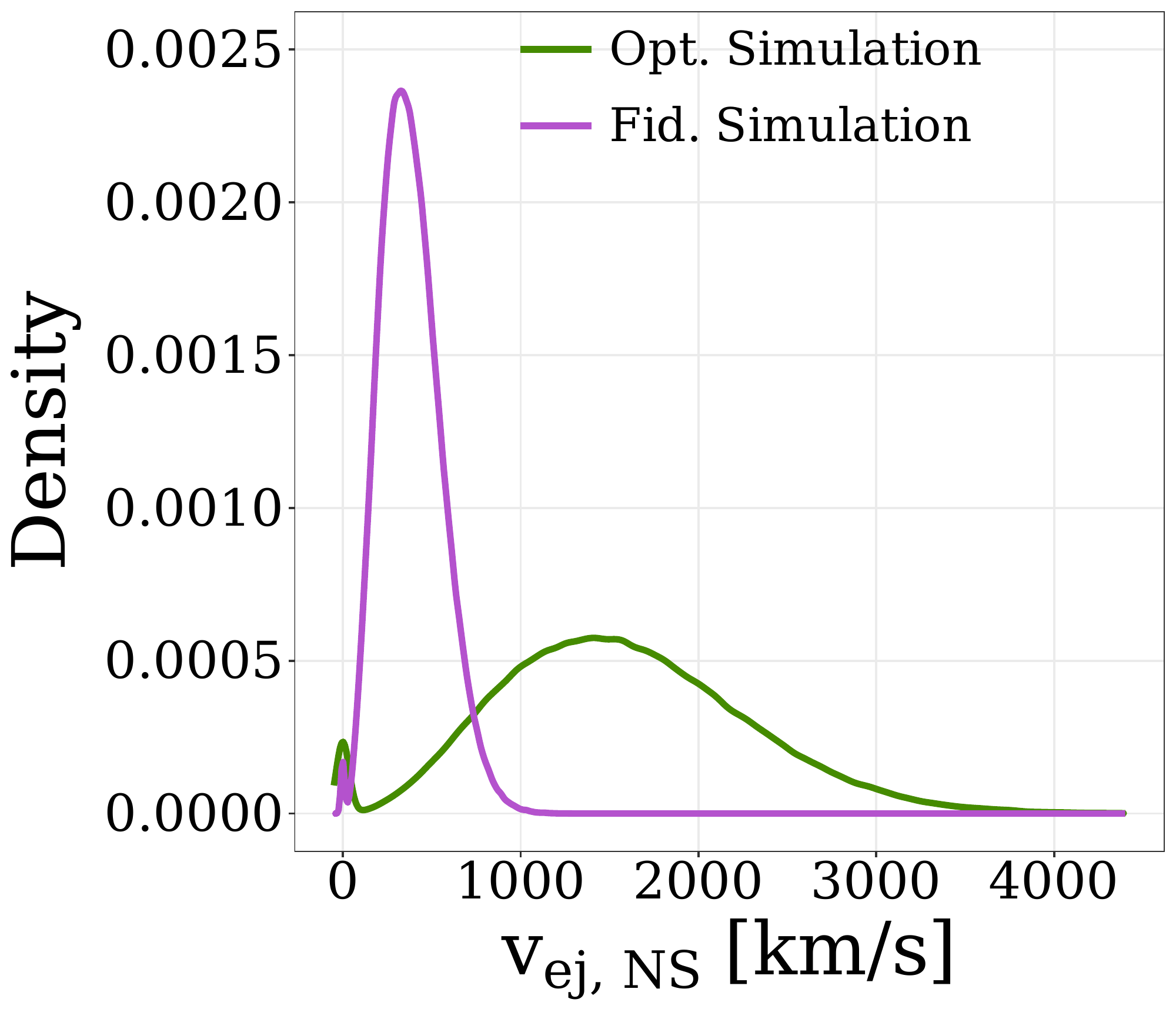}
    \includegraphics[width=1.\columnwidth]{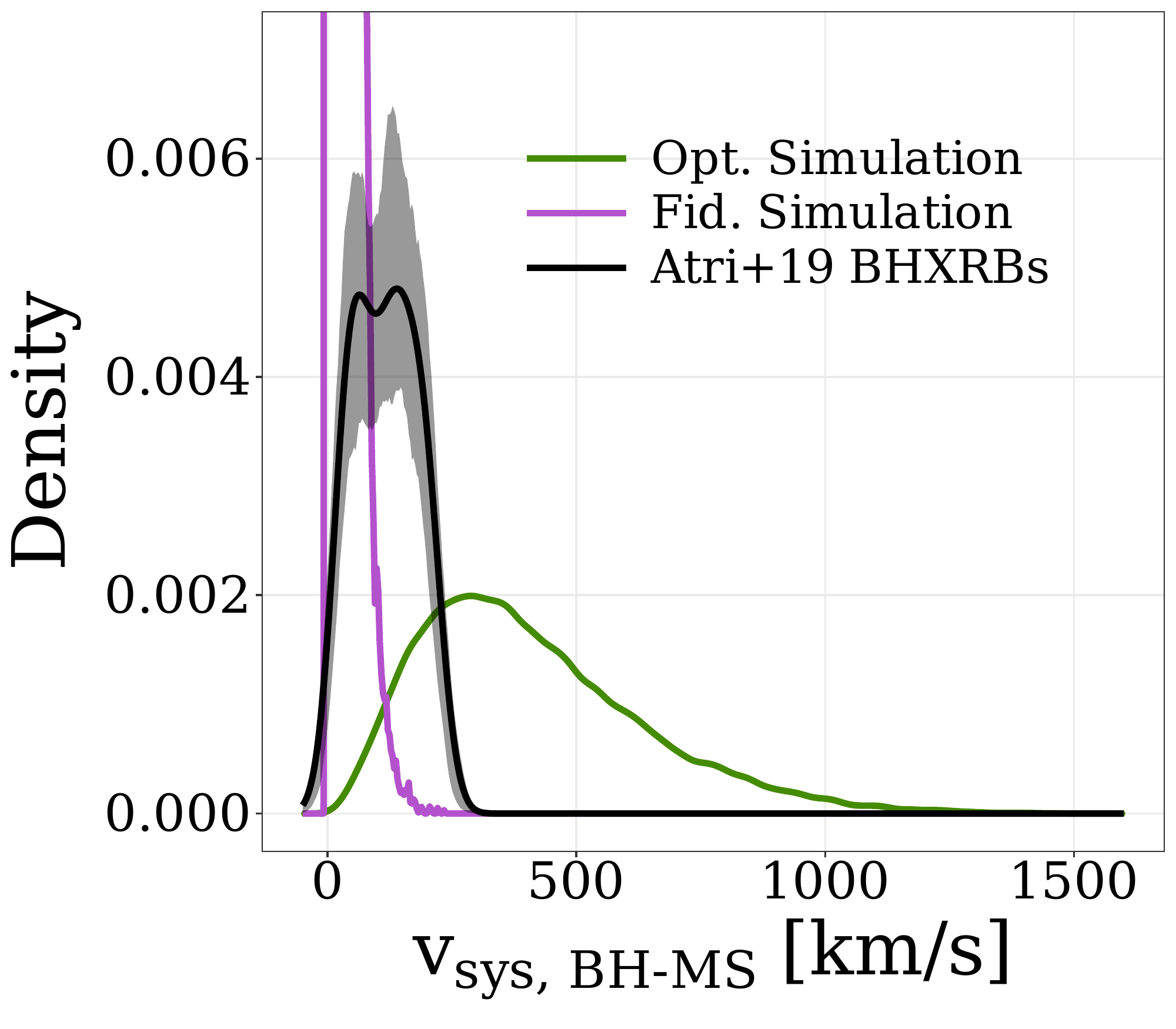}
    \caption{\textit{Left:}  probability-weighted ejection velocity distribution for neutron stars ejected from unbound binaries in the fiducial (violet) and optimized (green) simulations. \textit{Right:} probability-weighted post-CC systemic velocity distributions for remaining bound black hole-main sequence binaries in the optimized (green) and fiducial (violet, cut off for visibility) simulations. Black curve shows distribution of disc-crossing velocities derived by \citet{Atri2019} for 16 Galactic black hole X-ray binaries. Shaded region shows $1\sigma$ bootstrapped confidence intervals.}
    \label{fig:vNS}
\end{figure*}
Fig. \ref{fig:vNS} shows velocity distributions for both neutron stars ejected from unbound binaries (left) and systemic velocities of black hole X-ray binaries which remain bound post-CC. Distributions are shown for both the fiducial and optimized simulations. See Sec. \ref{sec:beyond} for more details.


\bsp	
\label{lastpage}
\end{document}